\documentclass[preprint, 3p, table]{elsarticle}

\makeatletter
\def\ps@pprintTitle{%
    \let\@oddhead\@empty
    \let\@evenhead\@empty
    \def\@oddfoot{}%
    \let\@evenfoot\@oddfoot}
\makeatother

\usepackage[english]{babel}
\usepackage[utf8]{inputenc}
\usepackage[dvipsnames]{xcolor}
\usepackage[colorinlistoftodos]{todonotes}
\usepackage{amsmath}
\usepackage{amssymb}
\usepackage{amsthm} 
\usepackage{mathtools}
\usepackage{mathrsfs}
\usepackage{upgreek}
\usepackage{siunitx}
\usepackage{gensymb}
\usepackage{comment}

\usepackage[hidelinks]{hyperref}
\usepackage{algorithm}
\usepackage[noend]{algpseudocode}

\usepackage{subfig}

\usepackage{tikz}

\usepackage{tabularx}
\usepackage{multirow}
\usepackage{booktabs}

\usepackage{empheq}

\theoremstyle{plain}
\theoremstyle{plain}
\theoremstyle{plain}
\theoremstyle{plain}
\theoremstyle{plain}
\theoremstyle{definition}
\theoremstyle{remark}

\everymath{\displaystyle}

\usepackage{hhline}
\usepackage{highlight}

\usepackage[toc,page]{appendix}
\usepackage[capitalise,nameinlink]{cleveref}
\crefname{supp}{Supplement}{Supplements}

\usepackage{wasysym} 
\newif\ifdouble

\newcommand{\papertitle}{
Defining myocardial fiber bundle architecture in atrial digital twins
}

\newcommand{\keywordOne}{Cardiac digital twin}
\newcommand{\keywordFive}{Precision medicine}
\newcommand{\keywordTwo}{Computational modeling}
\newcommand{\keywordFour}{Numerical simulations}
\newcommand{\keywordThree}{Cardiac fiber architecture}

\definecolor{lifex}{HTML}{f60248}
\newcommand{\lifex}{\texttt{life\textsuperscript{\color{lifex}{x}}}}

\newcommand{\fZero}{\mathbf{f}}
\newcommand{\sZero}{\mathbf{s}}
\newcommand{\nZero}{\mathbf{n}}

\newcommand{\EPdiffTens}{\boldsymbol{D}}

\newcommand{\EPpot}{u_{\text{A}}}

\graphicspath{{./images/}}
\doublefalse

\begin{document}	
    
    \begin{frontmatter}
        \title{{\papertitle}}
        

\author[mox,advance]{Roberto~Piersanti\corref{cor1}}\ead{roberto.piersanti@polimi.it}
\author[advance,lehigh]{Ryan~Bradley}
\author[jhu]{Syed~Yusuf~Ali}
\author[mox,epfl]{Alfio~Quarteroni}
\author[mox]{Luca~Dede'}
\author[advance,jhu]{Natalia~A.~Trayanova}

\address[mox]{MOX - Laboratory of Modeling and Scientific Computing, Dipartimento di Matematica, Politecnico di Milano, Milano, Italy}
\address[advance]{ADVANCE - Alliance for Cardiovascular Diagnostic and Treatment Innovation, Johns Hopkins University, Baltimore, USA}
\address[lehigh]{Research Computing, Lehigh University, Bethlehem, Pennsylvania, USA}
\address[jhu]{Department of Biomedical Engineering, Johns Hopkins University, Baltimore, USA}
\address[epfl]{Mathematics Institute, \'{E}cole Polytechnique F\'{e}d\'{e}rale de Lausanne, Lausanne, Switzerland (\textit{Professor Emeritus})}

\cortext[cor1]{Corresponding author.}
        
        \begin{abstract}
            A key component in developing atrial digital twins (ADT) - virtual representations of patients' atria - is the accurate prescription of myocardial fibers which are essential for the tissue characterization. Due to the difficulty of reconstructing atrial fibers from medical imaging, a widely used strategy for fiber generation in ADT relies on mathematical models. Existing methodologies utilze semi-automatic approaches, are tailored to specific morphologies, and lack rigorous validation against imaging fiber data. 
In this study, we introduce a novel atrial Laplace-Dirichlet-Rule-Based Method (LDRBM) for prescribing highly detailed myofiber orientations and providing robust regional annotation in bi-atrial morphologies of any complexity. The robustness of our approach is verified in eight extremely detailed bi-atrial geometries, derived from a sub-millimiter Diffusion-Tensor-Magnetic-Resonance Imaging (DTMRI) human atrial fiber dataset. We validate the LDRBM by quantitatively recreating each of the DTMRI fiber architectures: a comprehensive comparison with DTMRI ground truth data is conducted, investigating differences between electrophysiology (EP) simulations provided by either LDRBM and DTMRI fibers. Finally, we demonstrate that the novel LDRBM outperforms current state-of-the-art fiber models, confirming the exceptional accuracy of our methodology and the critical importance of incorporating detailed fiber orientations in EP simulations.
Ultimately, this work represents a fundamental step toward the development of physics-based digital twins of the human atria, establishing a new standard for prescribing fibers in ADT.
        \end{abstract}
        
        
    \end{frontmatter}
    
    \clearpage
\section{Introduction}
\label{sec:introduction}
Atrial fibrillation (AF), characterized by uncoordinated cardiac activation, is the most common sustained electrical dysfunction of the heart~\cite{korantzopoulos2018inflammation,yamamoto2022atrial}, and is associated with prominent morbidity and mortality worldwide~\cite{bhat2020drivers,lippi2021global,bizhanov2023atrial}. 
Despite significant technological and medical advancements, current clinical treatments for AF remain suboptimal~\cite{goette2019ehra,guerra2024current} due to a limited understanding of the complex atrial anatomical substrates which directly sustain AF~\cite{wang2019robust,trayanova2024computational}. This is partly because treatments are not personalized to individual patients~\cite{aronis2019role}. 
Recently, the development of atrial digital twins (ADT) $-$ virtual representations of patients' atria that integrate computational models with patient speciﬁc anatomical and functional data $-$ has provided important insight in the mechanisms underlying AF~\cite{boyle2019computationally,roney2020silico,pagani2021computational,trayanova2023computational,sakata2024assessing}. With recent advancements in high-performance computing, ADT are currently in the early stage of clinical translation~\cite{aronis2019role,boyle2019computationally,roney2020silico,pagani2021computational,trayanova2023computational,sakata2024assessing,morrison2018advancing,gharaviri2020epicardial,luongo2021machine}. ADT are beginning to play a pivotal role in personalized risk assessment, showing promising capabilities in predicting optimal ablation targets for AF~\cite{boyle2019computationally,sakata2024assessing,morrison2018advancing,luongo2021machine}.

A crucial aspect of developing ADT revolves around accurately representing the arrangement of myocardial fibers, also known as myofibers, which are essential for the tissue characterization~\cite{whittaker2019investigation,piersanti2021modeling,falkenberg2022identifying,kamali2023contribution}. 
Aggregations of myofibers dictates how the electric potential propagates within the muscle~\cite{roberts1979influence,zhao2012image,maesen2013rearrangement}, exhibiting a propagation velocity three-four times faster along the fiber direction than along its orthogonal plane~\cite{kanai1995optical}. Moreover, also the muscle mechanical contraction, induced by electrical activation, heavily relies on the fiber architecture~\cite{carreras2012left,papadacci2017imaging,piersanti20223d,zappon2024integrated}.
Therefore, it is imperative to incorporate the most accurate fiber information possible into ADT. Extremely precise fiber architecture in personalized ADT would enhance the accuracy of these modeling efforts intended for clinical translation.

Over the years, many histo-anatomical studies have explored the fiber arrangement in the atria, revealing a highly intricate texture musculature~\cite{papez1920heart,thomas1959muscular,wang1995architecture,ho2002atrial,ho2009importance,sanchez2013standardized}. 
Atrial fibers are characterized by the presence of multiple overlapping bundles crossing and running along different directions throughout the cardiac chambers~\cite{ho2009importance,sanchez2013standardized,pashakhanloo2016myofiber}.
Nowadays, comprehensive myofiber information can be obtained through advanced imaging modalities, such as diffusion tensor magnetic resonance imaging (DTMRI)~\cite{pashakhanloo2016myofiber,zhao2017three}, micro-computed tomography~\cite{gonzalez2017whole}, shear wave imaging~\cite{lee2011mapping}, 
or optical mapping~\cite{giardini2021mesoscopic}. These techniques have proven to be valuable in determining myofiber patterns in ex-vivo hearts~\cite{pashakhanloo2016myofiber,zhao2017three}. However, in-vivo fiber identification~\cite{nielles2013vivo,froeling2014diffusion,nguyen2016vivo} are still constrained by a relatively coarse spatial resolution~\cite{toussaint2013vivo}. 
To date, the most comprehensive resource of information regarding the human atrial myofiber structure is an ex-vivo sub-millimeter resolution DTMRI fiber dataset~\cite{pashakhanloo2016myofiber}.
Nonetheless, this imaging technique required approximately 50 hours for scanning each atrium~\cite{pashakhanloo2016myofiber}.
For these considerations, contemporary fiber-imaging techniques are practically unusable in the construction of patient-specific ADT, which are models built from geometrical data acquired in-vivo.

In the past decade, multiscale ADT have achieved a high level of biophysical modeling details, incorporating precision in atrial anatomy, tissue characterization, and fibrosis distribution~\cite{boyle2019computationally,aronis2019role,dossel2012computational,mcdowell2012methodology,ferrer2015detailed,zahid2016patient,nagel2021biatria,roney2023constructing,azzolin2023augmenta,zingaro2024comprehensive}. 
However, due to the difficulties of acquiring patient-specific fiber data, different methodologies have been proposed to prescribe realistic myocardial fibers for ADT~\cite{piersanti2021modeling,ferrer2015detailed,roney2023constructing,ruiz2022physics}. 
These are classified in two main groups: atlas-based-methods (ABMs)~\cite{mcdowell2012methodology,roney2023constructing,satriano2013feature,labarthe2014bilayer, roney2019universal,hoermann2019automatic,roney2021constructing} and rule-based-methods (RBMs)~\cite{piersanti2021modeling,ferrer2015detailed,krueger2011modeling,labarthe2012semi,tobon2013three,wachter2015mesh,fastl2018personalized,saliani2019visualization}. ABMs are based on establishing a mapping between the patient's geometry and an oversimplified atlas morphology, previously reconstructed using imaging data or histological information~\cite{roney2019universal,hoermann2019automatic}.
Exploiting this mapping procedure, myofibers are directly transferred from the atlas onto the specific geometry. 
Nevertheless, ABMs strongly rely on the original data upon which the atlas were constructed, and they are tailored for specific morphologies without considering the highly variability of the atrial anatomy. 
Conversely, RBMs prescribe fiber orientations through mathematical rules inferred from histological-imaging observations, requiring information solely about the cardiac geometry~\cite{piersanti2021modeling,labarthe2012semi}. 
Classical atrial RBMs rely on manual~\cite{ferrer2015detailed,saliani2019visualization} or semi-automatic~\cite{labarthe2012semi,fastl2018personalized} approaches. The former require a considerable amount of hands-on intervention, by introducing several landmarks, seed-points and auxiliaries lines~\cite{krueger2011modeling,fastl2018personalized}.
Modern RBMs, known as Laplace-Dirichlet RBMs (LDRBMs)~\cite{piersanti2021modeling,bayer2012novel} have been specifically developed for the atrial chambers~\cite{piersanti2021modeling,zheng2021automate,rossi2022rule,mountris2022meshless}. 
In LDRBMs, both the fiber fields and atrial regions are determined by solving suitable Laplace boundary-value problems~\cite{piersanti2021modeling}.
These methods showed very promising results by representing atrial fibers in diverse morphologies~\cite{nagel2021biatria,azzolin2023augmenta,rossi2022rule},  extending their applicability to encompass geometries at the scale of the whole heart~\cite{piersanti2022modeling,fedele2023comprehensive,zingaro2023electromechanics}. 
However, they often neglect important fiber bundles and primarily treat the left (LA) and right (RA) atrium as two separate entities, frequently assuming homogeneous fiber transmurality or a simplified monolayer structure~\cite{piersanti2021modeling}. Additionally, LDRBMs lack rigorous validation against the same geometry used for acquiring imaging fiber data.

Motivated by the unresolved issues formally described, here we present a novel atrial LDRBM for prescribing myofiber orientations and providing annotation of anatomical regions in both atria. We begin by introducing the novel atrial LDRBM (Section~\ref{subsec:ldrbm_model}), capable of modeling
a highly biophysically detailed fiber architecture in volumetric bi-atrial morphologies of any complexity with a highly automated procedure.
Building on the foundation of the first atrial LDRBM~\cite{piersanti2021modeling, piersanti2021phd},
and incorporating key improvements for both LA~\cite{rossi2022rule} and RA~\cite{fedele2023comprehensive}, this method leverages several inter/intra-atrial distances, by solving suitable Laplace-Dirichlet problems, effectively decomposing the bi-atrial anatomy into characteristic anatomical bundle-regions. 
Solution gradients of these distances establish a local orthonormal coordinate axis system in each bundle. The latter is then utilized to construct and appropriately modulate a volumetric two-layer myofiber field. 
The method requires the definition of common boundary sets (e.g., endocardium, epicardium, atrioventricular valves and veins rings) and only four landmark points. The assignment of boundary labels is accomplished through an automatic pipeline, ensuring precision, reproducibility and high usability.

To demonstrate the robustness of our approach, we replicate the fiber orientation in eight highly intricate bi-atrial geometries, derived from the original DTMRI human atrial fiber dataset presented in~\cite{pashakhanloo2016myofiber}. We establish a systematic measurement procedure (Section~\ref{subsec:fiber_measurement}), exploiting the LDRBM reference axis system, to quantify the local myocardial fiber angle across the atrial wall. Then, we validate the novel LDRBM by quantitatively recreating the eight DTMRI fiber architectures. A comprehensive comparison between LDRBM fiber reconstructions and DTMRI ground truth data is conducted, analyzing differences in fiber orientations (Section~\ref{subsec:fiber_generation}). Then, numerical electrophysiology (EP) simulations of the electrical wave propagation, provided by both LDRBM and DTMRI fibers, are evaluated on the same geometries (Section~\ref{subsec:ep_simulations}).
Finally, the results obtained by our novel bi-atrial LDRBM, both for the fiber directions and EP simulations, are compared with two state-of-the-art atrial fiber generation models~\cite{piersanti2021modeling, roney2021constructing} against ground truth DTMRI fibers (Section~\ref{subsec:models_comparison}).

    \section{Methods}
\label{sec:model}
In this section, we provide a detailed description of the novel bi-atrial LDRBM for the generation of atrial myofibers (Section~\ref{subsec:ldrbm_model}).
\subsection{The bi-atrial Laplace-Dirichlet rules-based method}
\label{subsec:ldrbm_model}
\begin{figure}[t!]
    \centering
    \includegraphics[width=1\textwidth]{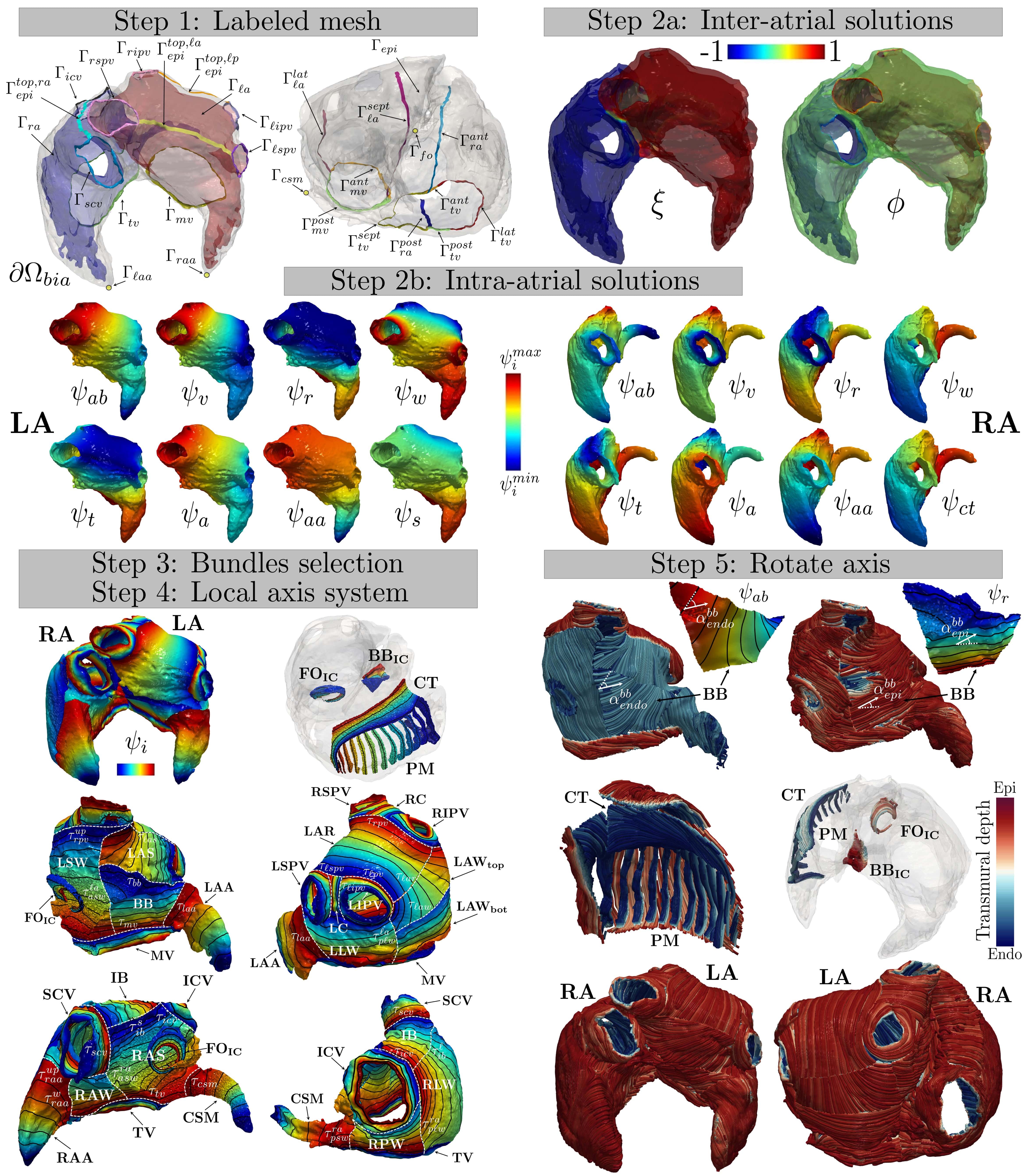}
    \caption{Schematic representation of the atrial LDRBM in a real bi-atrial geometry (derived from the DTMRI human atrial fiber dataset~\cite{pashakhanloo2016myofiber}).}
    \label{fig:Figure_ldrbm_pipeline}
\end{figure}

The propesed bi-atrial LDRBM is defined by the following steps which are hereby reported:
\begin{description}
    \item[1. Labeled mesh:] provide a labeled mesh of the atrial domain $\Omega_{bia}$ to identify specific partitions of the atrial boundary $\partial \Omega_{bia}$, see Step 1 in Figure~\ref{fig:Figure_ldrbm_pipeline}; 
    \item[2. Laplace solutions:] define several inter-atrial and intra-atrial distances by solving specific Laplace-Dirichlet boundary value problems, see Steps 2a$-$2b in Figure~\ref{fig:Figure_ldrbm_pipeline};	
    \item[3. Bundles selection:] divide the atrial domain $\Omega_{bia}$ in several anatomical subregion, named bundles, by establishing their dimensions according to the rules reported in Algorithms~\ref{BIA-LDRBM}--\ref{RA-LDRBM}, see Step 3 in Figure~\ref{fig:Figure_ldrbm_pipeline}. During this step the gradients of the inter/intra-atrial distances are used to build a {\sl transmural} $\boldsymbol{\gamma}$ and {\sl normal} $\boldsymbol{k}$ directions;
    \item[4. Local coordinate axis system:] construct an orthonormal coordinate axis system for each point of the atrial domain. This system comprises unit vectors representing the ``flat'' myofiber field, including the {\sl transmural} $\widehat{\boldsymbol{e}}_t$, {\sl normal} $\widehat{\boldsymbol{e}}_n$, and {\sl longitudinal} $\widehat{\boldsymbol{e}}_l$ direction (orthogonal to the former ones), see Step~4 in Figure~\ref{fig:Figure_ldrbm_pipeline}; 
    \item[5. Rotate axis:] rotate the reference frame with the purpose of defining the myofiber orientations, composed by the {\sl fiber} $\boldsymbol{f}$, the {\sl sheet-normal} $\boldsymbol{n}$ and the {\sl sheet} $\boldsymbol{s}$ directions, see Step 5 in Figure~\ref{fig:Figure_ldrbm_pipeline}. Rotations are chosen in order to match histology observations and/or DTMRI measurements.
\end{description} 

In what follows, we fully detail the five steps of the bi-atrial LDRBM. We refer to Figure~\ref{fig:Figure_ldrbm_pipeline} for a schematic representation of the method in a real geometry (retrieved from the DTMRI fiber dataset~\cite{pashakhanloo2016myofiber}).

\vspace{5pt}
\noindent
\textbf{1. Labeled mesh:} provide a labeled mesh of the bi-atrial computational domain $\Omega_{bia}$ to define the following boundary partitions $\partial\Omega_{bia}$ (see Step 1 in Figure~\ref{fig:Figure_ldrbm_pipeline})
\begin{equation*}
    \partial \Omega_{bia} = \Gamma_{\ell a} \cup \Gamma_{ra} \cup \Gamma_{epi} \cup \Gamma_{rpv} \cup \Gamma_{\ell pv} \cup \Gamma_{scv} \cup \Gamma_{icv} \cup \Gamma_{mv} \cup \Gamma_{tv} \cup \Gamma_{\ell aa} \cup \Gamma_{raa} \cup \Gamma_{fo} \cup \Gamma_{csm},
\end{equation*}
where $\Gamma_{\ell a}$, $\Gamma_{ra}$ are the LA and RA endocardium, $\Gamma_{epi}$ the atrial epicardium, $\Gamma_{rpv}$, $\Gamma_{\ell pv}$ the right (RPV) and left (LPV) pulmonary vein rings, $\Gamma_{scv}$, $\Gamma_{icv}$ the superior (SCV) and inferior (ICV) caval vein rings, $\Gamma_{mv}$, $\Gamma_{tv}$ the mitral (MV) and tricuspid (TV) valve rings, $\Gamma_{raa}$, $\Gamma_{\ell aa}$ the left (LAA) and right (RAA) atrial appendage apices, $\Gamma_{fo}$ the fossa ovalis (FO) centre, and $\Gamma_{csm}$ the coronary sinus muscle (CSM) apex. In particular, $\Gamma_{\ell a}$ ($\Gamma_{ra}$) is divided into the septal $\Gamma^{sept}_{\ell a}$ (anterior $\Gamma^{ant}_{ra}$) band, the lateral $\Gamma^{lat}_{\ell a}$ (posterior $\Gamma^{post}_{ra}$) band, and the remaining endocardial part $\Gamma^{endo}_{\ell a}$ ($\Gamma^{endo}_{ra}$) such that $\Gamma_{\ell a} = \Gamma_{\ell a}^{endo} \cup \Gamma_{\ell a}^{sept} \cup \Gamma_{\ell a}^{lat}$ ($\Gamma_{ra} = \Gamma_{ra}^{endo} \cup \Gamma_{ra}^{ant} \cup \Gamma_{ra}^{post}$). Moreover, $\Gamma_{epi}$ is splitted into $\Gamma_{epi} = \Gamma^{-}_{epi} \cup \Gamma_{epi}^{top,\ell a} \cup \Gamma_{epi}^{top,lp} \cup \Gamma_{epi}^{top,ra}$, where $\Gamma_{epi}^{top,\ell a}/\Gamma_{epi}^{top,lp}$ ($\Gamma_{epi}^{top,ra}$) is a boundary label connecting the upper region of anterior/posterior LPV (SCV) to  anterior/posterior RPV (ICV) rings, and $\Gamma^{-}_{epi}$ is the remaining epicardial surface. Furthermore, $\Gamma_{*pv}$ (with $*=\ell,r$) encloses the left/right superior pulmonay vein ring (LSPV/RSPV) $\Gamma_{*spv}$ and the left/right inferior pulmonary vein ring (LIPV/RIPV) $\Gamma_{*ipv}$, such that $\Gamma_{*pv} =  \Gamma_{*spv} \cup \Gamma_{*ipv}$. Finally, for LA $\Gamma_{mv} = \Gamma_{mv}^{ant} \cup \Gamma_{mv}^{post}$, where $\Gamma_{mv}^{ant}$ and $\Gamma_{mv}^{post}$ are the ring sections facing the anterior and posterior wall, respectively; whereas, for RA $\Gamma_{tv} = \Gamma_{tv}^{ant} \cup \Gamma_{tv}^{post} \cup \Gamma_{tv}^{sept} \cup \Gamma_{tv}^{lat}$, where $\Gamma_{tv}^{ant}$, $\Gamma_{tv}^{post}$, $\Gamma_{tv}^{sept}$, and $\Gamma_{tv}^{lat}$ are the ring portions related to the anterior, posterior, septal and lateral wall, respectively, see Step 1 in Figure~\ref{fig:Figure_ldrbm_pipeline}. A detailed description about the labeling procedure on a generic bi-atrial geometry is given in Appendix~\ref{supp:tagging}.

\vspace{5pt}
\noindent
\textbf{2. Laplace solutions:} define several inter/intra-atrial distances obtained by solving Laplace problems with proper Dirichlet boundary conditions on the atrial boundaries (see Steps 2a$-$2b in Figure~\ref{fig:Figure_ldrbm_pipeline}) in the form 
\begin{equation}
    \label{eqn:laplace}
    \begin{cases}
        -\Delta \chi=0 &\qquad{\text{in }}\Omega_{bia},
        \\
        \chi = \chi_a &\qquad{\text{on }}\Gamma_a,
        \\
        \chi= \chi_b &\qquad{\text{on }}\Gamma_b,
        \\
        \nabla \chi \cdot \boldsymbol{n}=0 &\qquad{\text{on }}\Gamma_n,
    \end{cases}
\end{equation}
for a generic unknown $\chi$ and suitable boundary data $\chi_a$, $\chi_b \in \mathbb{R}$ set on generic partitions of the atrial boundary $\Gamma_a$, $\Gamma_b$, $\Gamma_n$, with $\Gamma_a \cup \Gamma_b \cup \Gamma_n = \partial\Omega_{bia}$ and $\Gamma_n=\partial\Omega_{bia}/(\Gamma_a \cup \Gamma_b)$. The values $\chi_a$, $\chi_b$ are set in order to evaluate specific inter/intra-atrial distances between boundary partitions $\Gamma_a$, $\Gamma_b$. Refer to Table~\ref{tab:bc_step2} for the specific choices in problem~\eqref{eqn:laplace} made by the bi-atrial LDRBM. Specifically, the inter-atrial $\xi$ (from LA to RA) and the transmural $\phi$ (from endocardium to epicardium) distances are introduced, see Step 2a in Figure~\ref{fig:Figure_ldrbm_pipeline}. 
To define a consistent transmural distance, two auxiliaries Laplace solutions $\phi_{\ell a}=1-\phi$ and $\phi_{ra}=1+\phi$ are introduced for LA and RA, respectively. 
Furthermore, several intra-atrial distances $\psi_i$ (with $i=ab, v, r, w, t, a, aa, s, ct$) are computed, see Step 2b in Figure~\ref{fig:Figure_ldrbm_pipeline}. Explicitly, $\psi_{ab}$ is a solution of the problem \eqref{eqn:laplace} with different boundary data prescribed on LPV, RPV, MV, LAA for LA and IVC, SVC, TV, RAA, and CSM for RA; $\psi_v$ represents the distance among the pulmonary veins for LA and between the caval veins for RA; $\psi_r$ stands for the distance between MV/TV ring and the union of the pulmonary/caval veins and the top epicardial bands of LA/RA; $\psi_w$ is the distance from the superior pulmonary veins and the anterior ring of MV to the inferior pulmonary veins and the posterior ring of MV for LA, and between the septal and the lateral ring of TV, passing from the top epicardial band for RA; $\psi_t$ is the distance between MV/TV ring and the top epicaradial bands of LA/RA; $\psi_a$ represent the distance from the septal/anterior to lateral/posterior wall of LA/RA; $\psi_{aa}$ stands for the distance from FO to LAA for LA and from the posterior wall to RAA for RA. Finally, $\psi_s$ and $\psi_{ct}$ are  computed solely for LA and RA, respectively: $\psi_s$ is solved by prescribing suitable boundary conditions for the anterior and posterior parts of MV ring, the anterior and posterior epicardial bands, and the lateral and septal endocardial bands; $\psi_{ct}$ is the distance from the anterior, posterior and septal part of TV ring to the lateral one.
\begin{table}[t!]
    \centering
    \scalebox{1.0}{\begin{tabular}{ |c|c|c|c|c|c| } 
            \hline
            Type & $\chi$ & $\chi_{a}$ & $\Gamma_{a}$ & $\chi_{b}$ & $\Gamma_{b}$ \\
            \hline
            \hline
            \multirow{3}{*}{BIA} & $\xi$ & 1 & $\Gamma_{\ell a}$ & -1 & $\Gamma_{ra}$  \\\cline{2-6} 
            & \multirow{2}{*}{$\phi$} & 1 & $\Gamma_{\ell a}$ & \multirow{2}{*}{0} & \multirow{2}{*}{$\Gamma_{epi}$} \\
            & & -1 & $\Gamma_{ra}$ & &  \\\cline{2-6} 	
            \hline
            \hline
            \multirow{10}{*}{LA}  
            & \multirow{2}{*}{$\psi_{ab}$} & 2 & $\Gamma_{rpv}$ & 1 & $\Gamma_{mv}$ \\ 
            & & 0 & $\Gamma_{\ell pv}$ & -1 & $\Gamma_{\ell aa}$ \\\cline{2-6} 
            & $\psi_v$ & 1 & $\Gamma_{rpv}$ & 0 & $\Gamma_{\ell pv}$ \\\cline{2-6} 
            & $\psi_{r}$ & 1 & $\Gamma_{mv}$ & 0 & $\Gamma_{\ell pv} \cup \Gamma_{rpv} \cup \Gamma^{top, \ell a}_{epi} \cup \Gamma^{top, \ell p}_{epi}$ \\\cline{2-6} 
            & $\psi_{w}$ & 1 & $\Gamma_{rspv} \cup \Gamma_{\ell spv} \cup \Gamma^{ant}_{mv}$  & -1 & $\Gamma_{ripv} \cup \Gamma_{\ell ipv} \cup \Gamma^{post}_{mv}$ \\\cline{2-6} 
            & $\psi_{t}$ & 1 & $\Gamma_{mv}$  & 0 & $\Gamma^{top,\ell a}_{epi} \cup \Gamma^{top,\ell p}_{epi}$ \\\cline{2-6} 
            & $\psi_{a}$ & 1 & $\Gamma^{sept}_{\ell a}$  & 0 & $\Gamma^{lat}_{\ell a}$ \\\cline{2-6} 
            & $\psi_{aa}$ & 1 & $\Gamma_{fo}$  & -2 & $\Gamma_{\ell aa}$ \\\cline{2-6} 
            & \multirow{2}{*}{$\psi_{s}$} & 1 & $\Gamma^{ant}_{mv}$ & -0.5 & $\Gamma^{top,\ell p}_{epi}$  \\ 
            & & 0 & $\Gamma^{top,\ell a}_{epi} \cup \Gamma^{sept}_{\ell a} \cup \Gamma^{lat}_{\ell a}$ & -1 & $\Gamma^{post}_{mv}$ \\\cline{2-6} 
            \hline
            \hline
            \multirow{12}{*}{RA}  
            & \multirow{2}{*}{$\psi_{ab}$} & 2 & $\Gamma_{icv}$ & 1 & $\Gamma_{tv}$ \\ 
            & & 0 & $\Gamma_{scv}$ & -2 & $\Gamma_{raa} \cup \Gamma_{csm}$ \\\cline{2-6} 
            & $\psi_v$ & 1 & $\Gamma_{ivc}$ & 0 & $\Gamma_{svc}$ \\\cline{2-6} 
            & $\psi_{r}$ & 1 & $\Gamma_{tv}$ & 0 & $\Gamma_{ivc} \cup \Gamma_{svc} \cup \Gamma^{top, ra}_{epi}$ \\\cline{2-6} 
            & \multirow{2}{*}{$\psi_{w}$} & 1 & $\Gamma^{sept}_{tv}$ & \multirow{2}{*}{0} & \multirow{2}{*}{$\Gamma^{top,ra}_{epi}$} \\ 
            & & -1 & $\Gamma^{lat}_{tv}$ & & \\\cline{2-6} 
            & $\psi_{t}$ & 1 & $\Gamma_{tv}$  & 0 & $\Gamma^{top,ra}_{epi}$ \\\cline{2-6} 
            & \multirow{2}{*}{$\psi_{a}$} & 1 & $\Gamma^{ant}_{ra} \cup \Gamma_{svc}$ & \multirow{2}{*}{2} & \multirow{2}{*}{$\Gamma_{raa}$} \\ 
            & & -1 & $\Gamma^{post}_{ra} \cup \Gamma_{ivc}$ & & \\\cline{2-6} 
            & \multirow{2}{*}{$\psi_{aa}$} & 1 & $\Gamma_{ivc} \cup \Gamma^{post}_{ra}$ & \multirow{2}{*}{0} & \multirow{2}{*}{$\Gamma_{svc}$}  \\ 
            & & -2 & $\Gamma_{raa}$ & & \\\cline{2-6} 
            & $\psi_{ct}$ & 1 & $\Gamma^{sept}_{ra} \cup \Gamma^{ant}_{ra} \cup \Gamma^{post}_{ra}$ & -1 & $\Gamma^{lat}_{ra}$ \\\cline{2-6} 
            \hline
    \end{tabular}}
    \caption[Boundary data chosen in the atrial LDRBM]{Boundary data chosen in the Laplace problem~\eqref{eqn:laplace} for the inter-atrial distances (BIA) $\xi$, $\phi$ and the intra-atrial distances (LA/RA) $\psi_i$ with $i=ab, v, r, w, t, a, aa, s, ct$.}
    \label{tab:bc_step2}
\end{table}

\vspace{4pt}
\noindent
\textbf{3. Bundles selection:} define the atrial bundles and their dimensions throughout the domain $\Omega_{bia}$, see Step~3 in Figure~\ref{fig:Figure_ldrbm_pipeline}. With this aim, the bi-atrial LDRBM first selects the inter-atrial connections (IC), following the rules reported in Algorithm~\ref{BIA-LDRBM} ($\textbf{compute}_{\textbf{BIA}}$) and then compute LA and RA bundles, exploiting the rules reported in Algorithms~\ref{LA-LDRBM} ($\textbf{compute}_{\textbf{LA}}$) and~\ref{RA-LDRBM} ($\textbf{compute}_{\textbf{RA}}$). Algorithms~\ref{BIA-LDRBM}--\ref{RA-LDRBM} identify the principal anatomical atrial regions: for IC, the bachmann’s bundle ($\text{BB}_{\text{IC}}$), the fossa ovalis ($\text{FO}_{\text{IC}}$), and the coronary sinus ($\text{CS}_{\text{IC}}$) inter-atrial connections; for LA, mitral valve (MV), left and right inferior and superior pulmonary veins (LIPV, RIPV, LSPV, RSPV),  left (LC) and right (RC) carina, left atrial appendage (LAA), left anterior septum (LAS), left lateral wall (LLW), left atrial septum (LSW), left top ($\text{LAW}_{\text{top}}$) and bottom ($\text{LAW}_{\text{bot}}$) atrial posterior wall, left atrial roof (LAR), and bachmann’s bundle (BB); for RA, tricuspid valve (TV), inferior (ICV) and superior (SCV) caval veins, right atrial appendage (RAA), coronary sinus musculature (CSM), right atrial septum (RAS), right lateral wall (RLW), right anterior wall (RAW), right posterior wall (RPW),
inter-caval bundle (IB), crista terminalis~(CT), and pectinate muscles (PM), see Step 3 in Figure~\ref{fig:Figure_ldrbm_pipeline}.

To specify the bundle dimensions, the threshold parameters $\tau_{i}$ are introduced: for IC, $\tau^{ic,r}_{bb}$, $\tau^{ic,\ell}_{bb}$, $\tau^{ic}_{bb}$, $\tau^{ic,r}_{fo}$, $\tau^{ic,\ell}_{fo}$, $\tau^{ic}_{fo}$, $\tau^{ic,in}_{fo}$, $\tau^{ic,r}_{cs}$, $\tau^{ic,\ell}_{cs}$, $\tau^{ic}_{cs}$ referring to $\text{BB}_{\text{IC}}$, $\text{FO}_{\text{IC}}$, and $\text{CS}_{\text{IC}}$ connections; for LA, $\tau_{mv}$, $\tau_{\ell pv}$, $\tau^{up}_{\ell pv}$, $\tau_{\ell ipv}$, $\tau_{\ell spv}$, $\tau_{rpv}$, $\tau^{up}_{rpv}$, $\tau_{ripv}$, $\tau_{rspv}$, $\tau_{\ell aa}$, $\tau_{\ell as}$, $\tau^{\ell a}_{p\ell w}$, $\tau^{\ell a,up}_{p \ell w}$, $\tau^{\ell a}_{a \ell w}$, $\tau^{\ell a}_{psw}$, $\tau^{\ell a,up}_{psw}$, $\tau^{\ell a}_{asw}$, $\tau_{\ell ar}$, $\tau_{\ell aw}$, $\tau_{bb}$ referring to MV, LIPV, LSPV, RIPV, RSPV, LAA, LAS, LLW, LSW, LAR, LAW, BB, respectively; for RA, $\tau_{tv}$, $\tau^{aa}_{tv}$, $\tau_{icv}$, $\tau^{up}_{icv}$, $\tau_{scv}$, $\tau^{up}_{scv}$, $\tau_{raa}$, $\tau^w_{raa}$, $\tau^{up}_{raa}$, $\tau_{csm}$, $\tau^v_{csm}$, $\tau_{ras}$, $\tau^{ra}_{asw}$, $\tau^{ra,up}_{asw}$, $\tau^{ra}_{a\ell w}$, $\tau^{ra}_{psw}$, $\tau^{ra,up}_{psw}$, $\tau^{ra}_{p\ell w}$, $\tau^{ra,up}_{p\ell w}$, $\tau^s_{ib}$, $\tau^{\ell}_{ib}$, $\tau^+_{ct}$, $\tau^-_{ct}$, and $\tau^\phi_{ct}$ referring to TV, ICV, SCV, RAA, CSM, RAS, RAW, RPW, IB, CT, respectively. Finally, the atrial LDRBM allows to embed PMs in RLW, which requires to specify the parameters $\text{pm}_{tk}$, $\text{pm}_{rg}$, $\text{pm}_{end}$, and $N_{\text{pm}}$ related to the thickness, range interval, final position and the numbers of PM, respectively, see Step 3 in Figure~\ref{fig:Figure_ldrbm_pipeline}. 
\begin{algorithm}[t!]
    \caption{$\textbf{compute}_{\textbf{BIA}}$: {\sl bundles selection for bi-atrial geometry}}
    \noindent Let $\tau_{*}$, $\alpha^{*}_{endo}$, $\alpha^{*}_{epi}$, be the threshold parameter, the epicardial and the endocardial angles related to IC bundles. Moreover, let $\phi$, $\xi$ be the inter-atrial and $\psi_{*}$ the intra-atrial distances.  
    \begin{algorithmic}
        \If{$\text{BB}_\text{IC}=\text{true}$ \textbf{and} $\xi \in[\tau^{ic,r}_{bb}, \tau^{ic,\ell}_{bb}]$ \textbf{and} $\psi_{v}^{ra} \le \tau^{ic}_{bb}$} 
        
        \textbf{set(}$\nabla \xi$, $\nabla \phi$, $\alpha^{bb,ic}_{epi}$,  $\alpha^{bb,ic}_{epi}$\textbf{)} $\longrightarrow$ $\text{BB}_{\text{IC}}$
        
        \textbf{flip(}$\widehat{\boldsymbol{e}}_l$,$\widehat{\boldsymbol{e}}_t$\textbf{)}  
        
        \ElsIf{$\text{FO}_\text{IC}=\text{true}$ \textbf{and} $\xi \in[\tau^{ic,r}_{fo}, \tau^{ic,\ell}_{fo}]$ \textbf{and} $\psi_{v}^{ra} >  \tau^{ic}_{bb}$ \textbf{and} $\psi_{aa}^{\ell a} \in[\tau^{ic}_{fo}, \tau^{ic,in}_{fo}]$}
        
        \textbf{set(}$\nabla \xi$, $\nabla \psi^{\ell a}_{aa}$, $\alpha^{fo,ic}_{epi}$,$\alpha^{fo,ic}_{epi}$\textbf{)} $\longrightarrow$ $\text{FO}_{\text{IC}}$
        
        \ElsIf{$\text{CS}_\text{IC}=\text{true}$  \textbf{and} $\xi \in[\tau^{ic,r}_{cs}, \tau^{ic,\ell}_{cs}]$ \textbf{and} $\psi_{v}^{ra} >  \tau^{ic}_{bb}$ \textbf{and} $\psi_{ab}^{ra} \le  \tau^{ic}_{cs}$}
        
        \textbf{set(}$\nabla \xi$, $\nabla \phi$, $0$,$0$\textbf{)} $\longrightarrow$ $\text{CS}_{\text{IC}}$
        
        \textbf{flip(}$\widehat{\boldsymbol{e}}_l$,$\widehat{\boldsymbol{e}}_t$\textbf{)}
        \ElsIf{$\xi > 0$}
        
        
        $\textbf{compute}_{\textbf{LA}}$
        
        \Else 
        
        
        $\textbf{compute}_{\textbf{RA}}$
        
        \EndIf	
        
        \noindent	
        Note: we use $\psi_{i}^{ra}$ and $\psi_{i}^{\ell a}$ to distinguish LA and RA distances. Moreover, the function  \textbf{flip}($\widehat{\boldsymbol{e}}_l$,$\widehat{\boldsymbol{e}}_t$) flips the longitudinal $\widehat{\boldsymbol{e}}_l$ and the transmural $\widehat{\boldsymbol{e}}_t$ directions after the $\textbf{axis}$ function \eqref{eqn:axis} evaluation (see Step 4).	
    \end{algorithmic}
    \label{BIA-LDRBM}
\end{algorithm} 	
\begin{algorithm}[t!]
    \caption{$\textbf{compute}_{\textbf{LA}}$: {\sl bundles selection for LA}}
    \noindent Let $\tau_{*}$, $\alpha^{*}_{endo}$, $\alpha^{*}_{epi}$ be the threshold parameter, the epicardial and the endocardial angles related to LA bundles. Moreover, let $\phi_{\ell a}=1-\phi$ (with $\phi$ the transmural distance) and $\psi_{*}$ the LA intra-atrial distances.  
    \begin{algorithmic}
        \If{$\psi_{r} \ge$ $\tau_{mv}$}    
        \textbf{set(}$\nabla \phi$, $\nabla \psi_{r}$, $\alpha^{mv}_{endo}$, $\alpha^{mv}_{epi}$ \textbf{)} $\longrightarrow$ MV     
        \ElsIf {$\psi_v \ge \tau_{rpv}$ \textbf{and} $\psi_r \le \tau^{up}_{rpv}$} 
            \If {$\psi_w \le \tau_{ripv}$}           
            \textbf{set(}$\nabla \phi$, $\nabla \psi_{v}$, $\alpha^{ripv}_{endo}$,  $\alpha^{ripv}_{epi}$\textbf{)} $\longrightarrow$ RIPV
            \ElsIf {$\psi_w \ge \tau_{rspv}$} 
            \textbf{set(}$\nabla \phi$, $\nabla \psi_{v}$, $\alpha^{rspv}_{endo}$,  $\alpha^{rspv}_{epi}$\textbf{)} $\longrightarrow$ RSPV
            \Else
            \textbf{ set(}$\nabla \phi$, $\nabla \psi_{w}$, $\alpha^{rc}_{endo}$,  $\alpha^{rc}_{epi}$\textbf{)} $\longrightarrow$ RC
            \EndIf        
        \ElsIf {$\psi_v \le \tau_{\ell pv}$ \textbf{and} $\psi_r \le \tau^{up}_{\ell pv}$} 
            \If {$\psi_w \le \tau_{\ell ipv}$}
            \textbf{set(}$\nabla \phi$, $\nabla \psi_{v}$, $\alpha^{\ell ipv}_{endo}$,  $\alpha^{\ell ipv}_{epi}$\textbf{)} $\longrightarrow$ LIPV        
            \ElsIf {$\psi_w \ge \tau_{\ell spv}$} 
            \textbf{set(}$\nabla \phi$, $\nabla \psi_{v}$, $\alpha^{\ell spv}_{endo}$,  $\alpha^{\ell spv}_{epi}$\textbf{)} $\longrightarrow$ LSPV
            \Else
            \textbf{ set(}$\nabla \phi$, $\nabla \psi_{w}$, $\alpha^{\ell c}_{endo}$,  $\alpha^{\ell c}_{epi}$\textbf{)} $\longrightarrow$ LC			
            \EndIf
        \ElsIf {$\psi_{aa} \le \tau_{\ell aa}$}
        \textbf{set(}$\nabla \phi$, $\nabla \psi_{aa}$, $\alpha^{\ell aa}_{endo}$,  $\alpha^{\ell aa}_{epi}$\textbf{)} $\longrightarrow$ LAA           
        \ElsIf {$\psi_s \le \tau_{\ell as}$} 
            \If {$\psi_a \le \tau^{\ell a}_{p\ell w}$ \textbf{and} $\psi_{t} \ge \tau^{\ell a,up}_{p \ell w}$}
            \textbf{set(}$\nabla \phi$, $\nabla \psi_{r}$, $\alpha^{\ell w}_{endo}$,  $\alpha^{\ell w}_{epi}$\textbf{)} $\longrightarrow$ LLW
            \ElsIf {$\psi_a \le \tau^{\ell a}_{p\ell w}$ \textbf{and} $\psi_{t} < \tau^{\ell a,up}_{p \ell w}$}
            \textbf{set(}$\nabla \phi$, $\nabla \psi_{ab}$, $\alpha^{\ell ar}_{endo}$,  $\alpha^{\ell ar}_{epi}$\textbf{)} $\longrightarrow$ LAR
            \ElsIf {$\psi_a \ge \tau^{\ell a}_{psw}$ \textbf{and} $\psi_{t} \ge \tau^{\ell a,up}_{psw}$}
            \textbf{set(}$\nabla \phi$, $\nabla \psi_{r}$, $\alpha^{sw}_{endo}$,  $\alpha^{sw}_{epi}$\textbf{)} $\longrightarrow$ LSW
            \ElsIf {$\psi_a \ge \tau^{\ell a}_{psw}$ \textbf{and} $\psi_{t} < \tau^{\ell a,up}_{psw}$}
            \textbf{set(}$\nabla \phi$, $\nabla \psi_{ab}$, $\alpha^{\ell ar}_{endo}$,  $\alpha^{\ell ar}_{epi}$\textbf{)} $\longrightarrow$ LAR
            \ElsIf {$\psi_w > 0$}
            \textbf{set(}$\nabla \phi$, $\nabla \psi_{ab}$, $\alpha^{\ell ar}_{endo}$,  $\alpha^{\ell ar}_{epi}$\textbf{)} $\longrightarrow$ LAR
            \ElsIf {$\psi_t \le \tau_{\ell ar}$}
            \textbf{set(}$\nabla \phi$, $\nabla \psi_{ab}$, $\alpha^{\ell ar}_{endo}$,  $\alpha^{\ell ar}_{epi}$\textbf{)} $\longrightarrow$ LAR
            \ElsIf {$\psi_{r} \le \tau_{\ell aw}$}
            \textbf{set(}$\nabla \phi$, $\nabla \psi_v (\phi_{\ell a} < 0) + \nabla \psi_{ab} (\phi_{\ell a} \ge 0)$, $\alpha^{\ell aw,t}_{endo}$,  $\alpha^{\ell aw,t}_{epi}$\textbf{)} $\longrightarrow$ $\text{LAW}_{\text{top}}$
            \Else 
            \textbf{ set(}$\nabla \phi$, $\nabla \psi_v (\phi_{\ell a} <0) + \nabla \psi_{r} (\phi_{\ell a} \ge 0)$, $\alpha^{\ell aw,b}_{endo}$,  $\alpha^{\ell aw,b}_{epi}$\textbf{)} $\longrightarrow$ $\text{LAW}_{\text{bot}}$					
            \EndIf
        \ElsIf {$\psi_a \ge \tau^{\ell a}_{asw}$} 
        \textbf{set(}$\nabla \phi$, $\nabla \psi_{r}$, $\alpha^{sw}_{endo}$,  $\alpha^{sw}_{epi}$\textbf{)} $\longrightarrow$ LSW
        \ElsIf {$\psi_a \le \tau^{\ell a}_{a\ell w}$}
        \textbf{set(}$\nabla \phi$, $\nabla \psi_{r}$, $\alpha^{\ell w}_{endo}$,  $\alpha^{\ell w}_{epi}$\textbf{)} $\longrightarrow$ LLW
        \ElsIf {$\psi_r \ge \tau_{bb}$}
        \textbf{ set(}$\nabla \phi$, $\nabla \psi_{ab} (\phi_{\ell a} <0) + \nabla \psi_{r} (\phi_{\ell a} \ge 0)$, $\alpha^{bb}_{endo}$,  $\alpha^{bb}_{epi}$\textbf{)} $\longrightarrow$ BB
        \Else
        \textbf{ set(}$\nabla \phi$, $\nabla \psi_{ab}$, $\alpha^{\ell as}_{endo}$,  $\alpha^{\ell as}_{epi}$\textbf{)} $\longrightarrow$ LAS
        \EndIf
        
        \noindent
        Note: in $\text{LAW}_\text{top}$, $\text{LAW}_\text{bot}$ and BB, we define the normal direction as $\boldsymbol{k}=\nabla \psi_{i} (\phi_{\ell a} < 0) + \nabla \psi_{j} (\phi_{\ell a} \ge 0)$. This implies that $\boldsymbol{k}=\nabla \psi_{i}$ for $\phi_{\ell a} < 0$ (sub-endocardium) and $\boldsymbol{k}=\nabla \psi_{j}$ for $\phi_{\ell a} \ge 0$ (sub-epicardium).   
    \end{algorithmic}
    \label{LA-LDRBM}
\end{algorithm}
\begin{algorithm}[t!]
    \caption{$\textbf{compute}_{\textbf{RA}}$: {\sl bundles selection for RA}}
    \noindent \noindent Let $\tau_{*}$, $\alpha^{*}_{endo}$, $\alpha^{*}_{epi}$ be the threshold parameter, the epicardial and the endocardial angles related to RA bundles. Moreover, let $N_{\text{pm}}$, $\text{pm}_{*}$ the parameters referring to PM bundle. Finally, let $\phi_{ra}=1+\phi$ (with $\phi$~the transmural distance) and $\psi_{*}$ the RA intra-atrial distances. 
    \begin{algorithmic}
        \If{$\psi_{r} \ge$ $\tau_{tv}$ \textbf{and} $\psi_{ab} \ge$ $\tau^{aa}_{tv}$} 
        \textbf{set(}$\nabla \phi$, $\nabla \psi_{r}$, $\alpha^{tv}_{endo}$,  $\alpha^{tv}_{epi}$\textbf{)}
        $\longrightarrow$ TV 
        \ElsIf{$\psi_v \le \tau_{scv}$ \textbf{and} $\psi_{r} \le$ $\tau^{up}_{scv}$}               
        \textbf{set(}$\nabla \phi$, $\nabla \psi_{v}$, $\alpha^{scv}_{endo}$,  $\alpha^{scv}_{epi}$\textbf{)}
        $\longrightarrow$ SCV		
        \ElsIf{$\psi_v \ge \tau_{icv}$ \textbf{and} $\psi_{r} \le$ $\tau^{up}_{icv}$}
        \textbf{set(}$\nabla \phi$, $\nabla \psi_{v}$, $\alpha^{icv}_{endo}$,  $\alpha^{icv}_{epi}$\textbf{)}
        $\longrightarrow$ ICV
        \ElsIf{$\psi_{aa} \le \tau_{raa}$ \textbf{and} $\psi_{w} \le$ $\tau^w_{raa}$ \textbf{and} $\psi_{t} \ge$ $\tau^{up}_{raa}$}
        \textbf{set(}$\nabla \phi$, $\nabla \psi_{aa}$, $\alpha^{raa}_{endo}$,  $\alpha^{raa}_{epi}$\textbf{)}
        $\longrightarrow$ RAA
        \ElsIf{$\text{CSM}=\text{true}$ \textbf{and} $\psi_{ab} \le $ $\tau_{csm}$ \textbf{and} $\psi_{v} \ge $ $\tau^v_{csm}$} 
        \textbf{set(}$\nabla \phi$, $\nabla \psi_{ab}$, $\alpha^{csm}_{endo}$,  $\alpha^{csm}_{epi}$\textbf{)}
        $\longrightarrow$ CSM
        \ElsIf {$\psi_w \ge \tau_ {ras}$}         
            \If {$\psi_{a} \le \tau^{ra}_{psw}$ \textbf{and} $\psi_{t} \ge $ $\tau^{ra,up}_{psw}$}
            \textbf{set(}$\nabla \phi$, $\nabla \psi_{r}$, $\alpha^{pw}_{endo}$, $\alpha^{pw}_{epi}$\textbf{)} $\longrightarrow$ RPW
            \ElsIf {$\psi_{a} \ge \tau^{ra}_{asw}$ \textbf{and} $\psi_{t} \ge $ $\tau^{ra,up}_{asw}$}
            \textbf{set(}$\nabla \phi$, $\nabla \psi_{r}$, $\alpha^{raw}_{endo}$, $\alpha^{raw}_{epi}$\textbf{)} $\longrightarrow$ RAW
            \ElsIf {$\psi_{t} \le \tau^s_{ib}$}
            \textbf{set(}$\nabla \phi$, $\nabla \psi_{v}$, $\alpha^{ib}_{endo}$, $\alpha^{ib}_{epi}$\textbf{)} $\longrightarrow$ IB
            \Else
            \textbf{ set(}$\nabla \phi$, $\nabla \psi_{t}$, $\alpha^{ib}_{endo}$, $\alpha^{ib}_{epi}$\textbf{)} $\longrightarrow$ RAS
            \EndIf
        \ElsIf {$\psi_{a} \le \tau^{ra}_{p\ell w}$ \textbf{and} $\psi_{t} \ge $ $\tau^{ra,up}_{p\ell w}$}
        \textbf{ set(}$\nabla \phi$, $\nabla \psi_{r}$, $\alpha^{rpw}_{endo}$, $\alpha^{rpw}_{epi}$\textbf{)} $\longrightarrow$ RPW
        \ElsIf {$\psi_{a} \ge \tau^{ra}_{a\ell w}$}
        \textbf{ set(}$\nabla \phi$, $\nabla \psi_{r}$, $\alpha^{raw}_{endo}$, $\alpha^{raw}_{epi}$\textbf{)} $\longrightarrow$ RAW 
        \ElsIf {$\psi_{t} \le \tau^{\ell}_{ib}$}
        \textbf{ set(}$\nabla \phi$, $\nabla \psi_{v}$, $\alpha^{ib}_{endo}$, $\alpha^{ib}_{epi}$\textbf{)} $\longrightarrow$ IB
        \ElsIf {$\psi_{ct} \in [\tau^-_{ct},\tau^+_{ct}]$ \textbf{and} $\phi_{ra} \le $ $\tau^{\phi}_{ct}$} \textbf{set(}$\nabla \phi$, $\nabla \psi_{ct}$, $\alpha^{ct}_{endo}$, $\alpha^{ct}_{endo}$\textbf{)} $\longrightarrow$ CT
        \ElsIf {$\psi_{ct} > \tau^+_{ct}$}
        \textbf{set(}$\nabla \phi$, $\nabla \psi_{ab}$, $\alpha^{r \ell w}_{endo}$, $\alpha^{r \ell w}_{epi}$\textbf{)} $\longrightarrow$ RLW
        \Else
            \If {$\text{PM}=\text{true}$}
    
                \If {$\phi_{ra} > \tau^{\phi}_{ct}$}
                \textbf{set(}$\nabla \phi$, $\nabla \psi_{ab}$, $\alpha^{r \ell w}_{endo}$, $\alpha^{r \ell w}_{epi}$\textbf{)} $\longrightarrow$ RLW
                \EndIf
                \For{n=1:$N_{\text{pm}}$} 
    
                \qquad \textbf{$\{$} 
                \\ $\qquad \qquad \:\:
                \text{PM}_i = \tau_{raa} + (n - 1)(\text{pm}_{tk} + \text{pm}_{rg})$
    
                $\qquad \:\:
                \text{PM}_f = \text{PM}_i + \text{pm}_{tk}$
    
                $\qquad \:\: 
                \text{PM}_s = \text{PM}_f + \text{pm}_{rg}$
                
                	\If {$\psi_{raa} \in (\text{PM}_f,\text{PM}_s)$ \textbf{or} $\psi_{raa} > \text{pm}_{end}$ \textbf{or} $\psi_{raa} < \tau_{raa}$}
                	\\
                	$\qquad \qquad \qquad$ \textbf{ set(}$\nabla \phi$, $\nabla \psi_{ab}$, $\alpha^{r \ell w}_{endo}$, $\alpha^{r \ell w}_{epi}$\textbf{)} $\longrightarrow$ RLW   
                	\ElsIf {$\psi_{raa} \le \text{pm}_{end}$ \textbf{and} $\psi_{raa} \ge \text{PM}_i$ \textbf{and} $\psi_{raa} \le \text{PM}_f$}
                	\\
                    $\qquad \qquad \qquad$ \textbf{ set(}$\nabla \phi$, $\nabla \psi_{raa}$, $\alpha^{pm}_{endo}$, $\alpha^{pm}_{endo}$\textbf{)} $\longrightarrow$ PM
                    \EndIf 
                \EndFor 
                \quad $\:$  \textbf{$\}$}
            \Else
                \textbf{ set(}$\nabla \phi$, $\nabla \psi_{ab}$, $\alpha^{r \ell w}_{endo}$, $\alpha^{r \ell w}_{epi}$\textbf{)} $\longrightarrow$ RLW
            \EndIf
        \EndIf
    \end{algorithmic}
    \label{RA-LDRBM}
\end{algorithm}

During the bundles selection procedure, the LDRBM defines, for each atrial bundle, a unique transmural~$\boldsymbol{\gamma}$ and normal $\boldsymbol k$ directions, by taking the gradient of a specific inter-atrial ($\nabla \phi$, $\nabla \xi$) or intra-atrial distance ($\nabla \psi_i$, with $i=ab, v, r, w, t, a, aa, ct, s$). The method also establishes, for each bundle, two rotation angles, denoted as $\alpha_{endo}$ for the sub-endocardial layer and $\alpha_{epi}$ for the sub-epicardial one, that are used in Step 5 to properly rotate the myofiber field. This selection process is performed, through Algorithms~\ref{BIA-LDRBM}--\ref{RA-LDRBM}, using the following function
\begin{equation}
    \label{eqn:set}
    [\boldsymbol{\gamma}, \boldsymbol{k}, \alpha_{endo}, \alpha_{epi}]=\textbf{set(} \nabla \varphi, \nabla \vartheta, \alpha^{j}_{endo} , \alpha^{j}_{epi} \textbf{)}=
    \begin{cases}
        \boldsymbol{\gamma}&= \nabla \varphi 
        \\
        \boldsymbol{k}&= \nabla \vartheta
        \\
        \alpha_{endo} &= \alpha^{j}_{endo}
        \\
        \alpha_{epi} &= \alpha^{j}_{epi}
    \end{cases},
\end{equation}
where $\varphi$ and $\vartheta$ represent generic inter/intra-atrial distances, while $\alpha^{j}_{endo}$ and $\alpha^{j}_{epi}$ are the prescribed fiber rotation angles for the generic $j$-th bundle. 

The complete bundles selection procedure for the atrial LDRBM is fully detailed in Algorithms~\ref{BIA-LDRBM}--\ref{RA-LDRBM}, see also Step 3 in Figure~\ref{fig:Figure_ldrbm_pipeline}.

\vspace{2pt}
\noindent
\textbf{4. Local coordinate axis system:} build for each point of the atrial domain $\Omega_{bia}$ an orthonormal local coordinate axial system $Q$ 
\begin{equation}
    \label{eqn:axis}
    Q=[\widehat{\boldsymbol{e}}_l, \widehat{\boldsymbol{e}}_n, \widehat{\boldsymbol{e}}_t]= \textbf{axis}(\boldsymbol{\gamma},\boldsymbol{k})=
    \begin{cases}
        \widehat{\boldsymbol{e}}_t &= \frac{\boldsymbol{\gamma}}{\left\lVert \boldsymbol{\gamma} \right\rVert}
        \\
        \widehat{\boldsymbol{e}}_n &= \frac{\boldsymbol{k} - (\boldsymbol{k} \cdot \widehat{\boldsymbol{e}}_t )\widehat{\boldsymbol{e}}_t}{\left\lVert \boldsymbol{k} - (\boldsymbol{k} \cdot \widehat{\boldsymbol{e}}_t )\widehat{\boldsymbol{e}}_t \right\rVert}
        \\
        \widehat{\boldsymbol{e}}_l &= \widehat{\boldsymbol{e}}_n \times \widehat{\boldsymbol{e}}_t 
    \end{cases},
\end{equation}
composed by the unit longitudinal $\widehat{\boldsymbol{e}}_l$, transmural $\widehat{\boldsymbol{e}}_t$ and normal $\widehat{\boldsymbol{e}}_n$ directions, which represent the ``flat" myofibers field. It is important to emphasize that each atrial bundle features a distinct local coordinate axial system $Q$, which nevertheless remains consistent within the bundle itself. By construction, the $\widehat{\boldsymbol{e}}_l$ fields (i.e. the ``flat" fiber directions) consists of vectors placed along the isochrones lines of the specific intra-atrial distance $\psi_i$ chosen for that bundle, see Step 4 in Figure~\ref{fig:Figure_ldrbm_pipeline}.   

\vspace{2pt}
\noindent
\textbf{5. Rotate axis:} rotate the reference frame $Q$, defined at the previous step for each point of the atrial domain $\Omega_{bia}$, with the purpose of defining the myofiber orientations, see Step 5 in Figure~\ref{fig:Figure_ldrbm_pipeline}. This is performed by rotating the longitudinal direction $\widehat{\boldsymbol{e}}_l$ around $\widehat{\boldsymbol{e}}_t$ by means of a suitable angle $\alpha$
\begin{equation}
    \label{eqn:orient}
    [\boldsymbol{f}, \boldsymbol{n}, \boldsymbol{s}]=\textbf{orient}(Q,\alpha)=[\widehat{\boldsymbol{e}}_l, \widehat{\boldsymbol{e}}_n, \widehat{\boldsymbol{e}}_t] R_{\widehat{\boldsymbol{e}}_t}(\alpha),
\end{equation}
where $R_{\widehat{\boldsymbol{e}}_t}(\alpha)$ and $\alpha$ are given by
\begin{equation*}
    R_{\widehat{\boldsymbol{e}}_t}(\alpha)=
    \begin{bmatrix}
        \mathrm{cos}(\alpha) & -\mathrm{sin}(\alpha) & 0
        \\
        \mathrm{sin}(\alpha) & \mathrm{cos}(\alpha) & 0
        \\
        0 & 0 & 1
    \end{bmatrix}, 	
    \qquad
    \alpha = 
    \begin{cases}
        \alpha_{endo} &\text{for } \vert \phi \vert > 0 
        \\
        \alpha_{epi} &\text{for } \vert \phi \vert \le 0
    \end{cases},
\end{equation*}
with $\alpha_{endo}$, $\alpha_{epi}$ the rotation angles, on the sub-endocardial ($\vert \phi \vert > 0$) and sub-epicardial ($\vert \phi \vert \le 0$) layers, prescribed in the \textbf{set} function \eqref{eqn:set}. The resulting three unit vectors correspond to the final fiber~$\boldsymbol{f}$, sheet $\boldsymbol{s}$ and sheet-normal $\boldsymbol{n}$ directions. In this way a volumetric transmural bilayer (with a sub-endocardial and sub-epicardial) structure is prescribed in each atrial bundle, see Step 5 in Figure~\ref{fig:Figure_ldrbm_pipeline}. Rotation angles are chosen in order to match histology observations and/or DTMRI measurements.
    \section{Results}
\label{sec:results}
This section is dedicated to present numerical results both for the fiber generation and EP simulations, employing the LDRBM discussed in Section~\ref{subsec:ldrbm_model}.
We organize this section as follows. Section~\ref{subsec:settings} describes the common settings for all the simulations. Section~\ref{subsec:fiber_measurement} presents the measurement procedure, exploiting the LDRBM axis system, to assess the local myocardial fiber angle in bi-atrial geometries embedded with DTMRI data~\cite{pashakhanloo2016myofiber}. Section~\ref{subsec:fiber_generation} illustrates the realization of digital twin atrial fiber architectures usign the LDRBM to quantitatively recreate the DTMRI myofiber bundle structures. A comprehensive comparison between LDRBM fiber reconstructions and DTMRI ground truth data is conducted. Section~\ref{subsec:ep_simulations} showcases EP simulations induced by both LDRBM and DTMRI fibers. Section~\ref{subsec:models_comparison} provides a comparison of the proposed LDRBM (presented in Section~\ref{subsec:ldrbm_model}) with state-of-the-art atrial fiber models (i.e., the universal atrial coordinates ABM~\cite{roney2021constructing} and the first LDRBM~\cite{piersanti2021modeling}) against DTMRI data: we compare the fiber orientations and we analyze their discrepancies in terms of EP activation times. 
\subsection{Simulation settings}
\label{subsec:settings}
All the simulations are performed on real bi-atrial geometries processed from the original ex-vivo DTMRI fiber-geometry dataset established in~\cite{pashakhanloo2016myofiber}. This includes eight segmented geometrical models of the human atria, embedding volumetric fiber orientations at a submillimeter resolution, thus providing an unprecedented level of information on both human atrial structure and fibers (see~\cite{pashakhanloo2016myofiber} for furhter details). Being an extrimely detailed models of the human atria, it demonstrates the applicability roboustness of the proposed bi-atrial LDRBM to arbitrary patient-specific scenarios.

To build the computational mesh associated with the bi-atrial geometries, we use the Vascular Modeling Toolkit software~\texttt{vmtk} (\url{http://www.vmtk.org}) by exploiting the semi-automatic cardiac meshing tools~\cite{fedele2021polygonal} in combination with the software~\texttt{meshmixer} (\url{http://www.meshmixer.com}). The mesh generation process begins with a pre-processing step in~\texttt{meshmixer}, focusing on minimal cleaning and smoothing the atrial surfaces: LA, RA endocardium, and epicardium. This step aims to meticulously separate them while preserving their morphological structures to the fullest extent possible. Then, the surface labeling and tetrahedral volumetric Finite Element (FE) mesh generation is performed in~\texttt{vmtk}. The labeling procedure carried out in this work, for the atrial LDRBM (see Step 1 in  Section~\ref{subsec:ldrbm_model}), is fully detailed in Appendix~\ref{supp:tagging}. Finally, volumetric DTMRI fibers, embedded in the orginal bi-atrial dataset, were assigned to each nodal point of the labeled computational domains by means of linear projection using~\texttt{vmtk}.  

For representing EP activity in the atrial tissue, we employ the Eikonal-diffusion model~\cite{franzone2014mathematical,quarteroni2019mathematical} (detailed in Appendix~\ref{supp:ep_model}). The numerical approximation of the eikonal model, requires the following physical data: the velocity parameter $c_f$ and the conductivities along the myofiber directions $\sigma_f$, $\sigma_s$ and $\sigma_n$. We set $c_f=100$~$\unit{s^{-1/2}}$, $\sigma_f=1\times10^{-4}$ $\unit{m^2s^{-1}}$, and $\sigma_s=\sigma_n=0.16\times10^{-4}$~$\unit{m^2s^{-1}}$, in order to achieve the conduction velocities of 1~$\unit{\metre\per\second}$ in the fiber direction $\fZero$ and $0.4$ $\unit{\metre\per\second}$ along the sheet $\sZero$ and normal $\nZero$ directions \cite{piersanti2021modeling,ferrer2015detailed,fastl2018personalized,augustin2020impact,monaci2019computational}. Finally, to initiate the EP signal propagation, a spherical stimulus, with radius $2\times10^{-3}$~$\unit{m}$, is applied at time $t=0$ $\unit{s}$ in the Sino-Atrial-Node (SAN), which lies in the musculature of CT at the anterolateral junction with SCV \cite{sakamoto2005interatrial}. Regarding the mesh element size $h$ and the time step $\Delta t$, related to the space and time discretizations of the pseudo-time eikonal equation  (see Appendix~\ref{supp:ep_model}), we used continuous FE of order 1 on tetrahedral meshes with an average element size of $h = 6 \times 10^{-4}$ $\unit{m}$ and the Backward Difference Formulae (BDF) approximation of order 2 with a time step of $\Delta t=10^{-3}$ $\unit{s}$. We used this setting values for all the simulations reported in Sections~\ref{subsec:ep_simulations} and~\ref{subsec:models_comparison}.

The novel atrial LDRBM (see Section~\ref{subsec:ldrbm_model}), the measuring procedure (see Section~\ref{subsec:fiber_measurement}) and the Eikonal-diffusion model (see Appendix~\ref{supp:ep_model}) have been implemented and solved using \lifex{}~\cite{africa2022lifex,africa2023lifex-ep,africa2023lifex-fiber} (\url{https://lifex.gitlab.io}), an in-house high-performance C++ FE library focused on cardiac applications based on \texttt{deal.II} FE core~\cite{arndt2023dealii} (\url{https://www.dealii.org}). 

The statistical data analysis performed to estimate the variability of both the regional bundle dimension parameters (see Section~\ref{subsec:bundle_classification}) and the atrial fiber angles (see Section~\ref{subsec:fiber_assesment}) were performed in \texttt{matlab} (\url{https://www.mathworks.com}). We employed the \texttt{CircStat} toolbox~\cite{berens2009circstat} to carry out the circular fiber angle statistics, and the \texttt{CircHist} toolbox (\url{https://github.com/zifredder/CircHist}) to perform the polar angle histograms (see Appendix~\ref{supp:circular-statistic} for further details).    

To visualize the results we used ParaView (\url{https://www.paraview.org}) an open-source, multi-platform data analysis and visualization application. 

All the numerical simulations were executed on the cluster iHeart (Lenovo SR950 8x24-Core Intel Xeon Platinum 8160, 2100 MHz and 1.7 TB RAM) at MOX, Dipartimento di Matematica, Politecnico di Milano.
\subsection{Measurement procedure for atrial fiber orientations}
\label{subsec:fiber_measurement}
We present hereafter the systematic measurement procedure used to quantify the local myocardial fiber angle in each atrial bundle applied to the eight geometries of the DTMRI fiber dataset~\cite{pashakhanloo2016myofiber}. This exploits both the LDRBM bundle subdivisions and the related local coordinate axis system (see Section~\ref{subsec:ldrbm_model}) and it is characterized by the
following steps (refer to Figure~\ref{fig:measuring} for a schematic representation of the measurement procedure): 
\begin{figure}[t!]
    \centering
    \includegraphics[width=1\textwidth]{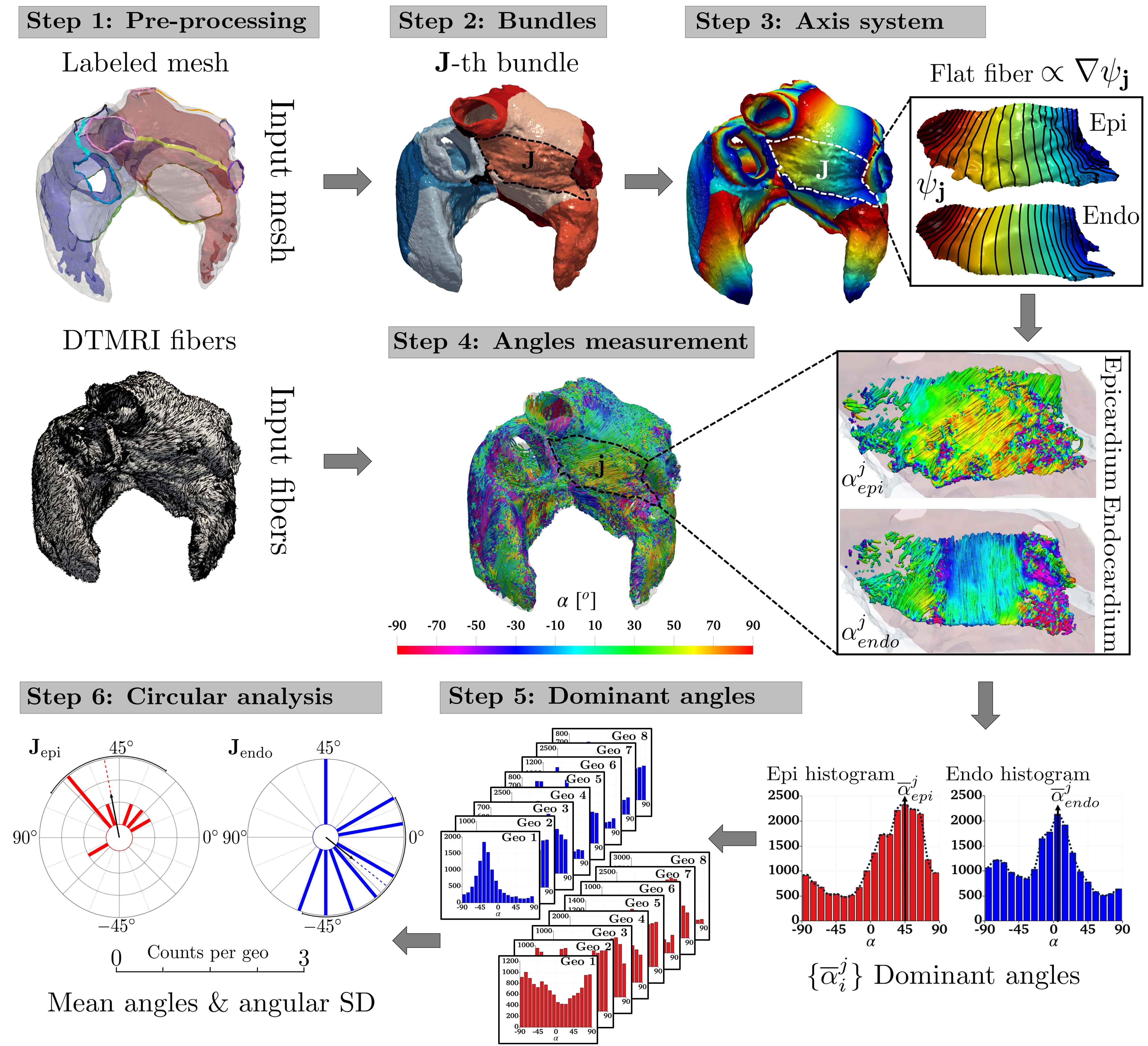}
    \caption{Schematic representation of the measurement procedure exploiting the LDRBM reference axis system to asses the local myocardial fiber angle in each atrial bundles across eight human bi-atrial geometries embedded with DTMRI fiber data~\cite{pashakhanloo2016myofiber}.}
    \label{fig:measuring}
\end{figure}
\begin{itemize}
    \item[\textbf{1.}] \textbf{Pre-processing:} Provide a labeled bi-atrial computational mesh and a related DTMRI fiber field. The labeling procedure is performed according to the LDRBM (see Step 1 in Method~\ref{subsec:ldrbm_model}), while the DTMRI fibers are linearly projected onto the labeled mesh;
    \item[\textbf{2.}] \textbf{Bundles:} Subdivide the atrial domain into characteristic anatomical regions, named bundles, employing the LDRBM bundle subdivisions (see Step 3 in Method~\ref{subsec:ldrbm_model});
    \item[\textbf{3.}] \textbf{Axis system:} Build a local coordinate axis system $Q=[\widehat{\boldsymbol{e}}_l, \widehat{\boldsymbol{e}}_n, \widehat{\boldsymbol{e}}_t]$ composed of the LDRBM flat myofiber vectors (see Step 4 in Method~\ref{subsec:ldrbm_model});    
    \item[\textbf{4.}] \textbf{Angles measurement:}  Embed the DTMRI fiber data $\boldsymbol{f}^m_{\text{DTMRI}}$ within the local coordinate axis system $Q$, using the following Gram-Schmidt process:
    \begin{equation*}	
        \begin{cases}
            \boldsymbol{s}_{\text{DTMRI}}=\boldsymbol{e}_t
            \\
            \boldsymbol{f}_{\text{DTMRI}}= \frac{\boldsymbol{f}^{m}_{\text{DTMRI}}-(\boldsymbol{f}^{m}_{\text{DTMRI}} \cdot \boldsymbol{e}_t)\boldsymbol{e}_t}{||\boldsymbol{f}^{m}_{\text{DTMRI}}-(\boldsymbol{f}^{m}_{\text{DTMRI}} \cdot \boldsymbol{e}_t)\boldsymbol{e}_t||} 
            \\
            \boldsymbol{n}_{\text{DTMRI}}= \boldsymbol{e}_t \times \mathbf{f}_{\text{MRI}} 
        \end{cases},
    \end{equation*}
    such that, starting from the experimentally derived DTMRI fiber field $\boldsymbol{f}^m_{\text{DTMRI}}$, three associated myofiber directions $\boldsymbol{f}_{\text{DTMRI}}$, $\boldsymbol{n}_{\text{DTMRI}}$, and $\boldsymbol{s}_{\text{DTMRI}}$ are retrieved and embedded in the same space spanned by the LDRBM axis system. Finally, for every node of the computational domain,  measure the DTMRI fiber orientation angles $\alpha^j_i \in (-\pi/2, \pi/2]$, in each bundle, relative to the LDRBM axis system (where $j$ refers to the generic $j$-th bundle and $i$ indicates the sub-epicardial $i=epi$ and sub-endocardial $i=endo$ layers). This is performed using the function $\alpha^j_i=\text{arcos}(\boldsymbol{f}_{\text{DTMRI}} \cdot \boldsymbol{e}_{\ell})$ corrected to account for the directional invariance of the fibers;
    \item[\textbf{5.}] \textbf{Dominant angles:} Compute histograms, for each bundle, of the measured DTMRI angles $\alpha^j_i$ and identify the dominant angle $\overline{\alpha}^j_i$ by selecting the modal values of the distributions;
    \item[\textbf{6.}] \textbf{Circular analysis:} Perform a circular statistical analysis~\cite{berens2009circstat,fisher1995statistical,jammalamadaka2001topics} (detailed in Appendix~\ref{supp:circular-statistic}) to quantitatively assess the fiber angle variability across the entire dataset.
\end{itemize}
 
For the human atrial DTMRI fiber dataset~\cite{pashakhanloo2016myofiber}, Step 2 is outlined in Section~\ref{subsec:bundle_classification}, while Steps 4$-$6 are described in Section~\ref{subsec:fiber_assesment}.
\subsubsection{Classification of atrial bundles}
\label{subsec:bundle_classification}
\begin{figure}[t!]
    \centering
    \includegraphics[width=1\textwidth]{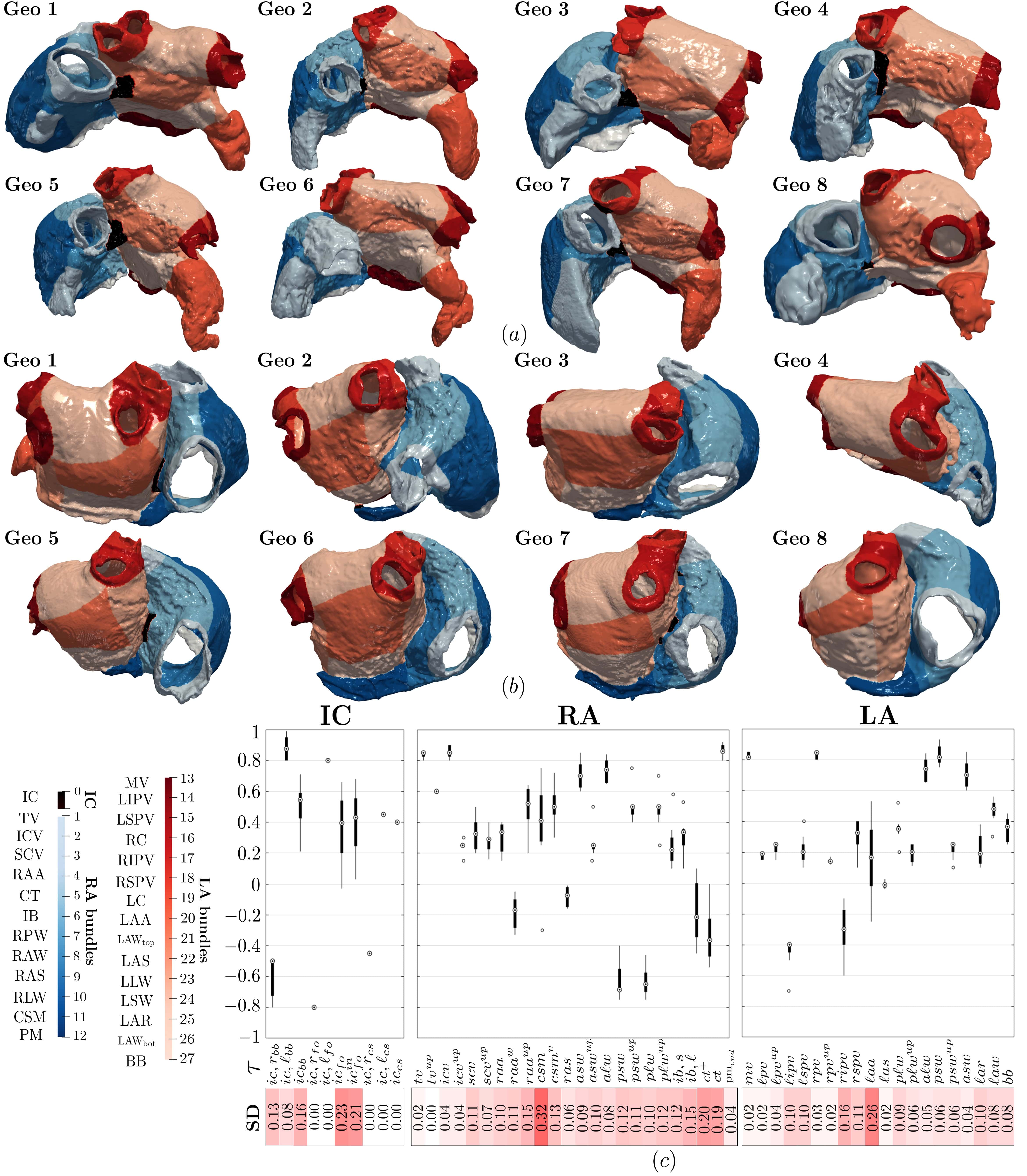}
    \caption{Bundles subdivision performed by the bi-atrial LDRBM for the DTMRI dataset geometries: (a) anterior view; (b) posterior view; (c) box-plots showing the bundle parameter variation for the left atrium (LA), right atrium (RA) and inter-atrial connections (IC); standard deviatian (SD) values of the bundle parameters, with SD color-coded on a scale from minimal (0) to maximal (0.32) values (see also Figure~\ref{fig:fiber_bundles} and Tables~\ref{tab:tau_bundles_BIA}$-$\ref{tab:tau_bundles_RA}).}
    \label{fig:bundle_classification}
\end{figure}

The partitioning of the atria into their characteristic anatomical subregions is carried out according to Step~2 of the measurement procedure (see Section~\ref{subsec:fiber_measurement}). This consists in applying the bundle subdivision of the novel bi-atrial LDRBM, across the eight DTMRI dataset geometries. Explicitly, according to the rules defined in Algorithm~\ref{BIA-LDRBM}$-$\ref{RA-LDRBM} (see Step~3 in Section~\ref{subsec:ldrbm_model}), the LDRBM first extracts the inter-atrial connections (IC) and then selects the LA and RA bundles. The following anatomical areas, illustrated in Figure~\ref{fig:bundle_classification}(a,b) (see also Figure~\ref{fig:fiber_bundles}), are identified:
\begin{itemize}
    \item{IC}: the Bachmann’s bundle connection ($\text{BB}_{\text{IC}}$) in the central anterior region of the atria; the fossa ovalis rim connection ($\text{FO}_{\text{IC}}$) across the atrial septum; the coronary sinus connection ($\text{CS}_{\text{IC}}$) in the posterior wall (extracted only within Geo 2, with other geometries presenting an almost fused CS structure into LA); connections between RA and the sleeves of the right pulmonary vein in Geo 7. 
    \item{LA}: the mitral valve (MV) vestibular region; the venous portion collecting left and right inferior and superior pulmonary veins (LIPV, LSPV, RIPV, RSPV); the right (RC) and left (LC) carina; the atrial appendage (LAA); the anterior septum (LAS) and Bachmann’s bundle (BB) in the anterior wall; the lateral wall (LLW) and atrial septum (LSW) in the lateral and septal regions of LA, respectively; the top ($\text{LAW}_{\text{top}}$) and bottom ($\text{LAW}_{\text{bot}}$) posterior wall; the atrial roof (LAR).
    \item{RA}: the tricuspid valve (TV) vestibular region; the venous portion composed of the inferior (ICV) and superior (SCV) caval veins; the roof wall between the orifices of caval veins, named the inter-caval bundle (IB); the coronary sinus musculature (CSM), joined to the adjacent $\text{LAW}_{\text{bot}}$ and MV regions (extracted in all geometries except Geo 4); the atrial appendage (RAA); the posterior wall (RPW) and anterior wall (RAW) below ICV and SCV, respectively; the atrial septum (RAS); the crista terminalis (CT), which is clearly detectable on the sub-endocardium and extends from SCV to ICV curving to the right of ICV; a series of bundles known as pectinate muscles (PM) that fan out from CT toward TV; the lateral wall (RLW), overlapping CT and PM structures.
\end{itemize}    
These anatomical regions (IC, LA and RA bundles) are clearly identified in all the eight morphologies by the LDRBM bundle subdivision procedure, see Step 3 in Section~\ref{subsec:ldrbm_model} and Algorithms~\ref{BIA-LDRBM}$-$\ref{RA-LDRBM}. 
The input values of the parameters $\tau_i$, used to define the bundle dimensions throughout the atrial domain across the DTMRI dataset geometries, are listed in Tables \ref{tab:tau_bundles_BIA}$-$\ref{tab:tau_bundles_RA} of Appendix~\ref{supp:measurement_dtmri_fiber}.
Figure~\ref{fig:bundle_classification}(c) reports the distribution (as box plots) of the bundle subdivision parameters $\tau_i$ across the DTMRI dataset. Higher parameter standard deviation (SD) values for these parameters are observed in LAA, CSM, CT, and  $\text{FO}_{\text{IC}}$, indicating greater anatomical variability, while the remaining bundles exhibit lower~SD revealing reduced morphological variation. Additional details are provided in appendix Figure~\ref{fig:fiber_bundles} and Tables~\ref{tab:tau_bundles_BIA}$-$\ref{tab:tau_bundles_RA}.
\subsubsection{Assessment of atrial fiber orientations}
\label{subsec:fiber_assesment}
The quantitative characterization of fiber angles across the atrial wall is revealed according to Steps~4$-$6 of the measuring procedure (presented at the beginning of Section~\ref{subsec:fiber_measurement}). Hereafter, we summarize the main findings for the DTMRI dataset~\cite{pashakhanloo2016myofiber}. Comprehensive details are available in Appendix~\ref{supp:measurement_dtmri_fiber}. Specifically, the histogram distributions of the measured fiber angles related to each atrial bundle are illustrated in Figures~\ref{fig:histo_LA_1}$-$\ref{fig:histo_LA_5} for LA and in Figures~\ref{fig:histo_RA_1}$-$\ref{fig:histo_RA_4} for RA. Moreover, the identified dominant fiber angles are reported in Tables~\ref{tab:tau_alpha_LA} and~\ref{tab:tau_alpha_RA}.   
Finally, the result of the circular data analysis~\cite{berens2009circstat,fisher1995statistical,jammalamadaka2001topics} (detailed in Appendix~\ref{supp:circular-statistic}) applied to the dominant fiber angles, is reported in Figures~\ref{fig:LA_angle_bundles} and~\ref{fig:RA_angle_bundles}. Resultant mean angles and the corresponding angular SD values are listed in Tables~\ref{tab:tau_alpha_LA} and~\ref{tab:tau_alpha_RA}.

\begin{figure}[ht!]
    \centering
    \includegraphics[width=1.0\textwidth]{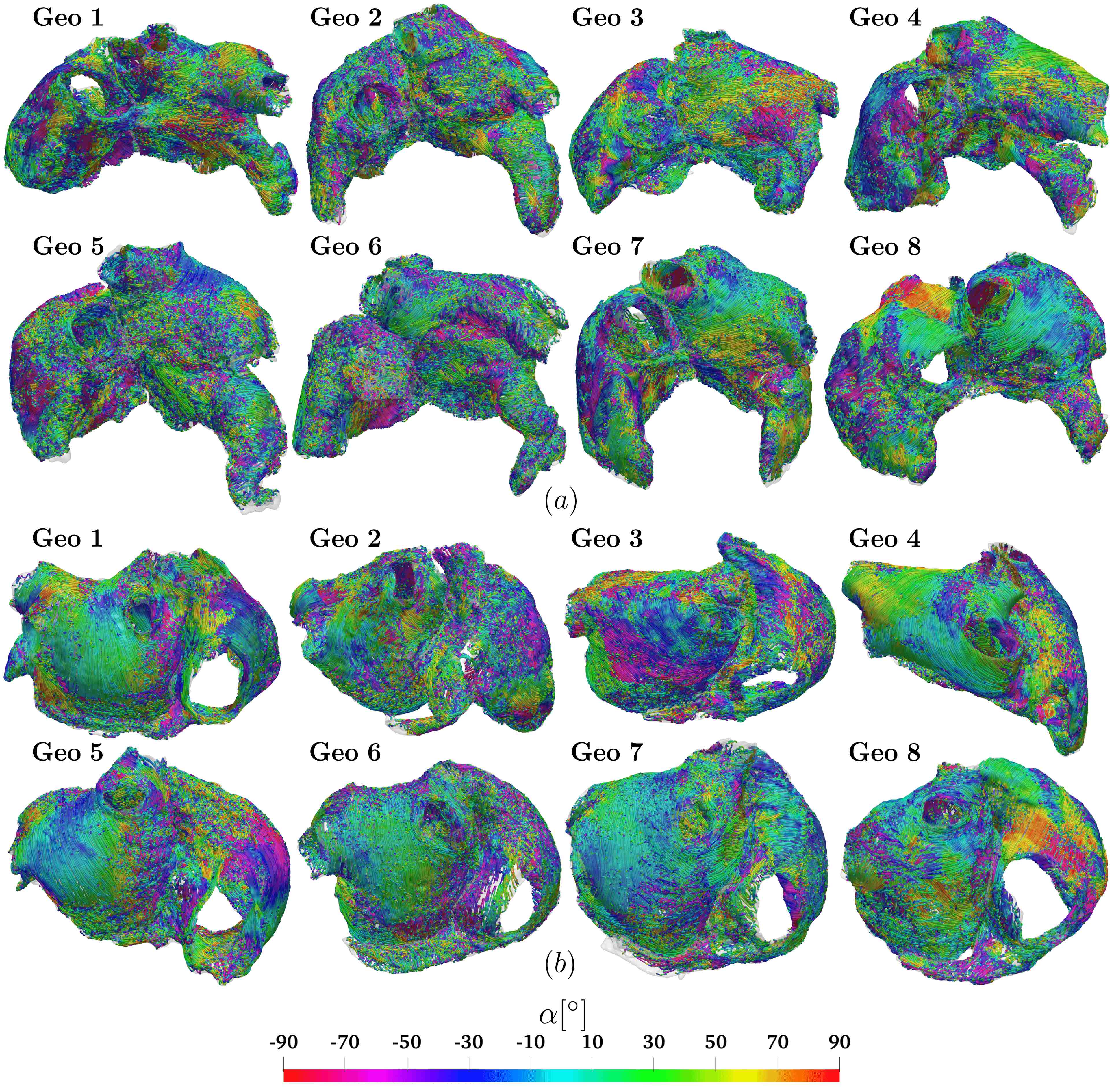}
    \caption{Measuring procedure applied to the eight geometries of DTMRI fiber dataset: streamlines represent DTMRI fiber directions, showing the measured angle $\alpha$ relative to LDRBM axis system. Anterior (a) and posterior (b) views.}
    \label{fig:fibers_dtmri}
\end{figure}

Figure~\ref{fig:fibers_dtmri} displays the global result of the measured fiber angles applied to the eight DTMRI geometries: streamlines show the fiber directions and the related angle $\alpha$ relative to the LDRBM local coordinate axis system.
Fiber angle distribution reveals a significant transmural heterogeneity from the sub-endocardium (ENDO) to the sub-epicardium (EPI), across almost every atrial bundles, with fibers intersecting and traveling in different directions. Despite variations in exact orientations among specimens, the primary features of atrial fiber architecture are mostly preserved across the entire dataset, see Figure~\ref{fig:fibers_dtmri}.

Following~\cite{ho2002atrial,sanchez2013standardized}, we describe the myofiber bundle structures as {\sl circumferential} if the fiber orientations are roughly parallel to the MV/TV orifice, {\sl longitudinal} if approximately perpendicular to MV/TV, and {\sl oblique} if otherwise oriented relative to MV/TV.

Concerning LA, MV exhibits circumferential fibers in both EPI and ENDO layers across all the dataset. LIPV and LSPV display longitudinal-circular arrangements, except for Geo 3 which has predominantly oblique orientations. RIPV and RSPV show larger variability: RIPV has circumferential and oblique configurations, while RSPV includes also crossing fibers. LC and RC exhibit significant variability, especially in $\text{LC}_{\text{EPI}}$, whereas $\text{LC}_{\text{ENDO}}$ and RC have a longitudinal pattern. LAR reveals longitudinal orientations in six of the eight specimens. Geo 2$-$3 are unique, containing a mixture of oblique directions, especially in $\text{LAR}_{\text{EPI}}$. LAW (both $\text{LAW}_{\text{top}}$ and $\text{LAW}_{\text{bot}}$) fibers were found to be posterior-to-anterior, with $\text{LAW}_{\text{bot}}$ showing more pronounced transmural variation compared to  $\text{LAW}_{\text{top}}$. A pattern of overlapping fibers is consistently observed in the LA anterior wall: $\text{BB}_{\text{IC}}$ presents circumferential fibers in half the dataset (Geo~1,~3,~7,~8) and oblique orientations in the remaining (Geo 2, 4, 5, 6); BB has oblique fibers more consistently in EPI compared to ENDO; LAS contains oblique EPI fibers in five specimens (Geo 2, 3, 4, 5, 7), transitioning to a longitudinal structure in ENDO; LAA shows a bimodal distribution of angles, with crossing directions from EPI to ENDO, significant variability in  $\text{LAA}_{\text{EPI}}$, and a preserved structure in  $\text{LAA}_{\text{ENDO}}$; LLW and LSW present well-established circumferential fibers, with  $\text{LSW}_{\text{ENDO}}$ featuring both oblique and longitudinal fibers; $\text{FO}_{\text{IC}}$ fibers circularly run around its center in five of the eight specimens.

Regarding RA, TV presents circumferential fibers in EPI, similar to MV, but with a pronounced transmural variation to oblique in ENDO. SCV features consistent oblique directions. Conversely, ICV fibers are longitudinal-circular only in ENDO and oblique in EPI. $\text{IB}_{\text{EPI}}$ shows circumferential orientations in four geometries (Geo 3, 4, 5, 8), longitudinal in two (Geo 2, 7), and oblique in the remaining (Geo 1, 6); whereas $\text{IB}_{\text{ENDO}}$ has circumferential fibers in five specimens (Geo 1, 2, 4, 6, 8) and perpendicular/oblique in the others (Geo 3, 5, 7). RAA has fibers encircling the appendage, with more variability in ENDO compared to EPI. CSM features a well-preserved circumferential structure. RAS shows longitudinal orientations in ENDO, while EPI reveals significant variability with circumferential (Geo 3, 6), longitudinal (Geo 4, 7), and oblique (Geo 1, 2, 5, 8) directions. RPW unveils considerable transmurality, transitioning from longitudinal (EPI) to oblique (ENDO) fibers. RAW is characterized by longitudinal (Geo 4, 5, 6, 7) and oblique (Geo~1,~2,~3,~8) orientations. RLW exhibits varying oblique directions and includes the distinct CT and PM structures: CT is mainly oriented longitudinally, while PM run perpendicularly following the ENDO trabeculated structure.
\subsection{Fiber generation}
\label{subsec:fiber_generation}
\begin{figure}[t!]
    \centering
    \includegraphics[width=1\textwidth]{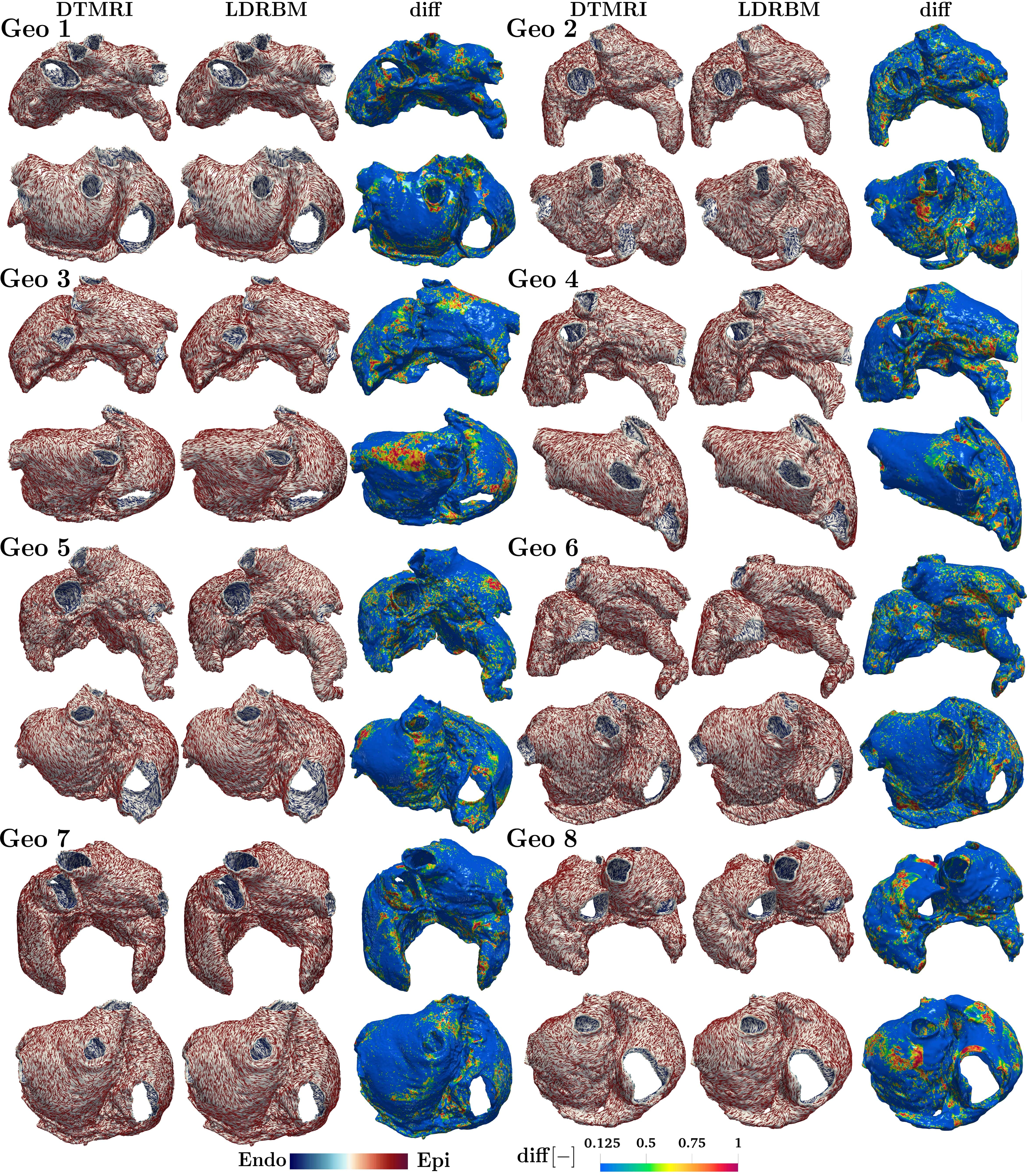}
    \caption{Comparison between the atrial LDRBM fibers and the DTMRI data, across eight DTMRI geometries. Glyph-rendered fiber vector fields are reported for each geometry (Geo 1$-$8), displayed in anterior (top) and posterior (bottom) views. The function diff, computed as $\text{diff}(\boldsymbol{x}) = 1 - |\boldsymbol{f}_\text{DTMRI}(\boldsymbol{x}) \cdot \boldsymbol{f}_\text{LDRBM}(\boldsymbol{x})|$, highlights the differences between LDRBM and DTMRI fibers, see also Figures~\ref{fig:fibers_LDRBM_epi}$-$\ref{fig:fibers_LDRBM_endo}.}
    \label{fig:fiber_generation}
\end{figure}
To verify the reliability of the bi-atrial LDRBM, we completely deploy the novel fiber generation model (see Section~\ref{subsec:ldrbm_model}), to reconstruct all the eight DTMRI human atrial myofiber architectures. Furthermore, we compare the generated LDRBM fiber fields to the ground truth DTMRI data, investigating the differences in fiber orientations across the entire DTMRI dataset.

For the LDRBM bundle subdivision, we consider the parameters (reported in Tables~\ref{tab:tau_bundles_BIA}$-$\ref{tab:tau_bundles_RA}) used for the regional classification presented in Section~\ref{subsec:bundle_classification}. Additionally, the input angular values (listed in Tables~\ref{tab:tau_alpha_LA} and~\ref{tab:tau_alpha_RA}) are chosen based on the observed dominant angles retrieved by the measurement procedure described in Section~\ref{subsec:fiber_assesment}.

Figure~\ref{fig:fiber_generation} (second and fifth columns) shows the fiber orientations reconstructed using the bi-atrial LDRBM for all the eight DTMRI geometries (see also Figures~\ref{fig:fibers_LDRBM_epi}$-$\ref{fig:fibers_LDRBM_endo}). The LDRBM captures the complex fiber arrangement in almost all the principal anatomical atrial regions, generally reproducing the DTMRI fiber orientations (first and fourth columns of Figure~\ref{fig:fiber_generation}) with visible differences only in limited areas.

\clearpage
Figure~\ref{fig:fiber_generation} (third and sixth columns) compares the generated LDRBM fibers with the ground truth DTMRI data, showing the fiber orientation differences evaluated using the function 
\begin{equation}
    \label{eqn:fiber_diff}
    \text{diff}(\boldsymbol{x}) = 1 - |\boldsymbol{f}_\text{DTMRI}(\boldsymbol{x}) \cdot \boldsymbol{f}_\text{LDRBM}(\boldsymbol{x})|,
\end{equation}
where $\boldsymbol{f}_\text{DTMRI}(\boldsymbol{x})$ and $\boldsymbol{f}_\text{LDRBM}(\boldsymbol{x})$ are the vector fiber fields associated with DTMRI data and LDRBM, respectively. 

To further quantify the amount of matching between DTMRI and LDRBM fibers, we analyzed the distribution of $\text{diff}(\boldsymbol{x})$ function values across the different morphologies (see Figures~\ref{fig:fibers_LDRBM_epi}$-$\ref{fig:fibers_LDRBM_endo}). 
The percentage of fibers in good agreement was approximately from $43\%$ to $48\%$ of the total fiber orientations, across all the geometries.

\subsection{Electrophysiology simulations}
\label{subsec:ep_simulations}
\begin{figure}[t!]
	\centering
	\includegraphics[width=1\textwidth]{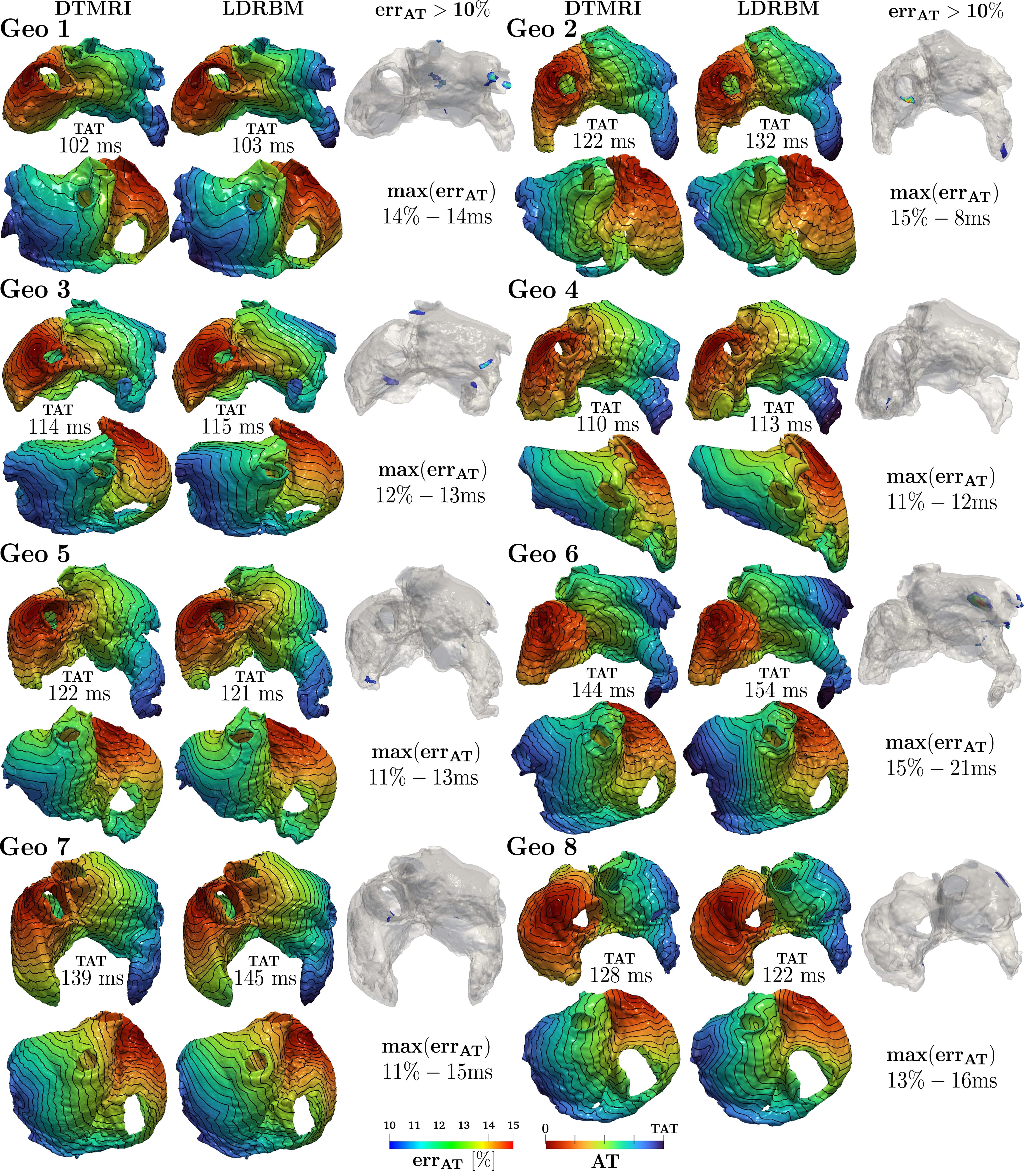}
	\caption{Comparison between the activation times (AT) derived from the numerical EP simulations endowed with LDRBM and DTMRI fibers, across the DTMRI dataset. Each geometry's anterior and posterior AT views are displayed, with isochrones spaced 10 \unit{ms} apart. $\text{TAT}$ is the total activation time obtained with DTMRI and LDRBM fibers; $\max(\text{err}_\text{AT})$ denotes the maximal AT (relative$-$absolute) error, while $\text{err}_\text{AT}>10\%$ highlights volumetric regions where AT relative difference exceeds 10\% of error. Further details are reported in Table~\ref{tab:eikonal_table}.
	}
	\label{fig:ep_simulations}
\end{figure} 
To assess the impact of using LDRBM and DTMRI fiber architectures on electric signal propagation, we perform two types of EP simulations employing the Eikonal-diffusion model (detailed in Appendix~\ref{supp:ep_model}) for each DTMRI geometry: one with DTMRI fiber data and the other with LDRBM fibers. To quantify the deviations in activation times (AT), resulting from the different fiber architectures, we evaluate:
\begin{itemize}
    \item the total activation time (TAT), defined as $$\text{TAT}_i=\max_{\boldsymbol{x}}{\left[\EPpot^{i}(\boldsymbol{x})\right]}, \qquad \qquad i=\text{DTMRI, LDRBM},$$ where $\EPpot^{\text{DTMRI}}(\boldsymbol{x})$ and $\EPpot^{\text{LDRBM}}(\boldsymbol{x})$ are the numerical AT retrieved by EP simulations endowed with DTMRI and LDRBM fibers, respectively;
    \item the error between TAT of LDRBM and DTMRI, as
    $$\text{err}_{\text{TAT}}=|\text{TAT}_\text{LDRBM}-\text{TAT}_\text{DTMRI}|;$$  
    \item the maximal AT error between LDRBM and DTMRI, namely $\max_{\boldsymbol{x}}{\left[\text{err}_{\text{AT}}(\boldsymbol{x})\right]}$ where $\text{err}_{\text{AT}}(\boldsymbol{x})$ is defined as  
    $$\text{err}_{\text{AT}}(\boldsymbol{x})=|\EPpot^\text{LDRBM}(\boldsymbol{x})-\EPpot^\text{DTMRI}(\boldsymbol{x})|;$$
    \item the volumetric compatibility index error $\text{Vol}_{>10\%}$, indicating the percentage of atrial volume where AT exceeds 10\% of error
    \begin{equation}
    \label{eqn:index_error}    
        \text{Vol}_{>10\%}[\%]=\frac{N_{tot}-N_{<10\%} }{N_{tot}}100,
    \end{equation}
    where $N_{tot}$ and $N_{<10\%}$ are the total number of EP solution degree of freedom (d.o.f) and the number of d.o.f where AT do not exceed $10\%$ of error, respectively.   
\end{itemize}

Figure~\ref{fig:ep_simulations} reports the comparison between AT coming from EP simulations obtained with either LDRBM and DTMRI fibers, across all the dataset (see also Table~\ref{tab:eikonal_table}). The simulations predict total activation times $\text{TAT}_i$ ($i=\text{DTMRI, LDRBM}$) that are perfectly compatible, with an absolute error of ranging from $1\unit{ms}$ to $10\unit{ms}$, corresponding to a relative one between $1\%$ to $9\%$ and a mean absolute/relative error, across all the eight geometries, of $5\unit{ms}$/4\%. The activation patters feature also a very similar morphology with marginal discrepancies only in limited regions. The latter is confirmed by the absolute/relative maximal AT error within $8\unit{ms}/11\%$ to $21\unit{ms}/15\%$, with a mean value of $14\unit{ms}/13\%$. Bundle-regions where the relative AT error exceeds the 10\% are almost equally observed in LA and RA: 5 LA (RPV, LPV, MV, LAW, LAA) and 4 RA (SCV, TV, RAA, SCV) bundles. However, the volumetric compatibility index error $\text{Vol}_{>10\%}$ is well below the 1\% across the entire dataset. This indicates that EP simulations are compatible within 99\% of the entire bi-atrial volume for all the specimens.
\begin{table}[t!]
	\centering
	\scalebox{1}{\begin{tabular}{ |c|c|c|c|c|c|c| } 
			\hline
			Geometry & $\text{TAT}_\text{DTMRI}$ & $\text{TAT}_\text{LDRBM}$ & $\text{err}_\text{TAT}$ & $\text{max}(\text{err}_\text{AT})$ & Bundle($\text{err}_\text{AT}>10\%$) & $\text{Vol}_{>10\%}$ \\
			\hline
			\hline
			GEO 1 & 102 \unit{ms} & 103 \unit{ms} & $1$ $\unit{ms}$/1\% & $14$  $\unit{ms}$/14\% & RPV/LPV/LAW & 0.40$\%$ \\ 
			GEO 2 & 122 \unit{ms} & 132 \unit{ms} & $10$ $\unit{ms}$/8\% & $8$  $\unit{ms}$/15\% & SCV/LAA           & 0.24$\%$ \\ 
			GEO 3 & 114 \unit{ms} & 115 \unit{ms} & $1$ $\unit{ms}$/1\% & $13$  $\unit{ms}$/12\% & RPV/LPV/LAA/TV    & 0.30$\%$ \\ 
			GEO 4 & 110 \unit{ms} & 113 \unit{ms} & $3$ $\unit{ms}$/6\% & $12$  $\unit{ms}$ /11\% & RAA               & 0.01$\%$ \\ 
			GEO 5 & 122 \unit{ms} & 121 \unit{ms} & $1$ $\unit{ms}$/1\% & $13$  $\unit{ms}$/11\% & RAA/LPV/MV       & 0.04$\%$ \\ 
			GEO 6 & 144 \unit{ms} & 154 \unit{ms} & $10$ $\unit{ms}$/9\% & $21$  $\unit{ms}$/15\% & LAW/RPV/CSM   & 0.82$\%$ \\ 
			GEO 7 & 139 \unit{ms} & 145 \unit{ms} & $6$ $\unit{ms}$/4\% & $15$  $\unit{ms}$/11\% & SCV/MV            & 0.02$\%$ \\ 
			GEO 8 & 128 \unit{ms} & 122 \unit{ms} & $6$ $\unit{ms}$/5\% & $16$  $\unit{ms}$/13\% & LAW           & 0.10$\%$ \\ 
			\hline      
	\end{tabular}}
	\caption{Differences in EP simulations between LDRBM and DTMRI fiber architectures across eight DTMRI geometries: $\text{TAT}_i$ is the total activation time (TAT) obtained with DTMRI ($i=\text{DTMRI}$) and LDRBM ($i=\text{LDRBM}$) fibers; $\text{err}_\text{TAT}$ denotes the TAT absolute (in \unit{ms}) and relative (in \%) error; $\text{max}(\text{err}_\text{AT})$ signifies the maximal absolute/relative activation time (AT) error; Bundle($\text{err}_\text{AT}>10\%$) lists regions where the AT exceeds 10\% error;  $\text{Vol}_{(>10\%)}$ is the volumetric error index indicating the percentage of atrial volume where the AT exceeds 10\% error.}    
	\label{tab:eikonal_table}
\end{table}
\begin{table}[t!]
    \centering
    \scalebox{1}{\begin{tabular}{ |c|c|c|c|c|c| } 
            \hline
            Type & $\text{TAT}$ & $\text{err}_\text{TAT}$ & $\text{max}(\text{err}_\text{AT})$ & Bundle($\text{err}_\text{AT}>10\%$) & $\text{Vol}_{(>10\%)}$ \\
            \hline
            \hline
            DTMRI      & 139 \unit{ms} & $-$    & $-$    & $-$                   & $-$       \\ 
            LDRBM-PQ24 & 145 \unit{ms} & $6$ $\unit{ms}$/$4\%$ & $15$ $\unit{ms}$/$11\%$ & SCV/MV                & 0.02$\%$  \\ 
            UAC        & 152 \unit{ms} & $13$ $\unit{ms}$/$9\%$  & $24$ $\unit{ms}$/$16\%$ & RAA/LAR/TV/RAS/LAS/MV & 3.45$\%$  \\ 
            LDRBM-PQ21 & 192 \unit{ms} & $53\unit{ms}$/$38\%$ & $78$ $\unit{ms}$/$41\%$ & almost half LA and RA & 44.97$\%$ \\ 
            \hline          
    \end{tabular}}
    \caption{Differences in EP simulations produced by different atrial fiber models (LDRBM-PQ24, UAC~\cite{roney2021constructing}, LDRBM-PQ21\cite{piersanti2021modeling}) against ground truth DTMRI fiber data (Geo 7 analyzed). $\text{TAT}$ is the total activation time; $\text{err}_\text{TAT}$ denotes the TAT absolute (in \unit{ms}) and relative (in \%) error; $\text{max}(\text{err}_\text{AT})$ signifies the maximal activation time (AT) absolute/relative error;  Bundle($\text{err}_\text{AT}>10\%$) lists regions where the AT exceeds 10\% error;  $\text{Vol}_{(>10\%)}$ is the volumetric error index, indicating the percentage of atrial volume where the AT exceeds 10\% error.}        
    \label{tab:eikonal_table_comparison}
\end{table}
\subsection{Comparison with state-of-the-art atrial fiber models}
\label{subsec:models_comparison}
The results obtained by our novel bi-atrial LDRBM, both for the fiber directions and the EP simulations, were compared with other fiber generation models. Specifically, we considered the Universal Atrial Coordinate (UAC) ABM, originally proposed in~\cite{roney2021constructing} and then extended to account for volumetric meshes in~\cite{roney2023constructing,ali2021b}, and the first atrial LDRBM presented in~\cite{piersanti2021modeling}. To differentiate bewteen the two LDRBM, we name hereafter the first LDRBM as LDRBM-PQ21, while the one presented in this work as LDRBM-PQ24. All the comparisons are performed on the same mesh, coming from Geo 7. The UAC input fibers is the one originally derived by the authors in~\cite{roney2021constructing}, corresponding to the morphology Geo 7 of the DTMRI dataset~\cite{pashakhanloo2016myofiber} (see also \url{https://zenodo.org/records/3764917}). Additional information are reported in Figure~\ref{fig:UAC_PQ21}. We refer the reader to~\cite{roney2023constructing,ali2021b} and \cite{piersanti2021modeling} for further details about UAC and LDRBM-PQ21 fiber models, respectively.

\begin{figure}[ht!]
    \centering
    \includegraphics[width=1\textwidth]{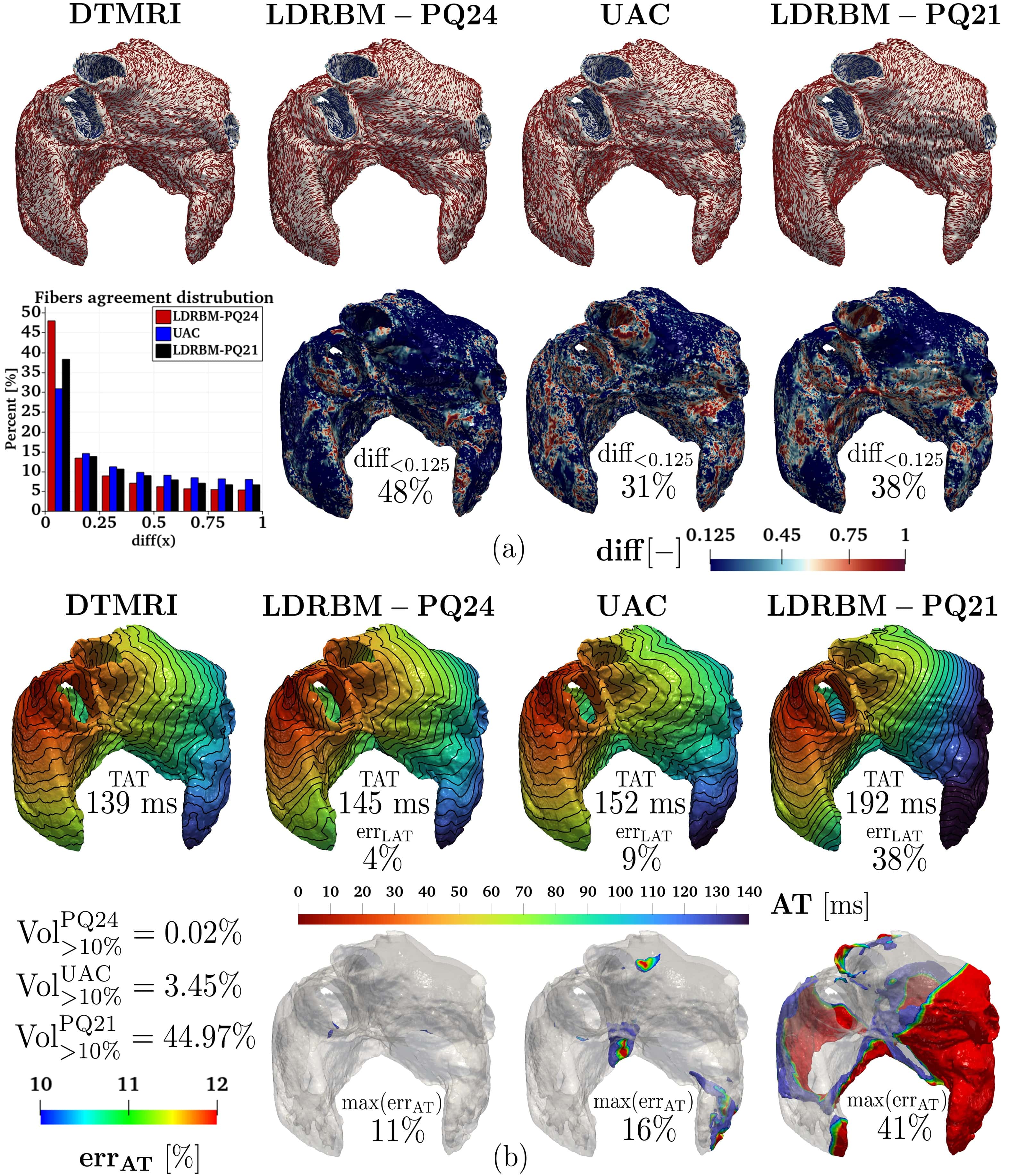}
    \caption{Comparative analysis of different fiber models (LDRBM-PQ24, UAC~\cite{roney2021constructing} and LDRBM-PQ21~\cite{piersanti2021modeling}) against DTMRI fibers (Geo 7 analyzed): (a-top) Glyph-rendered representations showcase the fiber architectures among models; (a-bottom) fiber disparities relative to DTMRI data, calculated as $\text{diff}(\boldsymbol{x}) = 1- |\boldsymbol{f}_\text{DTMRI}(\boldsymbol{x}) \cdot \boldsymbol{f}_i(\boldsymbol{x})|$, with $i=\text{LDRBM-PQ24}$ on the left, $i=\text{UAC}$ in the center, and $i=\text{LDRBM-PQ21}$ on the right; (b-top) Activation maps (with isochrones spaced 10 \unit{ms} apart) from EP simulations produced using different fiber models $\text{TAT}$ is the total activation time and $\text{err}_\text{TAT}$ denotes the TAT relative error; (b-bottom) $\text{err}_\text{AT}$ highlights volumetric regions where AT relative difference exceeds 10\% error and $\text{max}(\text{err}_\text{AT})$ signifies the maximal relative AT error, while $\text{Vol}^i_{(>10\%)}$ (with $i=\text{PQ24, UAC and PQ21}$) indicates the percentage of atrial volume where the AT exceeds 10\% error.
    }
    \label{fig:models_comparison}
\end{figure}

Figure~\ref{fig:models_comparison}(a) showcases the fiber comparison results. The architecture generated by LDRBM-PQ24 is in excellent agreement with DTMRI data, reproducing nearly the same fiber orientations in all the different atrial bundles. In contrast, both UAC and LDRBM-PQ21 exhibit several discrepancies compared to DTMRI fibers. The latter arise mostly in LSPV, RSPV, LAA, RAA, RAS and LAS for UAC, and in LAA, RAA, RAS, LAS, BB, CT and PM for LDRBM-PQ21. Furthermore, both LA and RA septal junctions and inter-atrial connections are completely misrepresented in LDRBM-PQ21. 

The fiber discrepancies are quantified by evaluating for each methodology the function~\eqref{eqn:fiber_diff} with respect to the DTMRI atrial fiber architecture, see Figure~\ref{fig:models_comparison}(a). The percentage of fibers in good agreement with respect to DTMRI data are $48\%$, $31\%$ and $37\%$ for LDRBM-PQ24, UAC and LDRBM-PQ21, respectively (see~Figures~\ref{fig:models_comparison}(a) and Figure~\ref{fig:UAC_PQ21}(c)).

To quantify the AT discrepancies predicted by the different fiber architectures (LDRBM-PQ24, UAC, LDRBM-PQ21 and DTMRI), we perform four EP simulations using the Eikonal-diffusion model (detailed in Appendix~\ref{supp:ep_model}). We evaluate the TAT/AT errors (with respect to DTMRI) and the volumetric compatibility index~\eqref{eqn:index_error}, following the same analysis presented in Section~\ref{subsec:ep_simulations}. 
\clearpage
\clearpage
Figure~\ref{fig:models_comparison}(b) reports the comparison among the different EP simulations predicted by DTMRI, LDRBM-PQ24, UAC, and LDRBM-PQ21 fibers. The results are also resumed in Table \ref{tab:eikonal_table_comparison}. Both AT and the propagation morphology between the simulations with LDRBM-PQ24 and DTMRI fibers are in very good agreement, with discrepancies exceeding 14 $\unit{ms}$ (i.e., 10\% of error with respect to AT produced by DTMRI fibers) only in restricted zones of SCV and TV, see Figure~\ref{fig:models_comparison}(b). These correspond to 0.02\% of the total myocardial volume. Conversely, AT predicted by UAC shows higher discrepancies in several LA and RA bundles (RAA, LAR, TV, RAS, LAS and MV), corresponding to 3.45\% of the atrial volume. Finally, the simulation with LDRBM-PQ21 produces remarkably different values for both TAT (with 38\% error with respect to DTMRI fibers) and local AT, with differences exceeding 10\% error extending over half of the LA and RA volume, see Figure~\ref{fig:models_comparison}(b) and Table~\ref{tab:eikonal_table_comparison}.
    \section{Discussion}
\label{sec:discussion}
In this work, we presented a novel LDRBM modeling approach (see~Section~\ref{subsec:ldrbm_model}) for prescribing myofiber orientations and providing robust regional annotation in bi-atrial morphologies of any complexity through a highly automated framework. The robustness of our method was verified using eight highly detailed bi-atrial geometries, processed from the original DTMRI human atrial fiber dataset presented in~\cite{pashakhanloo2016myofiber}. Furthermore, we developed a systematic measurement procedure, leveraging the LDRBM reference axis system, to assess local myocardial fiber angles across the atrial wall in geometries embedded with experimental DTMRI data (see~Section~\ref{subsec:fiber_measurement}). Next, we validated the atrial LDRBM by quantitatively reproducing all the eight DTMRI fiber architectures (see~Section~\ref{subsec:fiber_generation}). Moreover, we demonstrated that numerical EP simulations of electrical wave propagation, using both LDRBM and DTMRI fibers on the same geometries, exhibit excellent agreement (see~Section~\ref{subsec:ep_simulations}). Finally, by comparing our modeling approach with state-of-the-art atrial fiber generation models~\cite{piersanti2021modeling,roney2021constructing} against ground-truth DTMRI fibers, we proved that the bi-atrial LDRBM outperforms current methodologies used for prescribing myofiber orientations in~ADT (see~Section~\ref{subsec:models_comparison}). Therefore, our novel rule-based modeling approach establishes a new standard to prescribe fibers in ADT, with the potential to significantly enhance their precision.

Compared to previous atrial fiber computational approaches~\cite{ferrer2015detailed,roney2023constructing,satriano2013feature,labarthe2014bilayer,hoermann2019automatic,roney2021constructing,krueger2011modeling,tobon2013three,wachter2015mesh,fastl2018personalized,saliani2019visualization} our methodology, specifically designed for bi-atrial anatomies, preserves the distinctive simplicity of the LDRBM framework~\cite{piersanti2021modeling}, unlike many of the existing challenging-to-implement methodologies~\cite{roney2023constructing,krueger2011modeling,fastl2018personalized}. 
Our method requires the definition of Laplace problems with suitable Dirichlet boundary conditions prescribed on common boundary sets, such as endocardium, epicardium, atrioventricular valves and veins rings. Boundary label assignment is accomplished through a streamlined pipeline, ensuring precision, reproducibility, and high usability. Our fiber model not only preserves the natural flexibility and morphological adaptability of LDRBMs proposed so far~\cite{piersanti2021modeling,azzolin2023augmenta,zheng2021automate,rossi2022rule,fedele2023comprehensive}, but also significantly improves them by accounting for an heterogeneous bundle-specific fiber architecture. By leveraging on newly introduced inter/intra-atrial distances, consistently defined for LA and RA (see~Section~\ref{subsec:ldrbm_model}), our model extensively refine the bi-atrial regional classification. With dedicated fiber definitions in each bundle for both sub-epicardial and sub-endocardial layers, the bi-atrial LDRBM constructs and modulates a volumetric two-layer myofiber field, resulting in an exceptionally detailed fiber architecture with unprecedented results relative to the existing literature. For the first time, we demonstrated that a RBM is capable of reproducing the DTMRI atrial muscular architectures. We showed that numerical EP tissue activations predicted by LDRBM fibers are almost identical to the one produced by the ground truth fibers (see Sections~\ref{subsec:fiber_generation}$-$\ref{subsec:ep_simulations}). 

The atrial wall has been qualitatively observed in previous studies, including histo-anatomical ex vivo observations~\cite{ho2002atrial,ho2009importance,sanchez2013standardized} and DTMRI tractography analysis~\cite{pashakhanloo2016myofiber}, to exhibit a bilayer bundle structure characterized by fibers crossing and running in various directions. However, quantitative measurement of myocardial architecture properties, such as fiber angles, requires the definition of a coordinate system (or systems) in the atria that is reproducible across subjects with various atrial morphologies. This has been previously proposed for the ventricles~\cite{schuler2021cobiveco,bayer2018universal}, but it is particularly challenging for the atria due to their complex shape~\cite{roney2019universal}. Here, we established a systematic measurement procedure expoiting the LDRBM reference system to quantitatively characterize the DTMRI architectures, uncovering the local atrial fiber angle (see Section~\ref{subsec:fiber_measurement}). Compared to the previous attempt proposed in~\cite{roney2021constructing}, which measured the DTMRI human atrial fiber dataset~\cite{pashakhanloo2016myofiber} globally on highly smoothed endocardial and epicardial surfaces, our analysis allowed us to see through the entire atrial wall. This enabled to perform quantitative measurement and characterization of the local transmural fiber distribution throughout all the atrial bundles (see Section~\ref{subsec:fiber_assesment}). We exploited the semi-automatic LDRBM regional classification algorithm to identify 27 distinct bundles from DTMRI data (see Section~\ref{subsec:bundle_classification}). Then, we measured the local atrial fiber angles within the LDRBM axis system. The results demonstrates a bilaminar (sub-epicardial and sub-endocardial) architecture with fiber orientations revealing a significant transmural heterogeneity across almost every bundles. Despite variations among specimens, the primary features of the fibers are mostly preserved across the atrial dataset (see Section~\ref{subsec:fiber_assesment}). 

We constructed a set of eight distinct fiber architectures, along with a mean fiber configuration, blending togheter the bi-atrial LDRBM and DTMRI measurements. Specifically, we completely deployed the LDRBM fiber generation pipeline (see Section~\ref{subsec:ldrbm_model}), leveraging on DTMRI-LDRBM measuremnt information (Section~\ref{subsec:fiber_measurement}), to reconstruct all the eight DTMRI human atrial myofiber architectures. We verified that LDRBM fiber-replicas accurately capture the complex arrangement in nearly all anatomical regions, generally reproducing the same fiber orientations with visible differences only in limited areas. Approximately from $43\%$ to $48\%$ of the total fibers, across all geometries, were in good agreement (see Section~\ref{subsec:fiber_generation}). Moreover, EP simulations using both LDRBM and DTMRI fibers predicted a highly compatible AT, with a mean TAT error of $5 \pm 3\unit{ms}$.
The activation morphology patterns were nearly indistinguishable, with a maximal absolute/relative AT error averaging $14\unit{ms}/13\%$ (see Section~\ref{subsec:ep_simulations}). The volumetric compatibility index error, $\text{Vol}_{>10\%}<1\%$, indicates 99\% compatibility between EP simulations using LDRBM and DTMRI fibers (see Table~\ref{tab:eikonal_table}). It is important to note that our EP simulations account for variations in atrial wall thickness. Unlike other existing fiber pipelines~\cite{piersanti2021modeling,ferrer2015detailed,roney2019universal}, our methodology allows the development of ADT that incorporates fiber architecture for thickness-variable simulations. The significance of wall thickness and fibrosis distribution in arrhythmia mechanisms has been demonstrated in~\cite{hansen2015atrial}. Variations in fiber modeling related to thickness may also impact EP activation patterns. Therefore, incorporating bundle-specific fiber transmurality into ADT could be crucial in studying rhythm disorders~\cite{maesen2013rearrangement,hocini2002electrical,klos2008atrial,berenfeld2002frequency,hamabe2003correlation,markides2003characterization}.

Compared to a recent work~\cite{rossi2022rule} that examined various LA fiber methodologies, this study is the first to compare state-of-the-art RBM~\cite{piersanti2021modeling} and ABM~\cite{roney2021constructing} on patient-specific bi-atrial anatomy against DTMRI ground truth fibers. We proved that our LDRBM outperforms current state-of-the-art atrial fiber generation models (namely UAC~\cite{roney2021constructing} and PQ21~\cite{piersanti2021modeling}) in representing DTMRI fiber data. The percentage of fibers in good agreement was 48\% for LDRBM, compared to 31\% for UAC and 37\% for PQ21 (see~Section~\ref{subsec:models_comparison}). Additionally,  when comparing EP simulations, we found that AT and propagation morphology using LDRBM and DTMRI fibers are highly consistent (with 0.02\% of total atrial volume discrepancy). Conversely, UAC and PQ21 exhibited larger differences (3.45\% and 44.97\%, respectively). These results underscore the exceptional accuracy of LDRBM and highlight the critical role of incorporating biophysically detailed fiber orientations in EP simulations. In constrast to~\cite{roney2021constructing,he2022fiber}, claiming that the choice of fiber fields has minimal impact in sinus rithm pacing, we discovered that AT differences can range from to $24\unit{ms}$ to $78\unit{ms}$ when comparing different fiber models to the ground truth fiber data (see Section~\ref{subsec:models_comparison}). In fibrillatory dynamics, the fiber field can have a dramatic effect on predicting reentrant regions, as noted in~\cite{kamali2023contribution,roney2023constructing}. However, significant gaps remain in our understanding of fiber directions and their role in the genesis and progression of arrhythmias.

In conclusion, we believe that this work stands out  as a unique instance in the literature where atrial fiber architectures are modeled with an exceptional level of biophysical detail. To the best of our knowledge, this is the pioneering work of validating a rule-based atrial model against fiber orientations obtained from DTMRI data. Additionally, our approach offers a robust fiber-framework for the rapid development of personalized model cohorts accounting for detailed fiber anatomy and facilitates bi-atrial EP simulations. Ultimately, this study marks a significant advancement in building physics-based ADT, conceivably enhancing their precision for personalized risk assessment and potentially leading to better diagnostic and therapeutic strategies for cardiac diseases.
    
    \section*{Acknowledgements}
R.P. and L.D. have received support from the project PRIN2022, MUR, Italy, 2023–2025, 202232A8AN  ``Computational modeling of the heart: from efficient numerical solvers to cardiac digital twins”. R.P. has also received support from the INdAM GNCS project CUP E53C23001670001 ``Mathematical models and numerical methods for the construction of cardiac digital twins”. R.P., L.D. and A.Q. acknowledge their membership to INdAM GNCS - Gruppo Nazionale per il Calcolo Scientifico (National Group for Scientific Computing, Italy). R.P., L.D. and A.Q. acknowledge the “Dipartimento di Eccellenza 2023-2027”,
MUR, Italy, Dipartimento di Matematica, Politecnico di Milano. S.Y.A. and N.T. have received support from the following awards: National Institutes of Health (NIH) grants R01HL166759 and R01HL174440 and a Leducq Foundation grant.
    \section*{Role of the funding source}
The authors declare that the study sponsors had no role in the design of the study, the collection, analysis, and interpretation of data, the writing of the manuscript, or the decision to submit the manuscript for publication. All sources of funding are duly acknowledged, and no external influence has affected the integrity or independence of this research.

\section*{Declaration of competing interest}
The authors declare the following financial interests/personal relationships which may be considered as potential competing interests: R. Piersanti and L. Dede’ report financial support was provided by Ministry of University and Research, Italy. R. Piersanti reports travel support provided by National Group for Scientific Computing, Italy. S.Y. Ali and N. Trayanova report financial support was provided by National Institutes of Health and Leducq Foundation, United State of America. All the other authors declare that they have no known competing financial interests or personal relationships that could have appeared to influence the work reported in this paper.

\section*{CRediT authorship contribution statement}
\textbf{R. Piersanti}: Conceptualization, Methodology, Software, Simulation, Data curation, Formal analysis, Investigation,  Validation, Visualization, Writing – original draft, Writing – review \& editing.
\textbf{R. Bradley}: Conceptualization, Data curation.
\textbf{S.Y. Ali}: Data curation, Writing – review \& editing.
\textbf{A. Quarteroni}: Supervision, Writing – review \& editing.
\textbf{L. Dede’}: Supervision, Funding acquisition, Project administration, Writing – review \& editing.
\textbf{N. Trayanova}: Conceptualization, Supervision, Funding acquisition, Project administration, Writing – review \& editing.
    
    \vspace{6mm}
    
    \appendixpageoff
    \appendixtitleoff
    \renewcommand{\appendixtocname}{Appendix}
    \begin{appendices}
        \setcounter{figure}{0}
        \setcounter{section}{0}
        \crefalias{section}{supp}
        \let\oldthefigure\thefigure               
        \renewcommand{\thefigure}{A\oldthefigure} 
        \noindent
        \textbf{\LARGE Appendix}
        \vspace{1mm}
        
        \section{Mesh labeling procedure}
\label{supp:tagging}
The labeling procedure carried out in this work, for the atrial LDRBM (detailed in Section~\ref{subsec:ldrbm_model}) consists of the following steps (refer to Figure~\ref{fig:tagging}):
\begin{figure}[t!]
    \centering
    \includegraphics[width=1.0\textwidth]{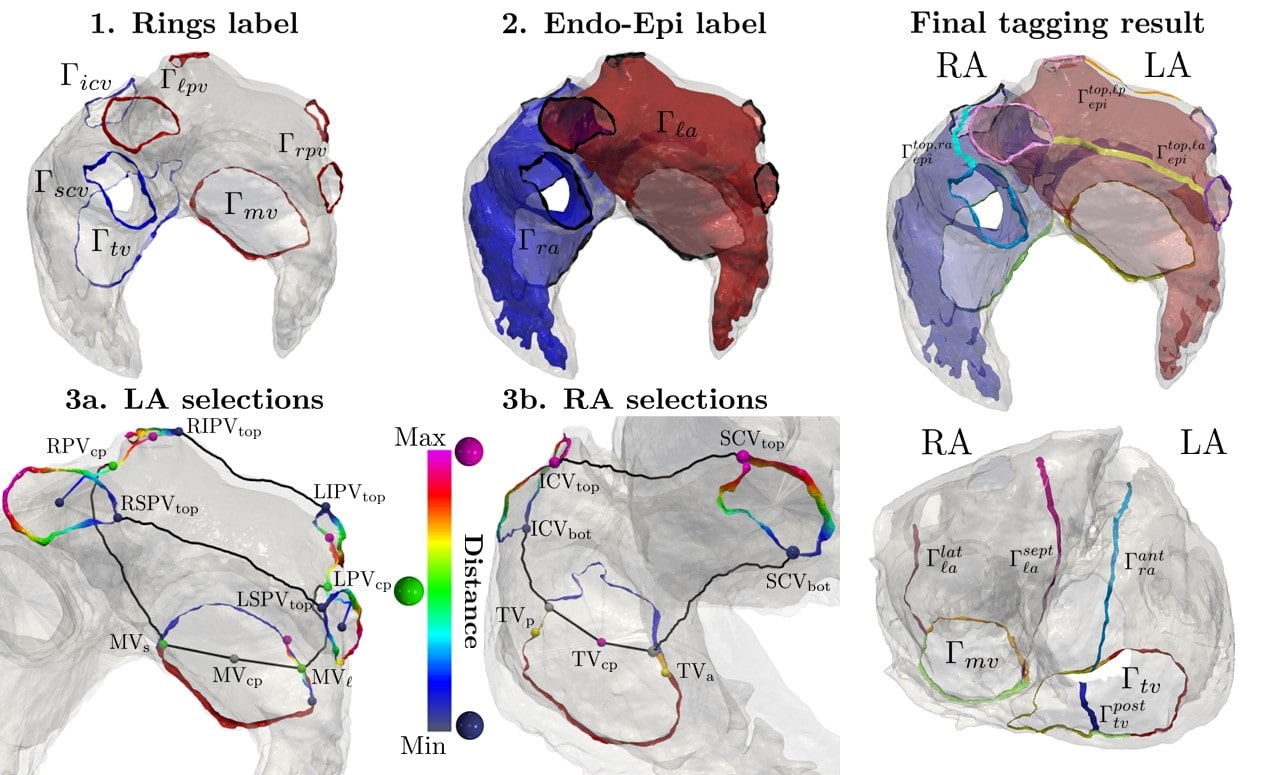}
    \caption{Atrial surface labeling procedure performed to impose the boundary conditions for the atrial LDRBM (presented in Section~\ref{subsec:ldrbm_model}): (1.) crafting the LA (in red) and RA (in blue) rings; (2.) extracting LA (red), RA (blue) endocardium and atrial epicardium (grey); (3a., 3b.) LA and RA point selection procedures applied to compute shortest distance path employed to perform the top-bands labeling ($\Gamma^{top,\ell a}_{epi}$, $\Gamma^{top,\ell p}_{epi}$ and $\Gamma^{top,ra}_{epi}$) and the annulus splitting (anterior, posterior, lateral and septal) for MV ($\Gamma_{mv}$) and TV ($\Gamma_{tv}$). On the right, the final result of the labeling procedure applied to Geo 7 of the DTMRI dataset.}    
    \label{fig:tagging}
\end{figure}
\begin{itemize}
    \item[1.] Draw the label rings for SCV, ICV, TV, LSPV, LIPV, RSPV, RIPV, MV;
    \item[2.] Identify the epicardium, LA and RA endocardia labels using a connectivity method;
    \item[3.] Find the following landmark points by means of an automatic selection procedure: $\text{LSPV}_\text{top}$/$\text{RSPV}_\text{top}$ as the minimal distance from RSPV/LSPV on LSPV/RSPV; $\text{LIPV}_\text{top}$/$\text{RIPV}_\text{top}$ as the minimal distance from RIPV/LIPV on LIPV/RIPV; $\text{SCV}_\text{top}$/$\text{ICV}_\text{top}$ as the minimal distance from ICV/SCV on SCV/ICV; $\text{LPV}_\text{cp}$/$\text{RPV}_\text{cp}$ as the barycenter of LPV/RPV considering the mean shortest-path distance point between the annulus center points of LSPV$-$LIPV for LPV and RSPV$-$RIPV for RPV;  $\text{MV}_\ell$ as the barycenter (considering the mean shortest-path distance) of two points identified by the minimal distance on MV from LSPV and LIPV, respectively; $\text{MV}_\text{cp}$/$\text{TV}_\text{cp}$ as the annulus center point of MV/TV; $\text{MV}_\text{s}$ as the diametral point from $\text{MV}_\ell$ passing from $\text{MV}_\text{cp}$; $\text{MV}_\text{s}$ as 
    the barycenter (considering the mean shortest-path distance) of two points identified by the minimal distance on MV from LSPV and LIPV, respectively; $\text{ICV}_\text{bot}$/$\text{SCV}_\text{bot}$ as the minimal distance from ICV/SCV on MV; $\text{TV}_\text{p}$/$\text{TV}_\text{a}$ as the first intersection point on MV in computing the shortest-path distance between $\text{ICV}_\text{bot}$/$\text{SCV}_\text{bot}$ and $\text{TV}_\text{cp}$
    \item[4.] Exploit a distance-tagging procedure, employing the shortest-path distance between two selected points, to identify: LA-top anterior/posterior band and RA-top band on epicardium between $\text{RSPV}_\text{top}-\text{LSPV}_\text{top}$ (LA-top anterior), $\text{RIPV}_\text{top}-\text{LIPV}_\text{top}$ (LA-top posterior) and $\text{SCV}_\text{top}-\text{ICV}_\text{top}$ (RA-top); the lateral/septal LA band on LA endocardium between $\text{LPV}_\text{cp}$/$\text{RPV}_\text{cp}$ and $\text{MV}_\ell$/$\text{MV}_\text{s}$; the anterior/posterior RA band on RA endocardium between $\text{SCV}_\text{bot}$/$\text{ICV}_\text{bot}$ and $\text{TV}_\text{cp}$; the MV annulus anterior and posterior portions between $\text{MV}_\ell$ and $\text{MV}_\text{s}$; the TV annulus anterior, posterior, lateral and septal portions of TV annulus.     
\end{itemize}
\clearpage
\section{Cardiac electrophysiology modeling}
\label{supp:ep_model}
We provide a concise overview of the modeling approach used for representing EP activity in atrial tissue, known as the Eikonal-diffusion model~\cite{franzone2014mathematical,quarteroni2019mathematical}. This model, a steady advection-diffusion equation, is employed to reconstruct the macroscopic propagation of action potential excitation wavefronts during the depolarization phase, thereby enabling the recovery of activation time as a spatial function within the myocardium~\cite{colli1990wavefront,franzone1993spreading,nagel2022comparison}.

The cardiac muscle is an orthotropic material, arising from its cellular organization into fibers, laminar sheets, and sheet normals, which is mathematically described by the conductivity tensor
\begin{equation}
    \label{eqn:Tensdiff}
    \EPdiffTens = \sigma_f \fZero \otimes \fZero + \sigma_s \sZero \otimes \sZero + \sigma_n \nZero \otimes \nZero,
\end{equation}
where $\sigma_f$, $\sigma_s$ and $\sigma_n$ are the conductivities along fiber $\fZero$, sheet $\sZero$, and sheet-normal $\nZero$ directions
\footnote{Notice that, in equation \eqref{eqn:Tensdiff}, the conductivity values, $\sigma_k$ with $k=f,s,n$, are measured in \unit{\meter^2 \per\second}.}.

We model the electrical activity, within atrial domain $\Omega_{bia}$, with the following Eikonal-diffusion equation
\begin{equation}
    \label{eqn:eikonal}
    \begin{cases}
        c_f \sqrt{\nabla \EPpot \cdot \EPdiffTens \nabla \EPpot} - \nabla \cdot \left( \EPdiffTens \nabla \EPpot \right) = 1 &\qquad{\text{in }}\Omega_{bia},
        \\
        \left(\EPdiffTens \nabla \EPpot\right) \cdot \boldsymbol{n} = 0  &\qquad{\text{on }}\partial \Omega_{bia} / S_{\text{san}} ,
        \\
        u_A = u_{0} &\qquad{\text{on }}S_{\text{san}},
    \end{cases}
\end{equation}
where the unknown $\EPpot=\EPpot(\boldsymbol{x}):\Omega_{bia} \rightarrow \mathbb{R}$, termed activation time, represents the time at which the depolarization wavefront reaches the point $\boldsymbol{x}$; $S_{\text{san}}$ denotes the portion of the physical boundary where the activation time $u_{0}$ originates, mimicking the onset of atrial activation near the sino-atrial node~\cite{ho2009importance,sakamoto2005interatrial}, and $\boldsymbol{n}$ is the outward directed unit vector normal to the boundary $\partial\Omega_{bia}$ of the domain $\Omega_{bia}$. The parameter $c_f$, uniform across the domain $\Omega_{bia}$, is the velocity of the action potential depolarization planar wavefront along the fiber direction in an infinite cable, under the assumption of a unit surface-to-volume ratio, membrane capacitance, and conductivity\footnote{The parameter $c_f$ is measured in in \unit{s^{-1/2}}.}~\cite{franzone2014mathematical,franzone1993spreading}. From the Eikonal equation, the following formula for the conduction velocity along the fiber direction can be derived~\cite{jacquemet2012eikonal,corrado2018conduction}
\begin{equation}
    \label{eqn:velocity}
    v_f=c_f\sqrt{\sigma_f}.
\end{equation}

We numerically solve the Eikonal-diffusion problem \eqref{eqn:eikonal}, by recovering the steady-state solution of the following parabolic pseudo-time problem
\begin{equation}
    \label{eqn:pseudo_time}
    \frac{\partial \EPpot}{\partial t} + c_f \sqrt{\nabla \EPpot \cdot \EPdiffTens \nabla \EPpot} - \nabla \cdot \left( \EPdiffTens \nabla \EPpot \right) = 1 \qquad{\text{in }}\Omega_{bia},
\end{equation}
with the same boundary and initial conditions as in \eqref{eqn:eikonal}, and $t$ representing the pseudo-time~\cite{stella2022fast}. 

We performed the time discretization of the pseudo-time problem~\eqref{eqn:pseudo_time} employing a fully implicit BDF, used in combination with the Newton algorithm~\cite{stella2022fast}. As for the space discretization, we used continuous FE on tetrahedral meshes~\cite{quarteroni2019mathematical}. The resulting linear systems arising at each pseudo-time step were solved by the GMRES method preconditioned with the AMG preconditioner~\cite{brenner2008mathematical,saad1986gmres}.

It's worth emphasizing that the Eikonal-diffusion problem \eqref{eqn:eikonal} offers superior accuracy compared to the pure Eikonal model (i.e., without considering any diffusive term in \eqref{eqn:eikonal}), as it takes into account the effects of wavefront curvature or the interaction between a wavefront with either the domain boundaries or with other fronts~\cite{franzone2014mathematical,gander2023accuracy}. For more comprehensive insights into the Eikonal-diffusion model, we refer the reader to~\cite{franzone2014mathematical,colli2006computational,quarteroni2019mathematical}.
\clearpage
\section{DTMRI atrial fiber measurements}
\label{supp:measurement_dtmri_fiber}
\begin{figure}[ht!]
    \centering
    \includegraphics[width=1.0\textwidth]{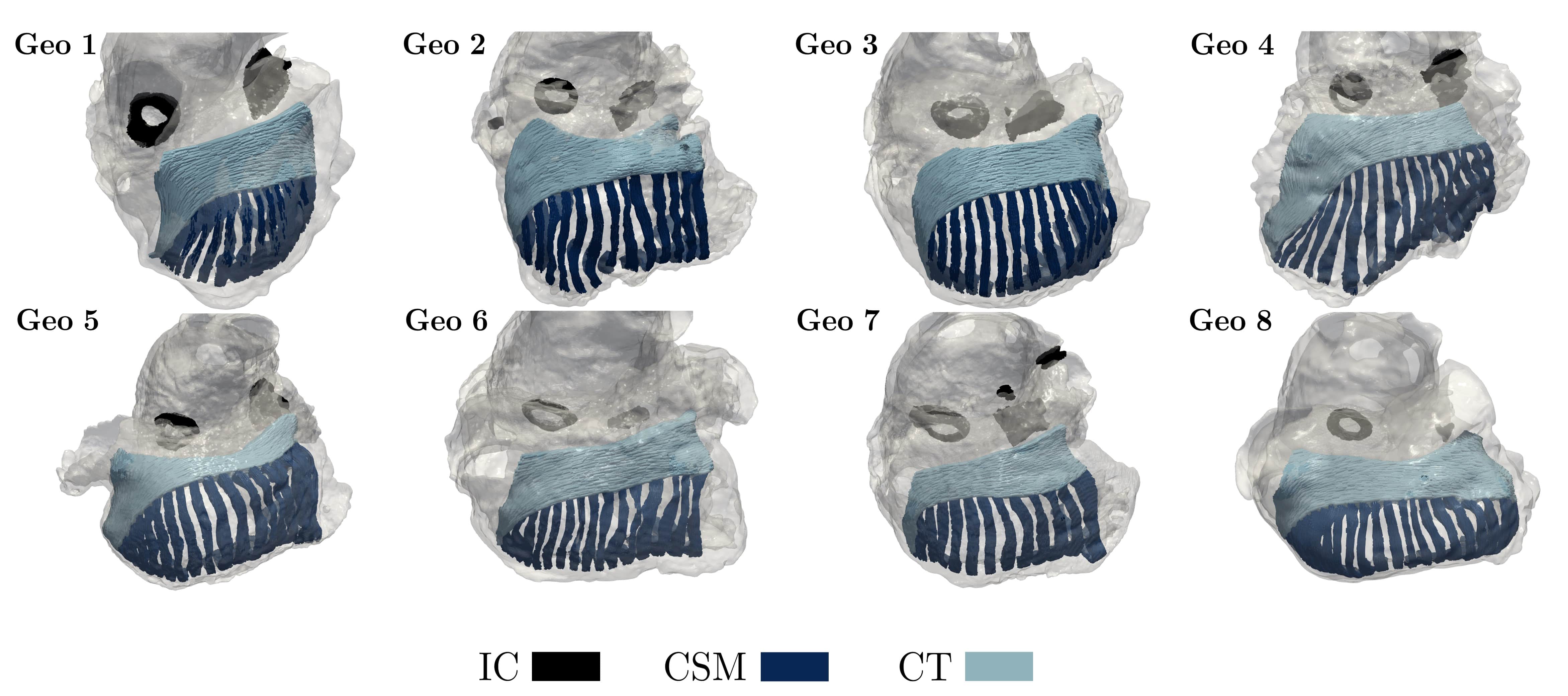}
    \caption{IC (black), PM (dark blue) and CT (light blue) bundles extraction performed by the bi-atrial LDRBM for DTMRI geometries. IC=Interatrial-Connections, PM=Pectinate Muscles, CT=Crista Terminalis.}
    \label{fig:fiber_bundles}
\end{figure}
\begin{table}[ht!]
    \centering
    \scalebox{1}{\begin{tabular}{ |c|c|c|c|c|c|c|c|c|c|c|c| } 
            \hline
            Type & $\tau$ & Geo 1 & Geo 2 & Geo 3 & Geo 4 & Geo 5 & Geo 6 & Geo 7 & Geo 8 & Mean & SD \\
            \hline
            \hline
            \multirow{10}{*}{IC} 
            & $\tau^{ic,r}_{bb}$    & -0.80 & -0.50 & -0.70 & -0.50 & -0.50 & -0.50 & -0.75 & -0.50 & \textbf{-0.59} & \gradientcell{0.13}{0}{0.32}{white}{red}{60} \\ 
            & $\tau^{ic,\ell}_{bb}$ & 0.80  &  0.95 &  0.80 &  0.90 &  0.80 &  0.99 &  0.85 &  0.95 & \textbf{0.88}  & \gradientcell{0.08}{0}{0.32}{white}{red}{60} \\ 
            & $\tau^{ic}_{bb}$      & 0.33  &  0.52 &  0.61 &  0.55 &  0.21 &  0.71 &  0.54 &  0.57 & \textbf{0.50}  & \gradientcell{0.16}{0}{0.32}{white}{red}{60} \\ 
            & $\tau^{ic,r}_{fo}$    & -0.80 & -0.80 & -0.80 & -0.80 & -0.80 & -0.80 & -0.80 & -0.80 & \textbf{-0.80} & \gradientcell{0.00}{0}{0.32}{white}{red}{60}    \\ 
            & $\tau^{ic,\ell}_{fo}$ & 0.80  &  0.80 &  0.80 &  0.80 &  0.80 &  0.80 &  0.80 &  0.80 & \textbf{0.80}  & \gradientcell{0.00}{0}{0.32}{white}{red}{60}    \\ 
            & $\tau^{ic}_{fo}$      & 0.41  &  0.11 &  0.66 &  0.54 &  0.54 &  0.38 &  0.29 & -0.03 & \textbf{0.36}  & \gradientcell{0.23}{0}{0.32}{white}{red}{60} \\ 
            & $\tau^{ic,in}_{fo}$   &  0.44 &  0.17 &  0.68 &  0.55 &  0.56 &  0.42 &  0.32 &  0.03 & \textbf{0.40}  & \gradientcell{0.21}{0}{0.32}{white}{red}{60} \\ 
            & $\tau^{ic,r}_{cs}$    & ---   & -0.45 &  ---  &  ---  &  ---  &  ---  &  ---  &  ---  & \textbf{-0.45} & \gradientcell{0.00}{0}{0.32}{white}{red}{60}    \\ 
            & $\tau^{ic,\ell}_{cs}$ & ---   &  0.45 &  ---  &  ---  &  ---  &  ---  &  ---  &  ---  & \textbf{0.45}  & \gradientcell{0.00}{0}{0.32}{white}{red}{60}    \\ 
            & $\tau^{ic}_{cs}$      & ---   &  0.40 &  ---  &  ---  &  ---  &  ---  &  ---  &  ---  & \textbf{0.40}  & \gradientcell{0.00}{0}{0.32}{white}{red}{60}    \\ 
            \hline          
    \end{tabular}}
    \caption{Bundle parameter values selected for the inter-atrial connections (BIA) across the eight DTMRI geometries. Mean and Standard Deviation (SD) values are provided, with SD color-coded on a scale from minimal (0) to maximal (0.32).}
    \label{tab:tau_bundles_BIA}
\end{table}
\begin{table}[ht!]
    \centering
    \scalebox{0.85}{\begin{tabular}{ |c|c|c|c|c|c|c|c|c|c|c|c| } 
            \hline
            Type & $\tau$ & Geo 1 & Geo 2 & Geo 3 & Geo 4 & Geo 5 & Geo 6 & Geo 7 & Geo 8 & Mean & SD \\
            \hline
            \hline
            \multirow{20}{*}{LA}
            & $\tau_{mv}$                   & 0.80  & 0.85   & 0.83   & 0.80   & 0.85   & 0.80   & 0.80  & 0.85   & \textbf{0.82}  & \gradientcell{0.02}{0}{0.32}{white}{red}{60}  \\ 
            & $\tau_{\ell pv}$              & 0.20  & 0.15   & 0.18   & 0.20   & 0.20   & 0.20   & 0.15  & 0.15   & \textbf{0.18}  & \gradientcell{0.02}{0}{0.32}{white}{red}{60}  \\ 
            & $\tau^{up}_{\ell pv}$         & 0.25  & 0.15   & 0.25   & 0.25   & 0.25   & 0.25   & 0.15  & 0.25   & \textbf{0.22}  & \gradientcell{0.04}{0}{0.32}{white}{red}{60}  \\ 
            & $\tau_{\ell ipv}$             & -0.40 & -0.70  & -0.40  & -0.40  & -0.40  & -0.50  & -0.40 & -0.40  & \textbf{-0.45} & \gradientcell{0.10}{0}{0.32}{white}{red}{60}  \\ 
            & $\tau_{\ell spv}$             & 0.20  & 0.20   & 0.20   & 0.10   & 0.10   & 0.40   & 0.30  & 0.20   & \textbf{0.21}  & \gradientcell{0.10}{0}{0.32}{white}{red}{60}  \\ 
            & $\tau_{rpv}$                  & 0.80  & 0.80   & 0.85   & 0.85   & 0.85   & 0.86   & 0.80  & 0.84   & \textbf{0.83}  & \gradientcell{0.03}{0}{0.32}{white}{red}{60}  \\ 
            & $\tau^{up}_{rpv}$             & 0.13  & 0.13   & 0.17   & 0.15   & 0.12   & 0.13   & 0.15  & 0.15   & \textbf{0.14}  & \gradientcell{0.02}{0}{0.32}{white}{red}{60}  \\ 
            & $\tau_{ripv}$                 & -0.30 & -0.10  & -0.15  & -0.60  & -0.20  & -0.30  & -0.40 & -0.40  & \textbf{-0.31} & \gradientcell{0.16}{0}{0.32}{white}{red}{60}  \\ 
            & $\tau_{rspv}$                 & 0.10  & 0.20   & 0.35   & 0.40   & 0.30   & 0.30   & 0.40  & 0.40   & \textbf{0.31}  & \gradientcell{0.11}{0}{0.32}{white}{red}{60}  \\ 
            & $\tau_{\ell aa}$              & 0.18  & -0.16  & 0.53   & 0.34   & 0.35   & 0.15   & 0.12  & -0.25  & \textbf{0.16}  & \gradientcell{0.26}{0}{0.32}{white}{red}{60}  \\ 
            & $\tau_{\ell as}$              & 0.015 & -0.030 & -0.010 & -0.015 & -0.010 & -0.040 & 0.025 & -0.030 & \textbf{-0.012}& \gradientcell{0.02}{0}{0.32}{white}{red}{60}  \\ 
            & $\tau^{\ell a}_{p \ell w}$    & 0.35  & 0.38   & 0.35   & 0.20   & 0.35   & 0.52   & 0.36  & 0.32   & \textbf{0.35}  & \gradientcell{0.09}{0}{0.32}{white}{red}{60}  \\ 
            & $\tau^{\ell a,up}_{p \ell w}$ & 0.25  & 0.25   & 0.11   & 0.25   & 0.15   & 0.12   & 0.25  & 0.25   & \textbf{0.19}  & \gradientcell{0.06}{0}{0.32}{white}{red}{60}  \\ 
            & $\tau^{\ell a}_{a \ell w}$    & 0.26  & 0.24   & 0.20   & 0.25   & 0.37   & 0.22   & 0.28  & 0.25   & \textbf{0.26}  & \gradientcell{0.05}{0}{0.32}{white}{red}{60}  \\ 
            & $\tau^{\ell a}_{psw}$         & 0.88  & 0.83   & 0.89   & 0.75   & 0.78   & 0.93   & 0.80  & 0.78   & \textbf{0.83}  & \gradientcell{0.06}{0}{0.32}{white}{red}{60}  \\ 
            & $\tau^{\ell a,up}_{psw}$      & 0.25  & 0.25   & 0.10   & 0.25   & 0.15   & 0.25   & 0.25  & 0.25   & \textbf{0.22}  & \gradientcell{0.06}{0}{0.32}{white}{red}{60}  \\ 
            & $\tau^{\ell a}_{asw}$         & 0.93  & 0.85   & 0.85   & 0.82   & 0.90   & 0.80   & 0.85  & 0.80   & \textbf{0.85}  & \gradientcell{0.04}{0}{0.32}{white}{red}{60}  \\ 
            & $\tau_{\ell ar}$              & 0.28  & 0.10   & 0.18   & 0.38   & 0.20   & 0.10   & 0.33  & 0.15   & \textbf{0.21}  & \gradientcell{0.10}{0}{0.32}{white}{red}{60}  \\ 
            & $\tau_{\ell aw}$              & 0.42  & 0.54   & 0.46   & 0.50   & 0.50   & 0.56   & 0.45  & 0.30   & \textbf{0.47}  & \gradientcell{0.08}{0}{0.32}{white}{red}{60}  \\ 
            & $\tau_{bb}$                   & 0.45  & 0.41   & 0.35   & 0.25   & 0.28   & 0.25   & 0.38  & 0.42   & \textbf{0.35}  & \gradientcell{0.08}{0}{0.32}{white}{red}{60}  \\ 
            \hline          
    \end{tabular}}
    \caption{Bundle parameter values selected for LA bundles across the eight DTMRI geometries. Mean and Standard Deviation (SD) values are provided, with SD color-coded on a scale from minimal (0) to maximal (0.32).}
    \label{tab:tau_bundles_LA}
\end{table}
\begin{table}[ht!]
    \centering
    \scalebox{0.85}{\begin{tabular}{ |c|c|c|c|c|c|c|c|c|c|c|c| } 
            \hline
            Type & $\tau$ & Geo 1 & Geo 2 & Geo 3 & Geo 4 & Geo 5 & Geo 6 & Geo 7 & Geo 8 & Mean & SD \\
            \hline
            \hline
            \multirow{28}{*}{RA} 
            & $\tau_{tv}$              & 0.80  & 0.85  & 0.80  & 0.85  & 0.85  & 0.85  & 0.85  & 0.85  & \textbf{0.84}  & \gradientcell{0.02}{0}{0.32}{white}{red}{60} \\ 
            & $\tau^{aa}_{tv}$         & 0.60  & 0.60  & 0.60  & 0.60  & 0.60  & 0.60  & 0.60  & 0.60  & \textbf{0.60}  & \gradientcell{0.00}{0}{0.32}{white}{red}{60} \\ 
            & $\tau_{icv}$             & 0.85  & 0.80  & 0.90  & 0.90  & 0.85  & 0.90  & 0.80  & 0.85  & \textbf{0.86}  & \gradientcell{0.04}{0}{0.32}{white}{red}{60} \\ 
            & $\tau^{up}_{icv}$        & 0.25  & 0.25  & 0.25  & 0.25  & 0.15  & 0.25  & 0.30  & 0.25  & \textbf{0.24}  & \gradientcell{0.04}{0}{0.32}{white}{red}{60} \\ 
            & $\tau_{scv}$             & 0.20  & 0.30  & 0.35  & 0.40  & 0.20  & 0.50  & 0.40  & 0.25  & \textbf{0.32}  & \gradientcell{0.11}{0}{0.32}{white}{red}{60} \\ 
            & $\tau^{up}_{scv}$        & 0.25  & 0.30  & 0.30  & 0.20  & 0.16  & 0.30  & 0.40  & 0.28  & \textbf{0.27}  & \gradientcell{0.07}{0}{0.32}{white}{red}{60} \\ 
            & $\tau_{raa}$             & 0.25  & 0.33  &  0.38 &  0.34 & 0.17  & 0.40  & 0.39  & 0.15  & \textbf{0.30}  & \gradientcell{0.10}{0}{0.32}{white}{red}{60} \\ 
            & $\tau^{w}_{raa}$         & -0.25 & -0.32 & -0.05 & -0.10 & -0.22 & -0.12 & -0.10 & -0.33 & \textbf{-0.18} & \gradientcell{0.11}{0}{0.32}{white}{red}{60} \\ 
            & $\tau^{up}_{raa}$        & 0.64  &  0.50 &  0.42 &  0.20 &  0.54 & 0.42  &  0.63 &  0.60 & \textbf{0.49}  & \gradientcell{0.15}{0}{0.32}{white}{red}{60} \\ 
            & $\tau_{csm}$             & 0.30  &  0.65 &  0.50 &  ---  & -0.30 & 0.25  &  0.75 &  0.45 & \textbf{0.37}  & \gradientcell{0.32}{0}{0.32}{white}{red}{60} \\ 
            & $\tau^{v}_{csm}$         & 0.40  &  0.55 &  0.72 &  ---  &  0.30 & 0.50  &  0.60 &  0.40 & \textbf{0.51}  & \gradientcell{0.13}{0}{0.32}{white}{red}{60} \\ 
            & $\tau_{ras}$             & -0.05 & -0.15 & -0.02 & -0.16 & -0.15 & -0.02 & -0.01 & -0.10 & \textbf{-0.08} & \gradientcell{0.06}{0}{0.32}{white}{red}{60} \\ 
            & $\tau^{ra}_{asw}$        & 0.85  & 0.65  & 0.60  & 0.65  & 0.75  & 0.75  & 0.80  & 0.60  & \textbf{0.71}  & \gradientcell{0.09}{0}{0.32}{white}{red}{60} \\ 
            & $\tau^{ra,up}_{asw}$     & 0.50  & 0.20  & 0.25  & 0.25  & 0.25  & 0.15  & 0.25  & 0.25  & \textbf{0.26}  & \gradientcell{0.10}{0}{0.32}{white}{red}{60} \\ 
            & $\tau^{ra}_{a\ell w}$    & 0.70  & 0.65  & 0.80  & 0.65  & 0.84  & 0.80  & 0.66  & 0.78  & \textbf{0.73}  & \gradientcell{0.08}{0}{0.32}{white}{red}{60} \\ 
            & $\tau^{ra}_{psw}$        & -0.55 & -0.70 & -0.70 & -0.70 & -0.75 & -0.55 & -0.40 & -0.67 & \textbf{-0.63} & \gradientcell{0.12}{0}{0.32}{white}{red}{60} \\ 
            & $\tau^{ra,up}_{psw}$     & 0.75  &  0.40 &  0.50 &  0.40 &  0.50 &  0.50 &  0.50 &  0.50 & \textbf{0.51}  & \gradientcell{0.11}{0}{0.32}{white}{red}{60} \\ 
            & $\tau^{ra}_{p\ell w}$    & -0.55 & -0.60 & -0.70 & -0.75 & -0.70 & -0.60 & -0.46 & -0.70 & \textbf{-0.63} & \gradientcell{0.10}{0}{0.32}{white}{red}{60} \\ 
            & $\tau^{ra,up}_{p\ell w}$ & 0.70  &  0.40 &  0.50 &  0.25 &  0.50 &  0.50 &  0.50 &  0.50 & \textbf{0.48}  & \gradientcell{0.12}{0}{0.32}{white}{red}{60} \\ 
            & $\tau^{\ell}_{ib}$       & 0.10  & 0.34  &  0.30 &  0.20 &  0.33 &  0.35 &  0.36 &  0.53 & \textbf{0.31}  & \gradientcell{0.12}{0}{0.32}{white}{red}{60} \\ 
            & $\tau^{s}_{ib}$          & 0.58  & 0.25  & 0.10  & 0.22  & 0.35  & 0.10  &  0.20 &  0.22 & \textbf{0.25}  & \gradientcell{0.15}{0}{0.32}{white}{red}{60} \\ 
            & $\tau^{+}_{ct}$          & -0.45 &  0.01 &  0.10 &  0.00 & -0.23 & -0.44 & -0.20 & -0.25 & \textbf{-0.18} & \gradientcell{0.20}{0}{0.32}{white}{red}{60} \\ 
            & $\tau^{-}_{ct}$          & -0.51 & -0.12 &  0.00 & -0.33 & -0.43 & -0.54 & -0.33 & -0.40 & \textbf{-0.33} & \gradientcell{0.19}{0}{0.32}{white}{red}{60} \\ 
            & $\tau^{\phi}_{ct}$       & 0.60  & 0.60  & 0.60  & 0.60  & 0.60  & 0.60  & 0.60  & 0.60  & \textbf{0.60}  & \gradientcell{0.00}{0}{0.32}{white}{red}{60} \\ 
            & $\text{pm}_{end}$        & 0.86  & 0.83  & 0.92  & 0.91  & 0.89  & 0.86  & 0.80  & 0.86  & \textbf{0.87}  & \gradientcell{0.04}{0}{0.32}{white}{red}{60} \\ 
            & $\text{pm}_{tk}$         & 0.015 & 0.015 & 0.015 & 0.015 & 0.015 & 0.015 & 0.015 & 0.015 & \textbf{0.015} & \gradientcell{0.00}{0}{0.32}{white}{red}{60} \\ 
            & $\text{pm}_{rg}$         & 0.015 & 0.015 & 0.015 & 0.015 & 0.015 & 0.015 & 0.015 & 0.015 & \textbf{0.015} & \gradientcell{0.00}{0}{0.32}{white}{red}{60} \\ \cline{2-12} 
            & $N_\text{pm}$            & 17    & 14    & 16    & 17    & 19    & 16    & 13    & 23    & \textbf{17}    & 3 \\ 
            \hline          
    \end{tabular}}
    \caption{Bundle parameter values selected for RA bundles across the eight DTMRI geometries. Mean and Standard Deviation~(SD) values are provided, with SD color-coded on a scale from minimal (0) to maximal (0.32).}
    \label{tab:tau_bundles_RA}
\end{table}
\clearpage
\begin{figure}[ht!]
    \centering
    \includegraphics[width=1.0\textwidth]{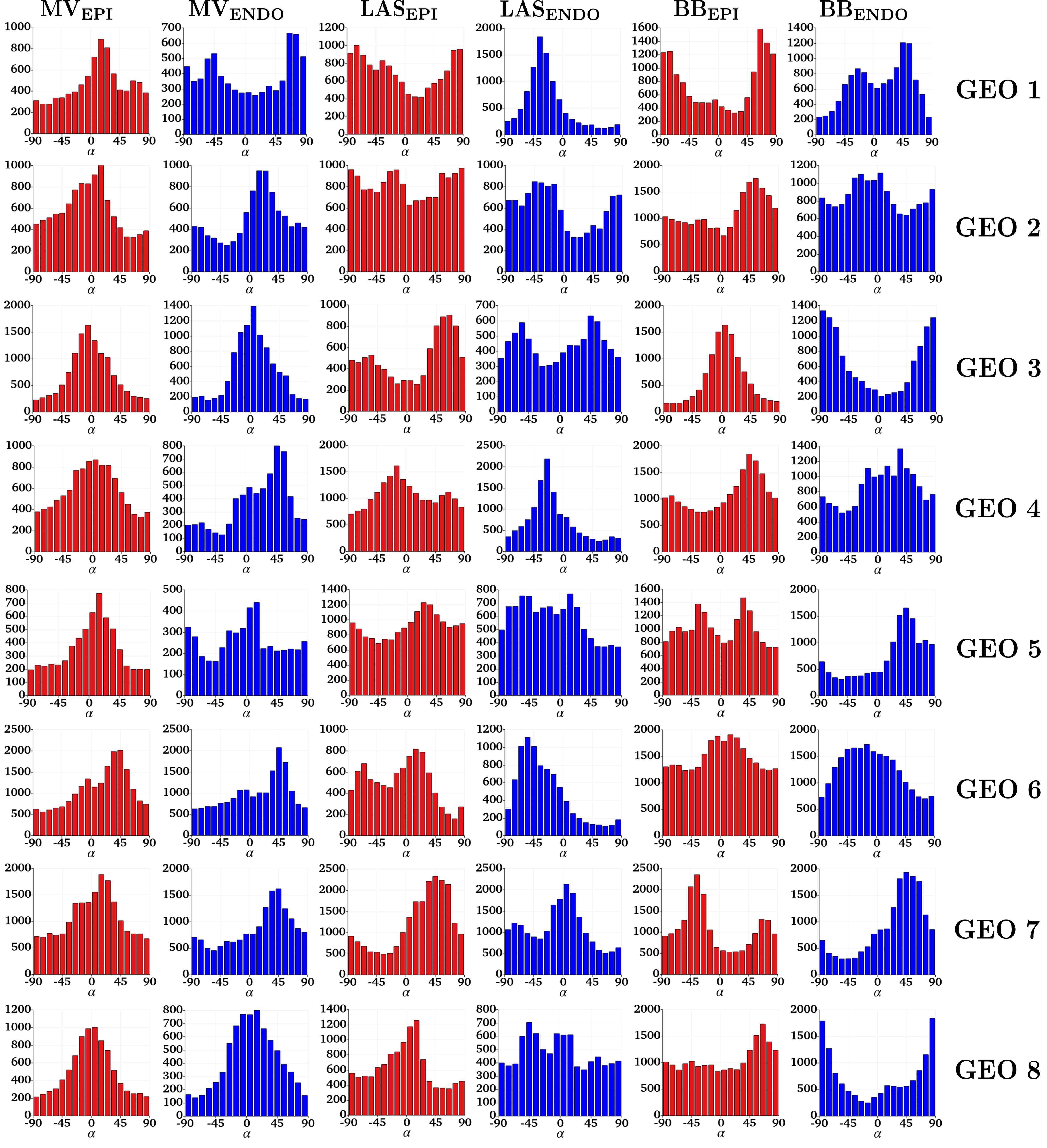}
    \caption{Histograms  displaying the measured alpha angle (with respect to LDRBM coordinate axis system) across the eight DTMRI geometries in LA bundles (MV, LAS, BB) in the sub-epicardial (red) and sub-endocardial (blue) layers.}
    \label{fig:histo_LA_1}
\end{figure}
\begin{figure}[ht!]
    \centering
    \includegraphics[width=1.0\textwidth]{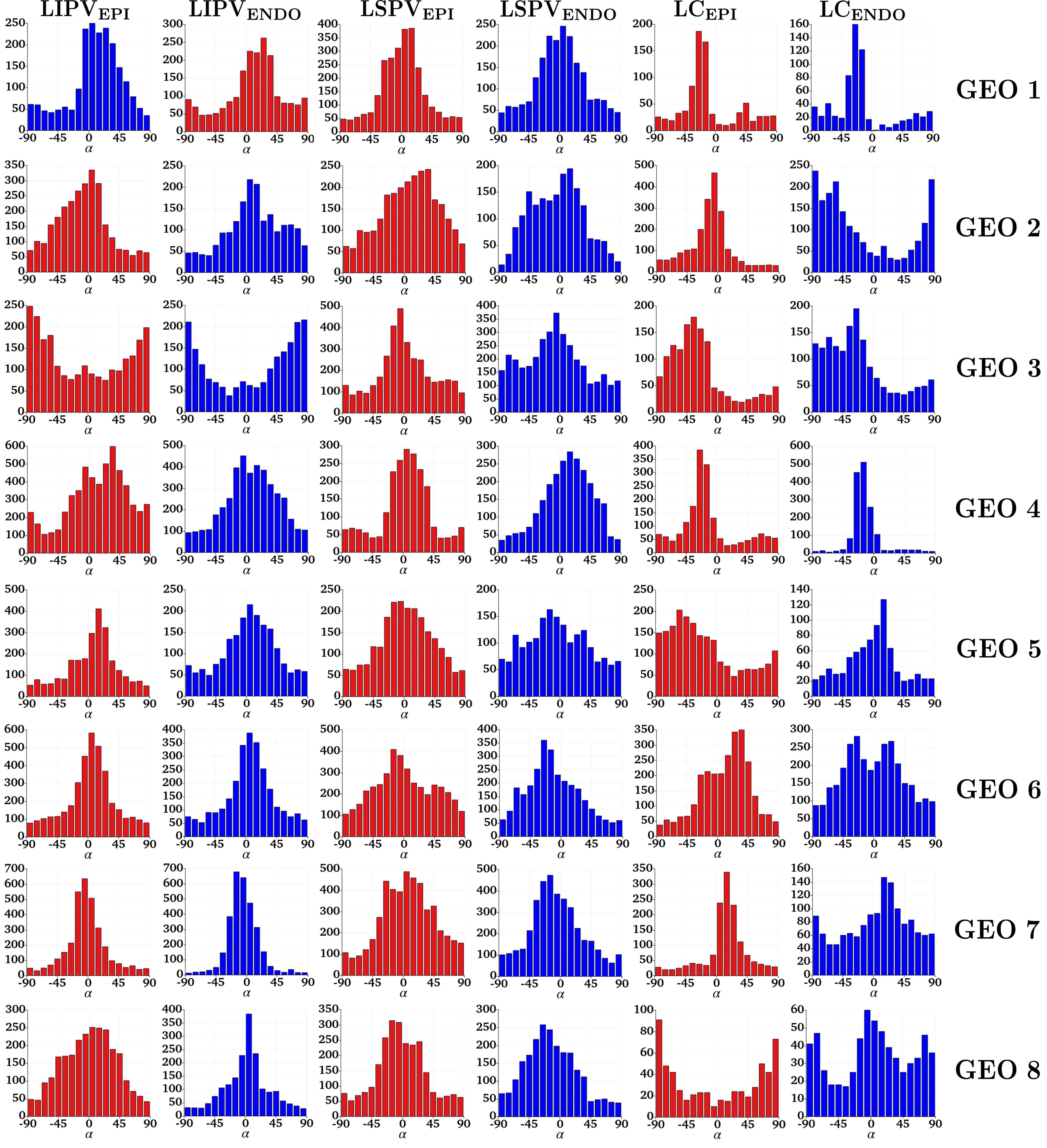}
    \caption{Histograms  displaying the measured alpha angle (with respect to LDRBM coordinate axis system) across the eight DTMRI geometries in LA bundles (LIPV, LSPV, LC) in the sub-epicardial (red) and sub-endocardial (blue) layers.}
    \label{fig:histo_LA_2}
\end{figure}
\begin{figure}[ht!]
    \centering
    \includegraphics[width=1.0\textwidth]{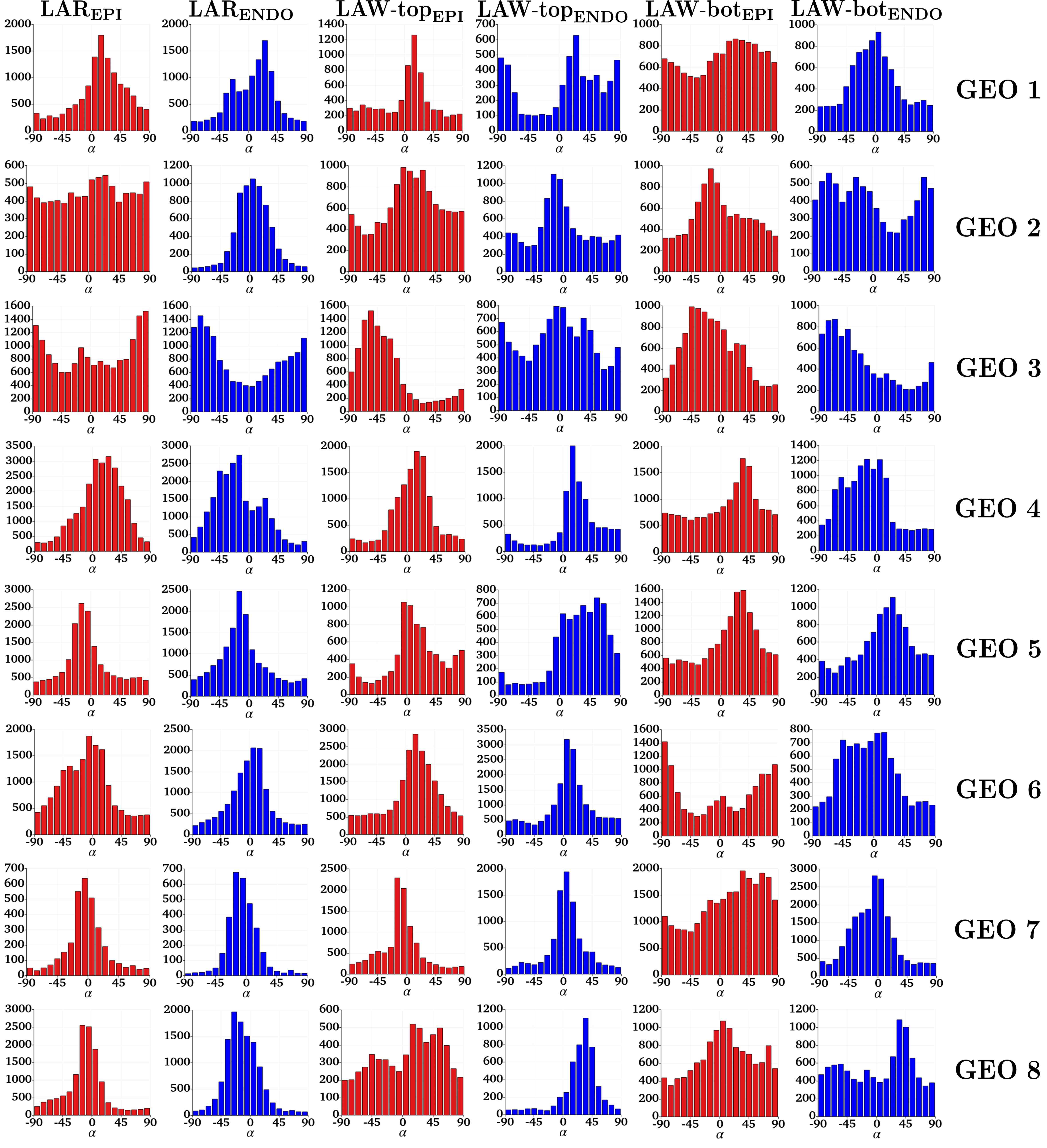}
    \caption{Histograms  displaying the measured alpha angle (with respect to LDRBM coordinate axis system) across the eight DTMRI geometries in LA bundles (LAR, LAW-top, LAW-bot) in the sub-epicardial (red) and sub-endocardial (blue) layers.}
    \label{fig:histo_LA_3}
\end{figure}
\begin{figure}[ht!]
    \centering
    \includegraphics[width=1.0\textwidth]{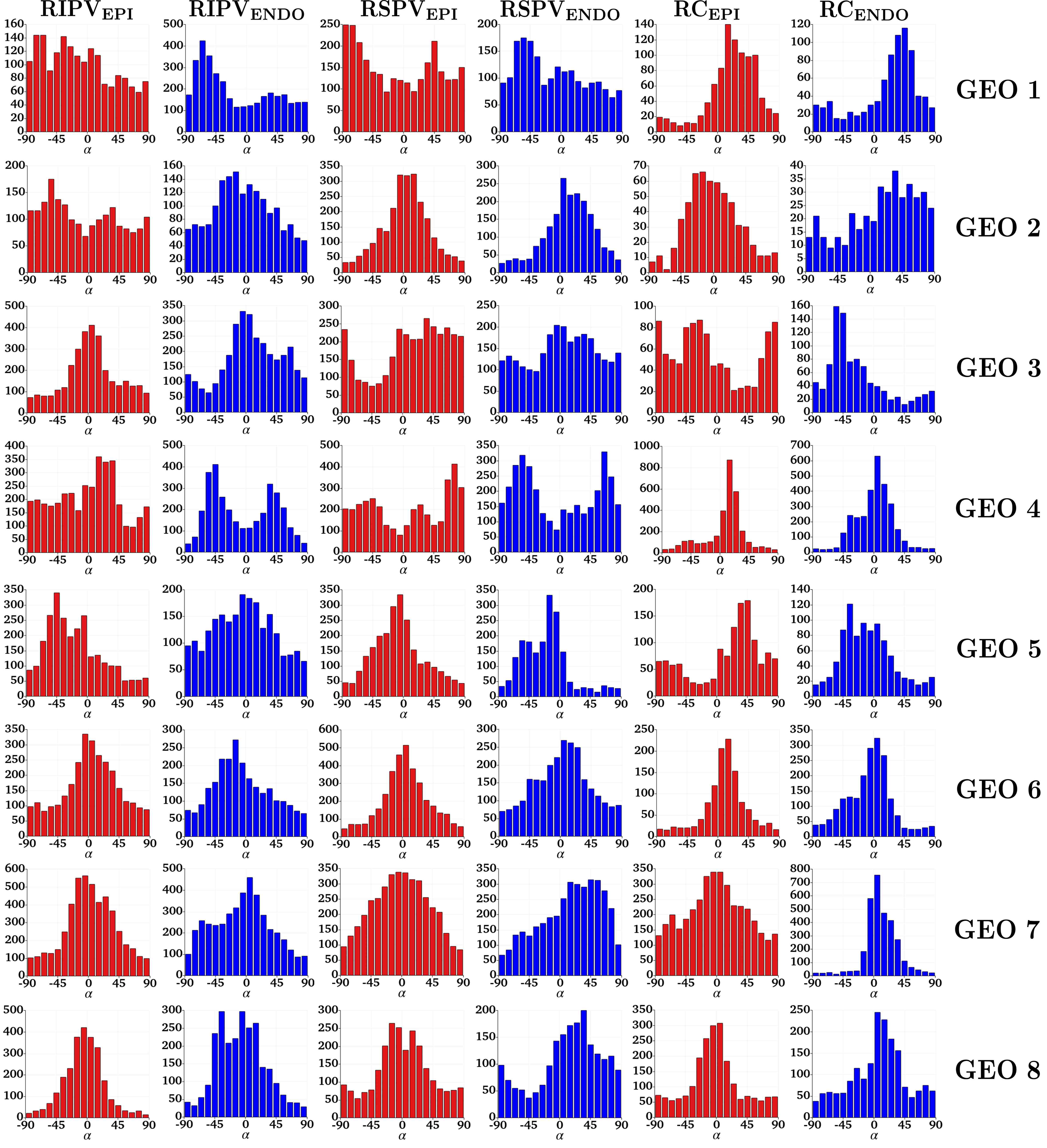}
    \caption{Histograms  displaying the measured alpha angle (with respect to LDRBM coordinate axis system) across the eight DTMRI geometries in LA bundles (RIPV, RSPV, RC) in the sub-epicardial (red) and sub-endocardial (blue) layers.}
    \label{fig:histo_LA_4}
\end{figure}
\begin{figure}[ht!]
    \centering
    \includegraphics[width=1.0\textwidth]{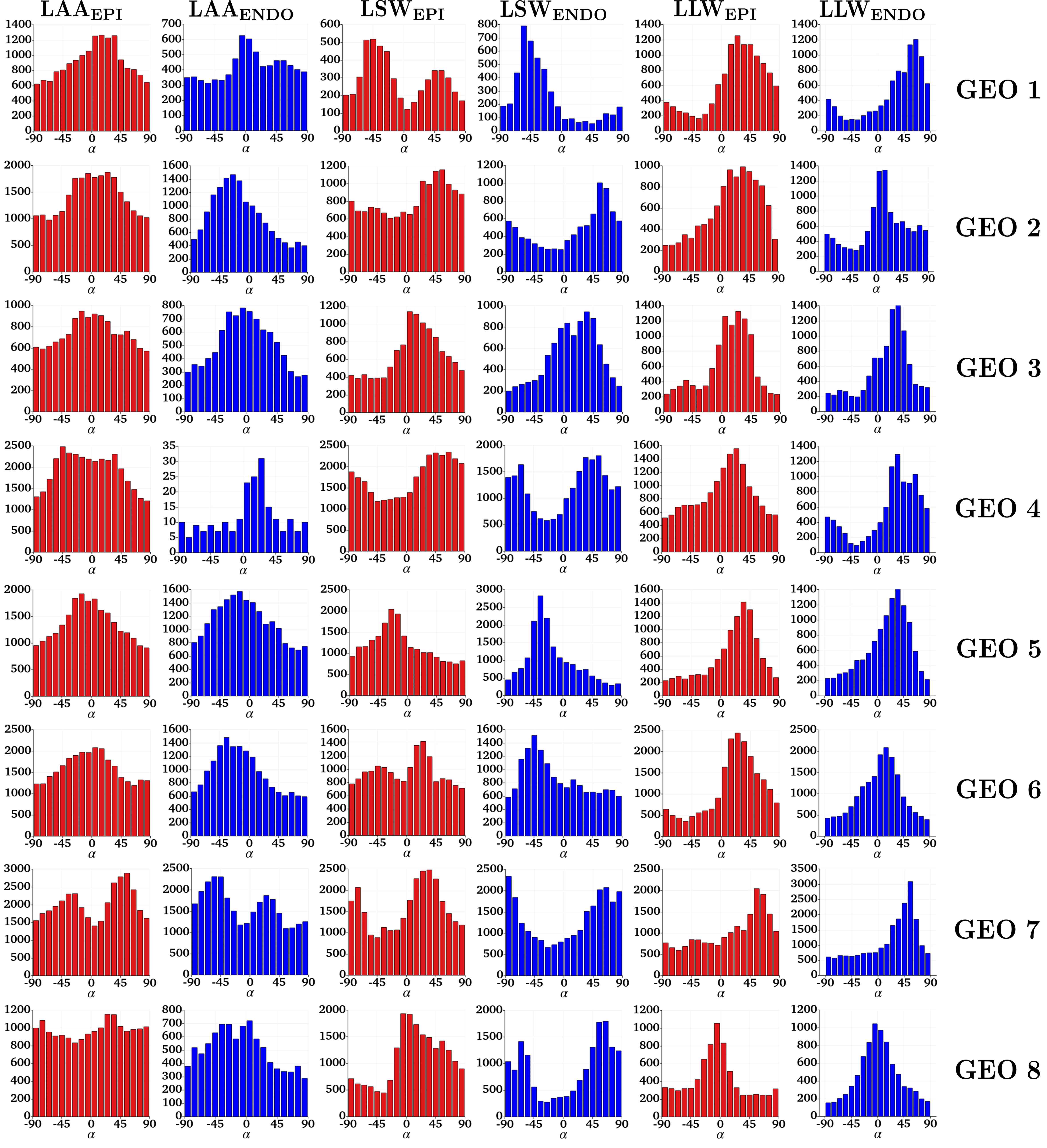}
    \caption{Histograms  displaying the measured alpha angle (with respect to LDRBM coordinate axis system) across the eight DTMRI geometries in LA bundles (LAA, LSW, LLW) in the sub-epicardial (red) and sub-endocardial (blue) layers.}
    \label{fig:histo_LA_5}
\end{figure}
\begin{figure}[ht!]
    \centering
    \includegraphics[width=1.0\textwidth]{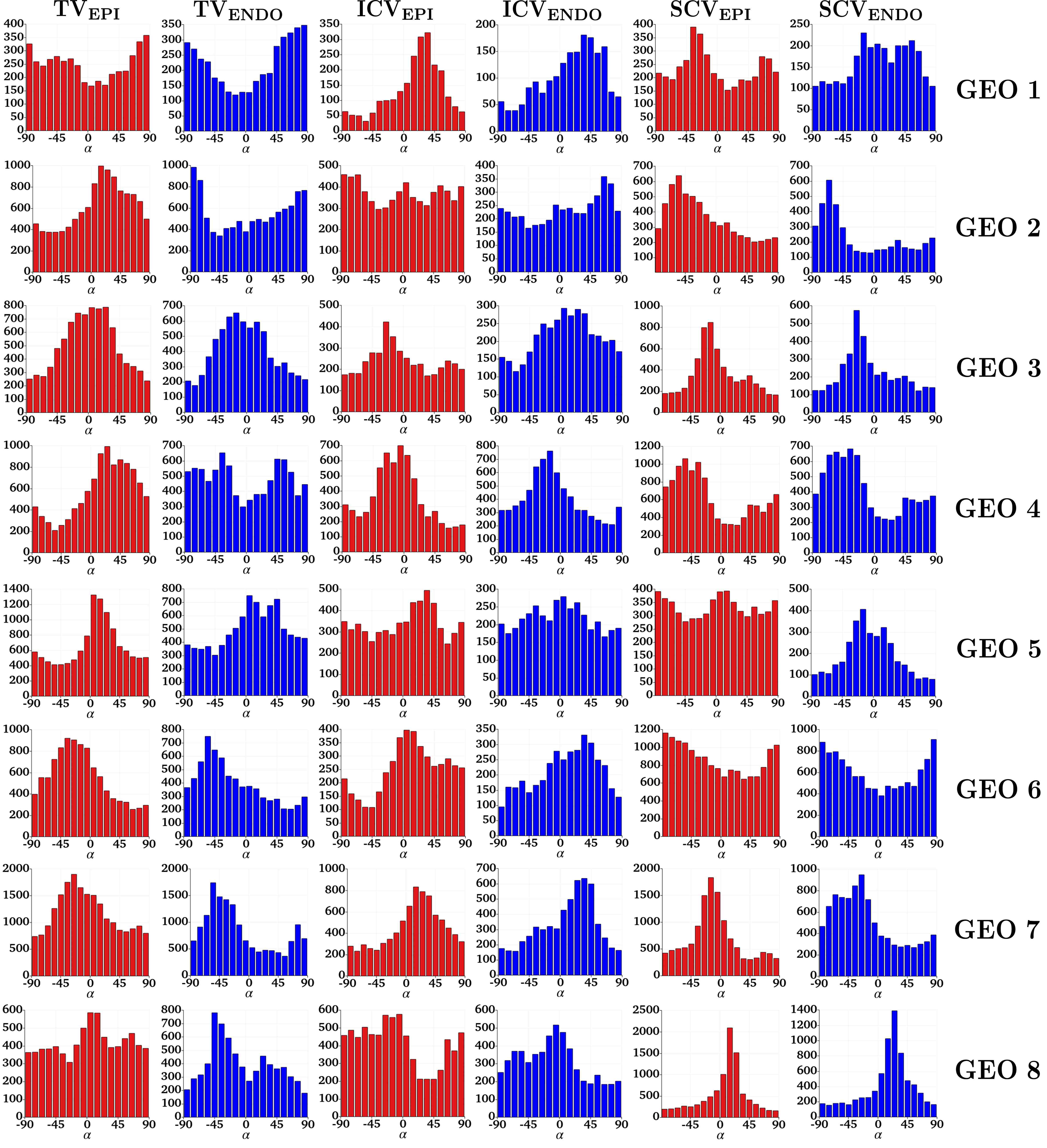}
    \caption{Histograms  displaying the measured alpha angle (with respect to LDRBM coordinate axis system) across the eight DTMRI geometries in RA bundles (TV, ICV, SCV) in the sub-epicardial (red) and sub-endocardial (blue) layers.}
    \label{fig:histo_RA_1}
\end{figure}
\begin{figure}[ht!]
    \centering
    \includegraphics[width=1.0\textwidth]{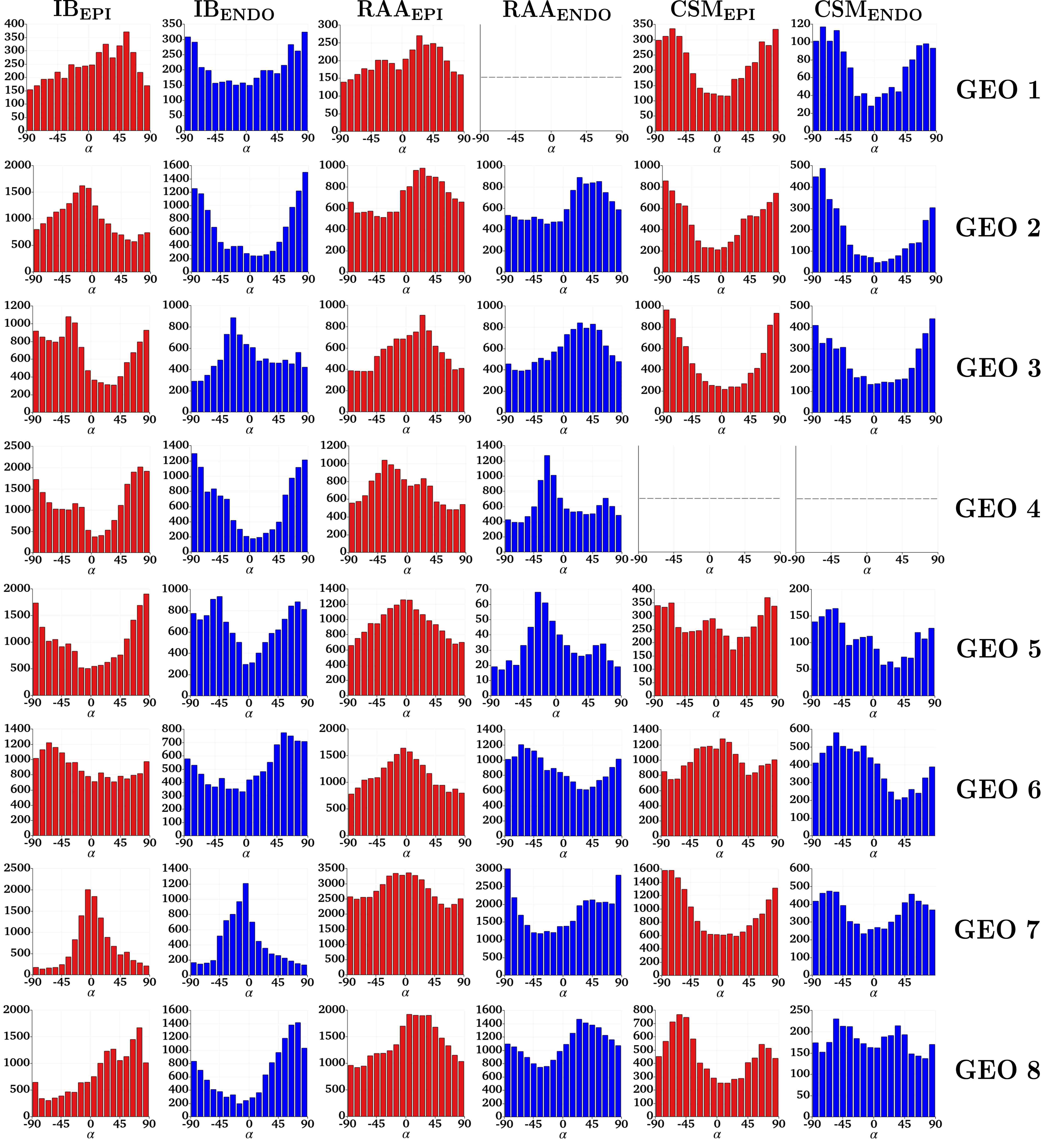}
    \caption{Histograms  displaying the measured alpha angle (with respect to LDRBM coordinate axis system) across the eight DTMRI geometries in RA bundles (IB, RAA, CSM) in the sub-epicardial (red) and sub-endocardial (blue) layers.}
    \label{fig:histo_RA_2}
\end{figure}
\begin{figure}[ht!]
    \centering
    \includegraphics[width=1.0\textwidth]{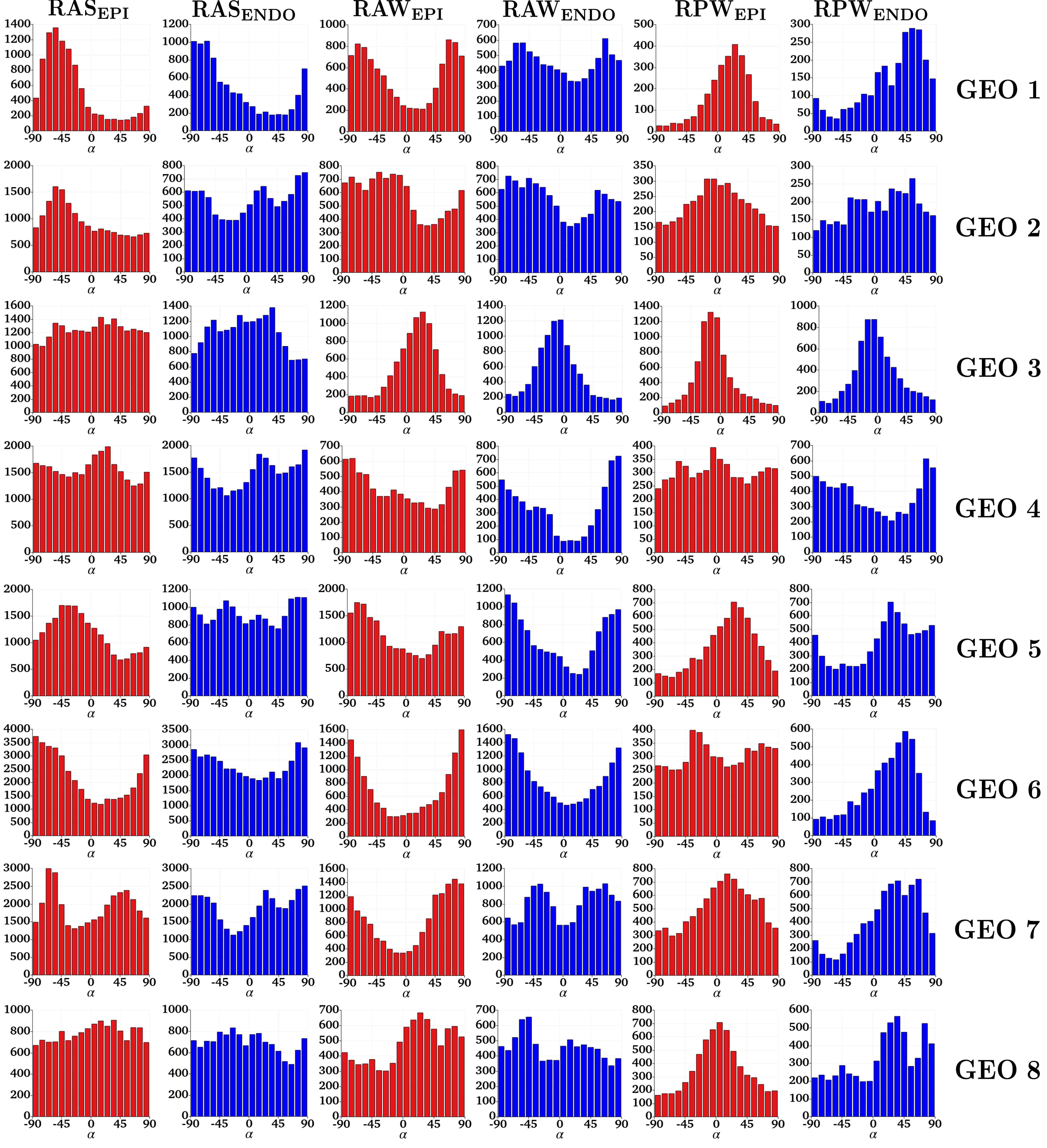}
    \caption{Histograms  displaying the measured alpha angle (with respect to LDRBM coordinate axis system) across the eight DTMRI geometries in RA bundles (RAS, RAW, RPW) in the sub-epicardial (red) and sub-endocardial (blue) layers.}
    \label{fig:histo_RA_3}
\end{figure}
\begin{figure}[ht!]
    \centering
    \includegraphics[width=1.0\textwidth]{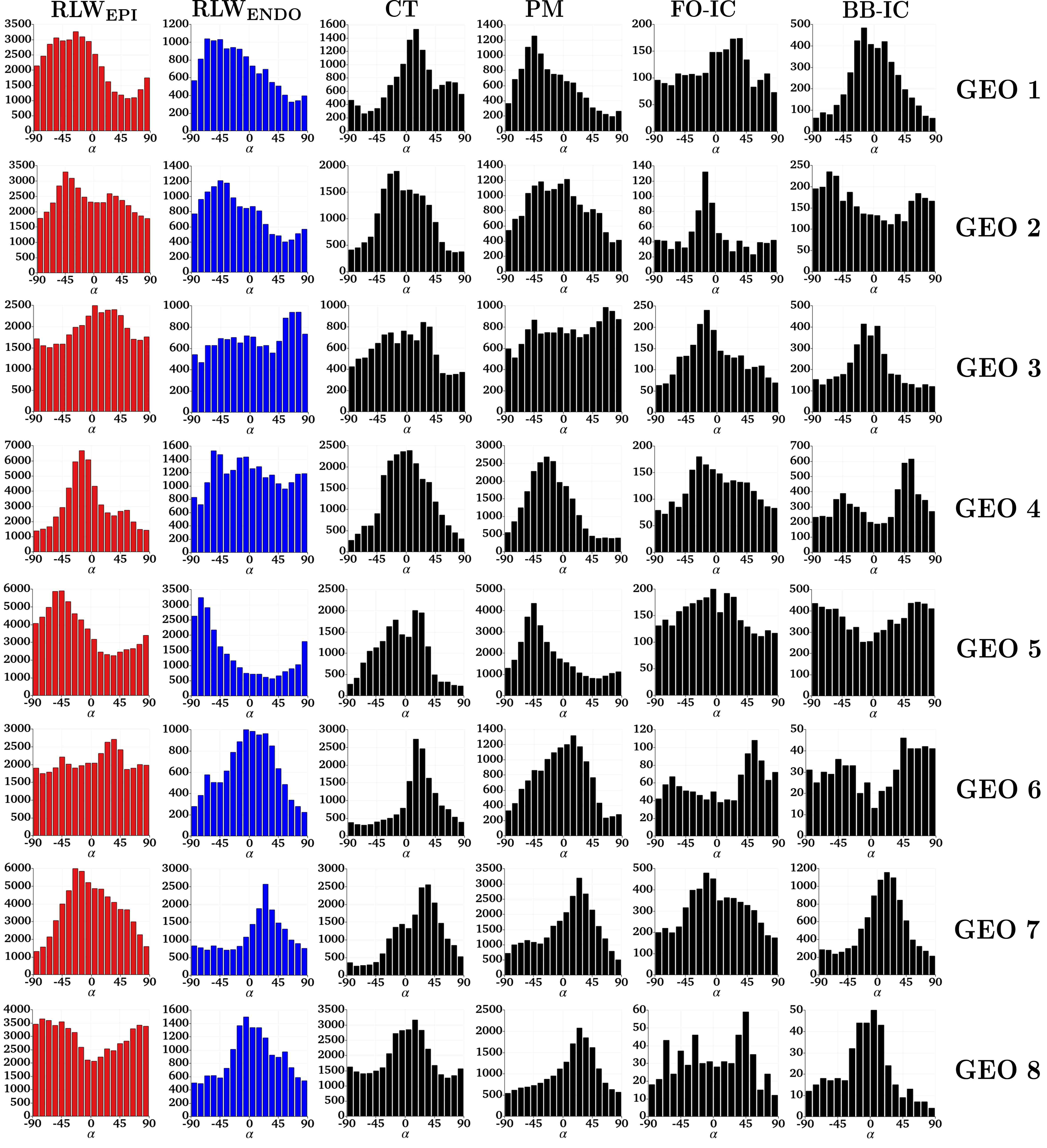}
    \caption{Histograms displaying the measured alpha angle (with respect to LDRBM coordinate axis system) across the eight DTMRI geometries in RLW bundle in the sub-epicardial (red) and sub-endocardial (blue) layers. CT, PM and inter-atrial connections (FO-IC, BB-IC) depicted in black.}
    \label{fig:histo_RA_4}
\end{figure}
\clearpage
\subsection{Circular statistics for directional data}
\label{supp:circular-statistic}
All the statistical data analysis performed to estimate the variability of both the regional bundle-dimension parameters and the atrial fiber angles inside each bundle were performed in \texttt{matlab} (\url{https://www.mathworks.com}). In particular, we employed the \texttt{CircStat} toolbox~\cite{berens2009circstat} for circular data analysis conducted for the fiber angle statistics, and the \texttt{CircHist} toolbox\footnote{\url{https://github.com/zifredder/CircHist}} to perform the polar angle histograms. 

We briefly summarize hereafter the main steps involved in evaluating the mean and standard deviation (SD) values employing the circular statistics. Moreover, we showcase a trivial example to highlight the importance of using such methodology when dealing with directional (angle) data. Due to its circular nature, directional data, like the fiber angles, cannot be analyzed with commonly used statistical techniques, that fail when applied to analyze circular/directional data~\cite{berens2009circstat,fisher1995statistical,jammalamadaka2001topics}. For further details regarding circular statistics, we refer to~\cite{berens2009circstat,fisher1995statistical,jammalamadaka2001topics}.

Considering $N$ directional observations (angles) $\alpha_i \in (0, 2\pi]$ as $(\alpha_1,\dots, \alpha_N)$, the steps to evaluate circular mean and SD, using circular statistics, are resumed hereafter:
\begin{itemize}
    \item[1.] Angles $\alpha_i$ are transformed to unit vectors $\mathbf{r}_i=\begin{pmatrix} \sin \alpha_i \\ \cos \alpha_i \end{pmatrix}$;
    \item[2.] Vectors are vectorial averaged $\overline{\mathbf{r}}=\frac{1}{N}\sum_{i=1}^N \mathbf{r}_i,$
    where $\overline{\mathbf{r}}$ represent the mean resultant vector;
    \item[3.] The mean angular direction is recovered using the four quadrant inverse tangent function \\
    $\overline{\alpha}=\text{atan2}\left( x_{\overline{\mathbf{r}}}, y_{\overline{\mathbf{r}}} \right),$
    where $x_{\overline{\mathbf{r}}}$ and $y_{\overline{\mathbf{r}}}$ are the $x$ and $y$ components of the mean resultant vector $\overline{\mathbf{r}}$; 
    \item[4.] The variance, defined as $\sigma=1-R$, with $R=||\overline{\mathbf{r}}||$, is linked to the length of the mean resultant vector~$R$. Hence, the closer $\sigma$ is to one, the more concentrated the data sample is around the mean direction. Finally, SD is evaluated using the formula
    $SD = \sqrt{2\sigma}$.
\end{itemize} 

\begin{figure}[t!]
    \centering
    \includegraphics[width=1.0\textwidth]{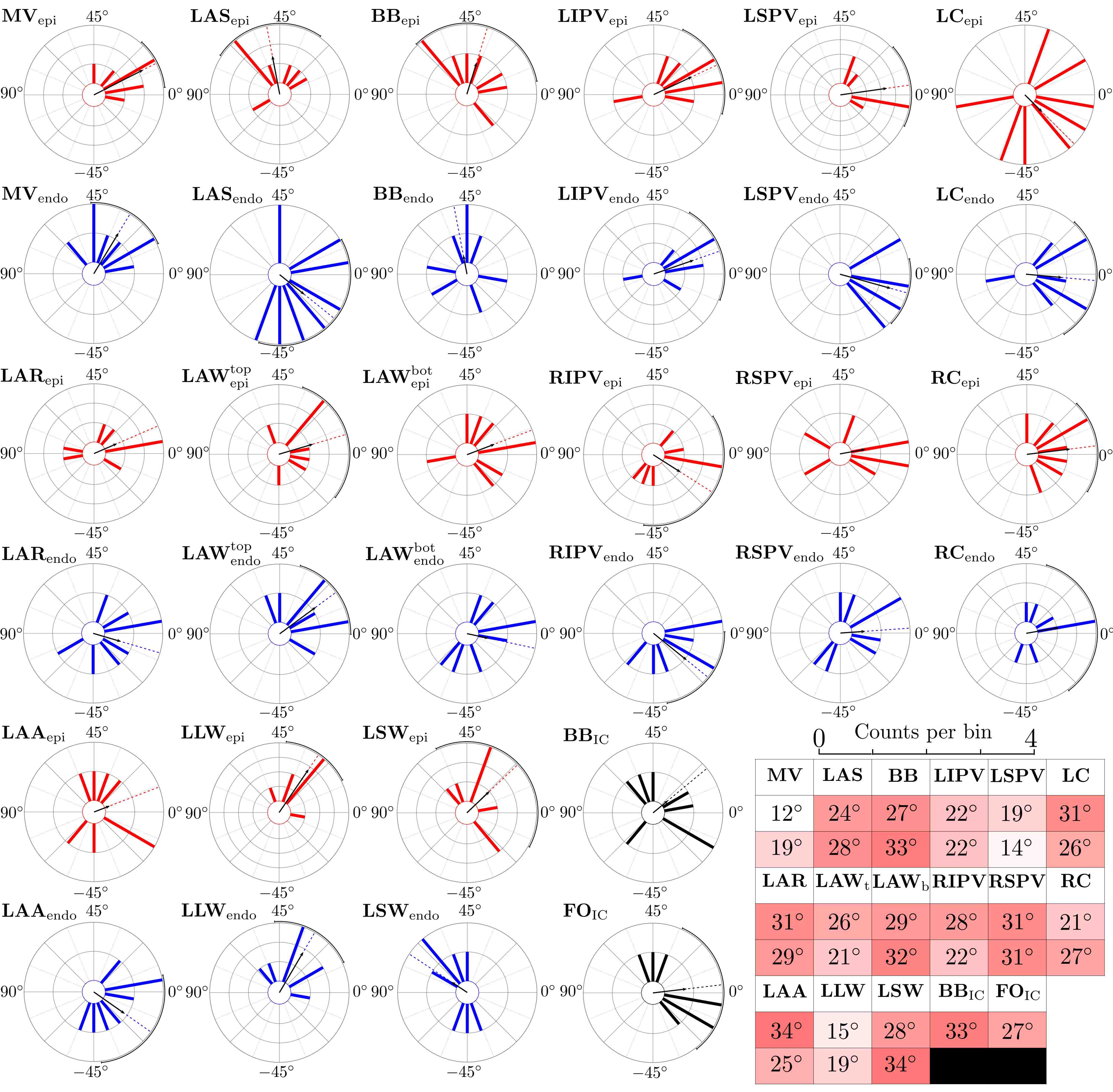}
    \caption{Circular histograms of dominant angle within LA bundles: sub-endocardium (blue), sub-epicardium (red) and inter-atrial connvections (black) across eight DTMRI dataset geometries. Mean angle represented by dotted line. Resultant mean vector depicted in black, with length indicative of angular variance. Corresponding SD values for each bundle shown in table at bottom right corner, with SD color-coded on a scale from minimal ($12^\circ$) to maximal ($37^\circ$).}
    \label{fig:LA_angle_bundles}
\end{figure}

\begin{figure}[t!]
    \centering
    \includegraphics[width=1.0\textwidth]{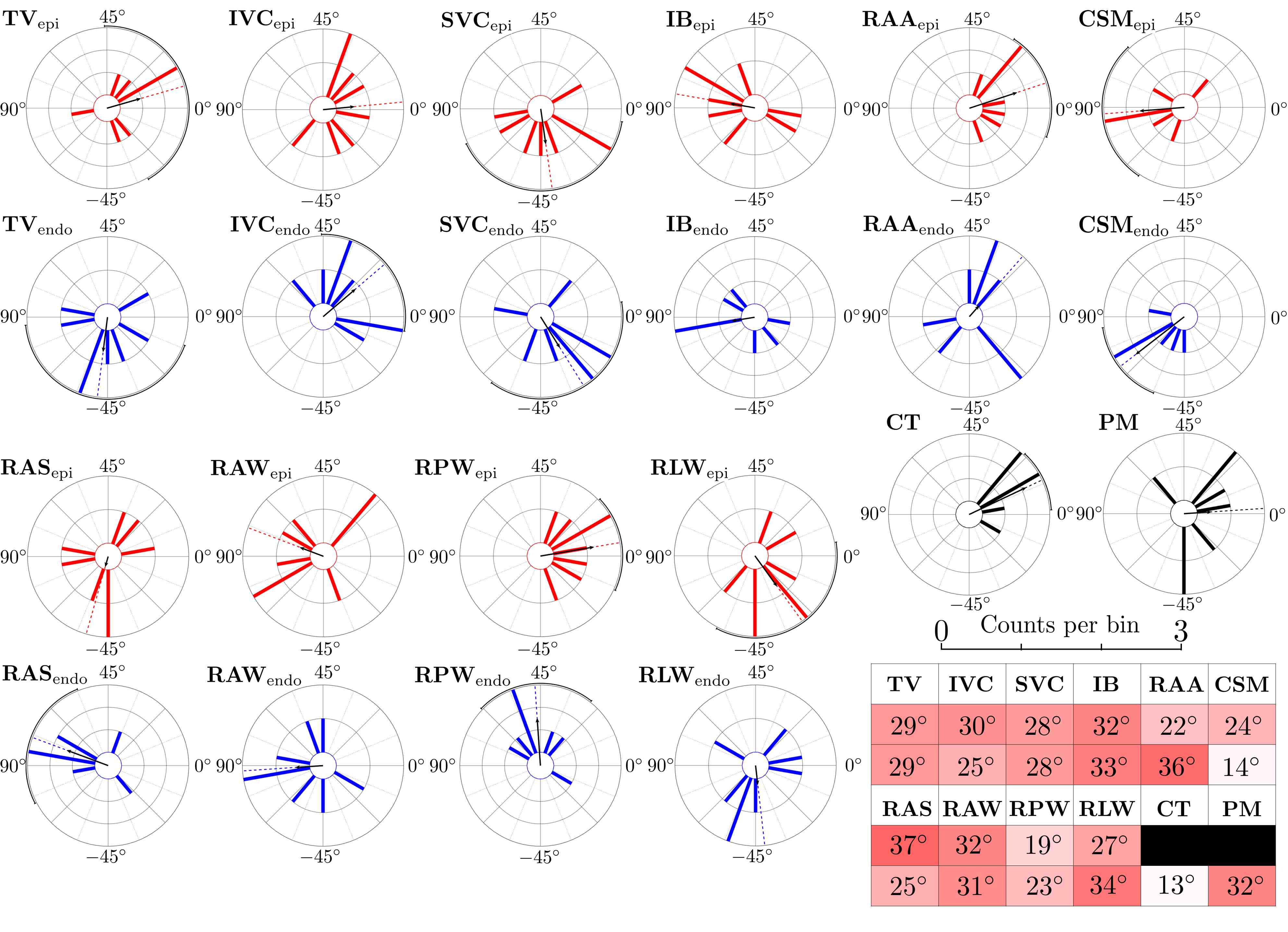}
    \caption{Circular histograms of dominant angle within RA bundles: sub-endocardium (blue), sub-epicardium (red) across eight DTMRI dataset geometries. Histograms of CT and PM bundles depicted in black. Mean angle represented by dotted line. Resultant mean vector depicted in black, with length indicative of angular SD. Corresponding SD values for each bundle shown in table at bottom right corner, with SD color-coded on a scale from minimal ($12^\circ$) to maximal ($37^\circ$).}
    \label{fig:RA_angle_bundles}
\end{figure}

Following the above procedure, we evaluated the mean and SD of the dominant fiber angle measurements for each bundle across the DTMRI dataset (see  Section~\ref{subsec:fiber_measurement}). The results are illustrated in Figures~\ref{fig:LA_angle_bundles} and~\ref{fig:RA_angle_bundles} and reported in Tables~\ref{tab:tau_alpha_LA} and~\ref{tab:tau_alpha_RA}. Notice that to account for the variability range of the fiber angle data, i.e. $\alpha \in (-\pi/2, \pi/2]$, we first mapped the angular values into $\beta \in (0, 2\pi]$, by means of $\beta=2\alpha + \pi$, then we recovered the mean angular vaule $\overline{\beta}$ and the related $\text{SD}_\beta$. Finally, we went back to the mean and SD angular values using $\overline{\alpha}=(\overline{\beta} - \pi)/2$ and $\text{SD}=\text{SD}_\beta /2$.  

To highlight that usual summary statistics, such as the sample mean and linear SD, cannot be used with angular values, let consider three angles $\beta = \left( 10^{\circ}, 30^{\circ}, 350^{\circ} \right)$. Classical statistical tools, for computing the mean value $\overline{\beta}_l$ and the $\text{SD}_{\beta,l}$ result in $130^\circ$ and $191^\circ$, respectively. However, all the data are roughly pointing towards $0^\circ$. Conversely, if we use circular statistics, we retrieve the correct value of $\overline{\beta}_{c}=10^\circ$ and $\text{SD}_{\beta,c}=16^\circ$.

\clearpage
\begin{table}[ht!]
    \centering
    \scalebox{0.7}{\begin{tabular}{ |c|c|c|c|c|c|c|c|c|c|c|c| } 
            \hline
            Type & $\alpha$ & Geo 1 & Geo 2 & Geo 3 & Geo 4 & Geo 5 & Geo 6 & Geo 7 & Geo 8 & Mean & SD \\
            \hline
            \hline
            \multirow{2}{*}{BIA}
            & $\alpha^{bb,ic}_{epi}$     & -15 & -65 & -15 &  55 &  65 &  45 &  15 &   5 & \textbf{20} & \gradientcell{33}{12}{37}{white}{red}{60}  \\ 
            & $\alpha^{fo,ic}_{epi}$     &  30 & -15 & -15 & -25 &  -5 &  55 & -10 &  45 & \textbf{3}  & \gradientcell{27}{12}{37}{white}{red}{60}  \\ 
            \hline
            \hline
            \multirow{30}{*}{LA}             
            & $\alpha^{mv}_{epi}$           &  15 &  15 &  -5 &   5 &  15 &  40 &  20 &   5 & \textbf{13}  & \gradientcell{12}{12}{37}{white}{red}{60}  \\ 
            & $\alpha^{mv}_{endo}$          &  65 &  20 &   5 &  45 &  15 &  45 &  35 &  10 & \textbf{29}  & \gradientcell{19}{12}{37}{white}{red}{60}  \\ 
            & $\alpha^{\ell ipv}_{epi}$     &  30 &   5 & -85 &   5 &  15 &  10 &  -5 &  20 & \textbf{12}  & \gradientcell{22}{12}{37}{white}{red}{60}  \\ 
            & $\alpha^{\ell ipv}_{endo}$    &  20 &  10 &  90 &  15 &   5 &  15 & -15 &   5 & \textbf{10}  & \gradientcell{22}{12}{37}{white}{red}{60}  \\ 
            & $\alpha^{\ell spv}_{epi}$     &  25 &  30 &  -5 &  35 & -10 & -15 & -10 & -10 & \textbf{4}   & \gradientcell{19}{12}{37}{white}{red}{60}  \\ 
            & $\alpha^{\ell spv}_{endo}$    &  15 &  15 &  -5 &  -5 & -15 & -25 & -15 & -25 & \textbf{-8}  & \gradientcell{14}{12}{37}{white}{red}{60}  \\ 
            & $\alpha^{\ell c}_{epi}$       & -20 &  -5 & -45 & -25 & -55 &  35 &  15 & -85 & \textbf{-22}  & \gradientcell{31}{12}{37}{white}{red}{60}  \\ 
            & $\alpha^{\ell c}_{endo}$      & -20 &  90 & -25 & -15 &  15 &  20 &  15 &  -5 & \textbf{-3}  & \gradientcell{26}{12}{37}{white}{red}{60}  \\ 
            & $\alpha^{ripv}_{epi}$         & -70 & -55 &   5 &  20 & -45 &  -5 &  -5 &  -5 & \textbf{-16} & \gradientcell{28}{12}{37}{white}{red}{60}  \\ 
            & $\alpha^{ripv}_{endo}$        & -65 & -15 &  -5 & -45 &   5 & -15 &   5 & -35 & \textbf{-19} & \gradientcell{22}{12}{37}{white}{red}{60}  \\ 
            & $\alpha^{rspv}_{epi}$         & -80 &   5 &  35 &  75 &  -5 &   5 &  -5 & -15 & \textbf{5}   & \gradientcell{31}{12}{37}{white}{red}{60}  \\ 
            & $\alpha^{rspv}_{endo}$        & -55 &  10 &  -5 & -65 & -15 &  10 &  40 &  35 & \textbf{2}   & \gradientcell{31}{12}{37}{white}{red}{60}  \\ 
            & $\alpha^{rc}_{epi}$           &  20 & -20 & -35 &  15 &  40 &  15 & -10 &   0 & \textbf{3}   & \gradientcell{21}{12}{37}{white}{red}{60}  \\ 
            & $\alpha^{rc}_{endo}$          &  45 &  35 & -55 &   5 & -35 &   0 &   5 &  10 & \textbf{5}   & \gradientcell{27}{12}{37}{white}{red}{60}  \\ 
            & $\alpha^{\ell aa}_{epi}$      &  30 &  40 & -15 & -45 & -15 &  20 &  55 & -70 & \textbf{10}  & \gradientcell{34}{12}{37}{white}{red}{60}  \\ 
            & $\alpha^{\ell aa}_{endo}$     &   0 & -25 &  -5 &  25 & -55 & -35 & -50 &   5 & \textbf{-17} & \gradientcell{25}{12}{37}{white}{red}{60}  \\ 
            & $\alpha^{\ell as}_{epi}$      & -75 &  60 &  65 &  65 &  35 &  25 &  50 &  15 & \textbf{50}  & \gradientcell{24}{12}{37}{white}{red}{60}  \\ 
            & $\alpha^{\ell as}_{endo}$     & -35 & -20 &  45 & -25 &  15 & -55 &   5 & -45 & \textbf{-19} & \gradientcell{28}{12}{37}{white}{red}{60}  \\ 
            & $\alpha^{\ell \ell w}_{epi}$  &  25 &  35 &  25 &  25 &  35 &  25 &  55 &  -5 & \textbf{28}  & \gradientcell{15}{12}{37}{white}{red}{60}  \\ 
            & $\alpha^{\ell \ell w}_{endo}$ &  60 &  10 &  35 &  35 &  35 &  15 &  50 &  -5 & \textbf{30}  & \gradientcell{19}{12}{37}{white}{red}{60}  \\ 
            & $\alpha^{\ell sw}_{epi}$      & -25 &  50 &  30 &  65 & -25 &  30 &  30 &   0 & \textbf{22}  & \gradientcell{28}{12}{37}{white}{red}{60}  \\ 
            & $\alpha^{\ell sw}_{endo}$     & -55 &  60 &  45 &  50 & -35 & -45 &  70 &  65 & \textbf{73}  & \gradientcell{34}{12}{37}{white}{red}{60}  \\ 
            & $\alpha^{\ell ar}_{epi}$      &  25 &  90 &  85 &  30 & -15 &   0 &   0 &   0 & \textbf{12}  & \gradientcell{31}{12}{37}{white}{red}{60}  \\ 
            & $\alpha^{\ell ar}_{endo}$     &  35 &   5 & -75 & -45 & -15 &  15 &   5 & -30 & \textbf{-8}  & \gradientcell{29}{12}{37}{white}{red}{60}  \\ 
            & $\alpha^{\ell aw,t}_{epi}$    &  20 &   0 & -50 &  20 &  -5 &  25 & -15 &  55 & \textbf{8}   & \gradientcell{26}{12}{37}{white}{red}{60}  \\ 
            & $\alpha^{\ell aw,t}_{endo}$   &  25 & -15 &   0 &  15 &  55 &  20 &   5 &  45 & \textbf{18}  & \gradientcell{21}{12}{37}{white}{red}{60}  \\ 
            & $\alpha^{\ell aw,b}_{epi}$    &  20 & -15 & -30 &  40 &  35 & -85 &   5 &   5 & \textbf{10}  & \gradientcell{29}{12}{37}{white}{red}{60}  \\ 
            & $\alpha^{\ell aw,b}_{endo}$   &   5 & -60 & -70 &   5 &  25 & -40 &  -5 &  35 & \textbf{-6}  & \gradientcell{32}{12}{37}{white}{red}{60}  \\ 
            & $\alpha^{bb}_{epi}$           &  65 &  55 &   5 &  45 &  35 &  10 & -25 &  65 & \textbf{36}  & \gradientcell{27}{12}{37}{white}{red}{60}  \\ 
            & $\alpha^{bb}_{endo}$          &  50 & -35 & -80 &  35 &  45 & -10 &  45 &  85 & \textbf{50}  & \gradientcell{33}{12}{37}{white}{red}{60}  \\ 
            \hline          
    \end{tabular}}
    \caption{Dominant angle $\alpha$ values measured for inter-atrial connections (BIA) and left atrial bundles (LA) in the sub-epicardium ($epi$) and sub-endocardium ($endo$) across the eight DTMRI geometries. Circular Mean and Standard Deviation (SD) values are provided, with SD color-coded on a scale from minimal ($12^\circ$) to maximal ($37^\circ$). All values are expressed in degrees.}    
    \label{tab:tau_alpha_LA}
\end{table}
\begin{table}[ht!]
    \centering
    \scalebox{0.7}{\begin{tabular}{ |c|c|c|c|c|c|c|c|c|c|c|c| } 
            \hline
            Type & $\alpha$ & Geo 1 & Geo 2 & Geo 3 & Geo 4 & Geo 5 & Geo 6 & Geo 7 & Geo 8 & Mean & SD \\
            \hline
            \hline
            \multirow{22}{*}{RA}
            & $\alpha^{tv}_{epi}$       &  90 &  15 &  25 &  30 &  10 & -35 & -25 &  10 & \textbf{8}    & \gradientcell{29}{12}{37}{white}{red}{60}  \\ 
            & $\alpha^{tv}_{endo}$      &  85 & -85 & -15 & -35 &  10 & -55 & -55 & -45 & \textbf{-49}  & \gradientcell{29}{12}{37}{white}{red}{60}  \\ 
            & $\alpha^{ivc}_{epi}$      &  35 & -70 & -25 &  -5 &  35 &  10 &  20 & -35 & \textbf{3}    & \gradientcell{30}{12}{37}{white}{red}{60}  \\ 
            & $\alpha^{ivc}_{endo}$     &  40 &  65 &  25 & -20 & -10 &  30 &  35 &  -5 & \textbf{20}   & \gradientcell{25}{12}{37}{white}{red}{60}  \\ 
            & $\alpha^{svc}_{epi}$      & -35 & -50 & -15 & -55 & -85 & -75 & -15 &  15 & \textbf{-41}  & \gradientcell{28}{12}{37}{white}{red}{60}  \\ 
            & $\alpha^{svc}_{endo}$     & -15 & -60 & -25 & -40 & -15 &  85 & -30 &  25 & \textbf{-29}  & \gradientcell{28}{12}{37}{white}{red}{60}  \\ 
            & $\alpha^{raa}_{epi}$      &  25 &  25 &  25 & -35 & -15 &  -5 &   5 &  35 & \textbf{9}    & \gradientcell{22}{12}{37}{white}{red}{60}  \\ 
            & $\alpha^{raa}_{endo}$     &  25 &  35 &  45 & -25 & -25 & -65 &  90 &  30 & \textbf{24}   & \gradientcell{36}{12}{37}{white}{red}{60}  \\ 
            & $\alpha^{csm}_{epi}$      &  90 & -85 & -80 & --- &  75 &  25 & -85 & -55 & \textbf{-87}  & \gradientcell{24}{12}{37}{white}{red}{60}  \\ 
            & $\alpha^{csm}_{endo}$     & -75 & -75 &  80 & --- & -60 & -65 & -75 & -50 & \textbf{-71}  & \gradientcell{14}{12}{37}{white}{red}{60}  \\ 
            & $\alpha^{ras}_{epi}$      & -55 & -50 &  80 &  20 & -45 & -85 &   0 &  35 & \textbf{-53}  & \gradientcell{37}{12}{37}{white}{red}{60}  \\ 
            & $\alpha^{ras}_{endo}$     & -85 &  80 &  35 &  85 &  75 &  75 &  80 & -30 & \textbf{80}   & \gradientcell{25}{12}{37}{white}{red}{60}  \\ 
            & $\alpha^{rpw}_{epi}$      &  25 &  10 & -15 &  -5 &  30 & -35 &  15 &   5 & \textbf{5}    & \gradientcell{19}{12}{37}{white}{red}{60}  \\ 
            & $\alpha^{rpw}_{endo}$     &  55 &  55 & -15 &  75 &  25 &  50 &  60 &  35 & \textbf{47}   & \gradientcell{23}{12}{37}{white}{red}{60}  \\ 
            & $\alpha^{raw}_{epi}$      &  65 & -35 &  25 & -80 & -75 &  90 &  75 &  25 & \textbf{80}   & \gradientcell{32}{12}{37}{white}{red}{60}  \\ 
            & $\alpha^{raw}_{endo}$     &  45 & -70 & -15 &  80 & -85 & -85 &  55 & -45 & \textbf{-88}  & \gradientcell{31}{12}{37}{white}{red}{60}  \\ 
            & $\alpha^{ib}_{epi}$       &  55 & -15 & -85 &  75 &  85 & -65 &  -5 &  75 & \textbf{85}   & \gradientcell{32}{12}{37}{white}{red}{60}  \\ 
            & $\alpha^{ib}_{endo}$      &  90 &  90 & -25 &  90 & -45 &  60 &  -5 &  70 & \textbf{-85}  & \gradientcell{33}{12}{37}{white}{red}{60}  \\ 
            & $\alpha^{r\ell w}_{epi}$  & -25 & -45 &  10 & -15 & -45 &  35 & -30 & -70 & \textbf{-27}  & \gradientcell{27}{12}{37}{white}{red}{60}  \\ 
            & $\alpha^{r\ell w}_{endo}$ & -60 & -45 &  70 & -55 & -70 &   0 &  25 &  -5 & \textbf{-42}  & \gradientcell{34}{12}{37}{white}{red}{60}  \\ 
            & $\alpha^{ct}_{endo}$      &  15 & -20 &  25 &   5 &  20 &  15 &  20 &  15 & \textbf{13}   & \gradientcell{13}{12}{37}{white}{red}{60}  \\ 
            & $\alpha^{pm}_{endo}$      & -45 &   5 &  65 & -25 & -45 &  15 &  25 &  25 & \textbf{2}    & \gradientcell{32}{12}{37}{white}{red}{60}  \\ 
            \hline          
    \end{tabular}}
    \caption{Dominant angle $\alpha$ values measured for right atrial bundles (RA) in the sub-epicardium ($epi$) and sub-endocardium ($endo$) across the eight DTMRI geometries. Circular Mean and Standard Deviation (SD) values are provided, with SD color-coded on a scale from minimal ($12^\circ$) to maximal ($37^\circ$). All values are expressed in degrees.}
    \label{tab:tau_alpha_RA}
\end{table}
\clearpage
\begin{figure}[t!]
    \centering
    \includegraphics[width=1.0\textwidth]{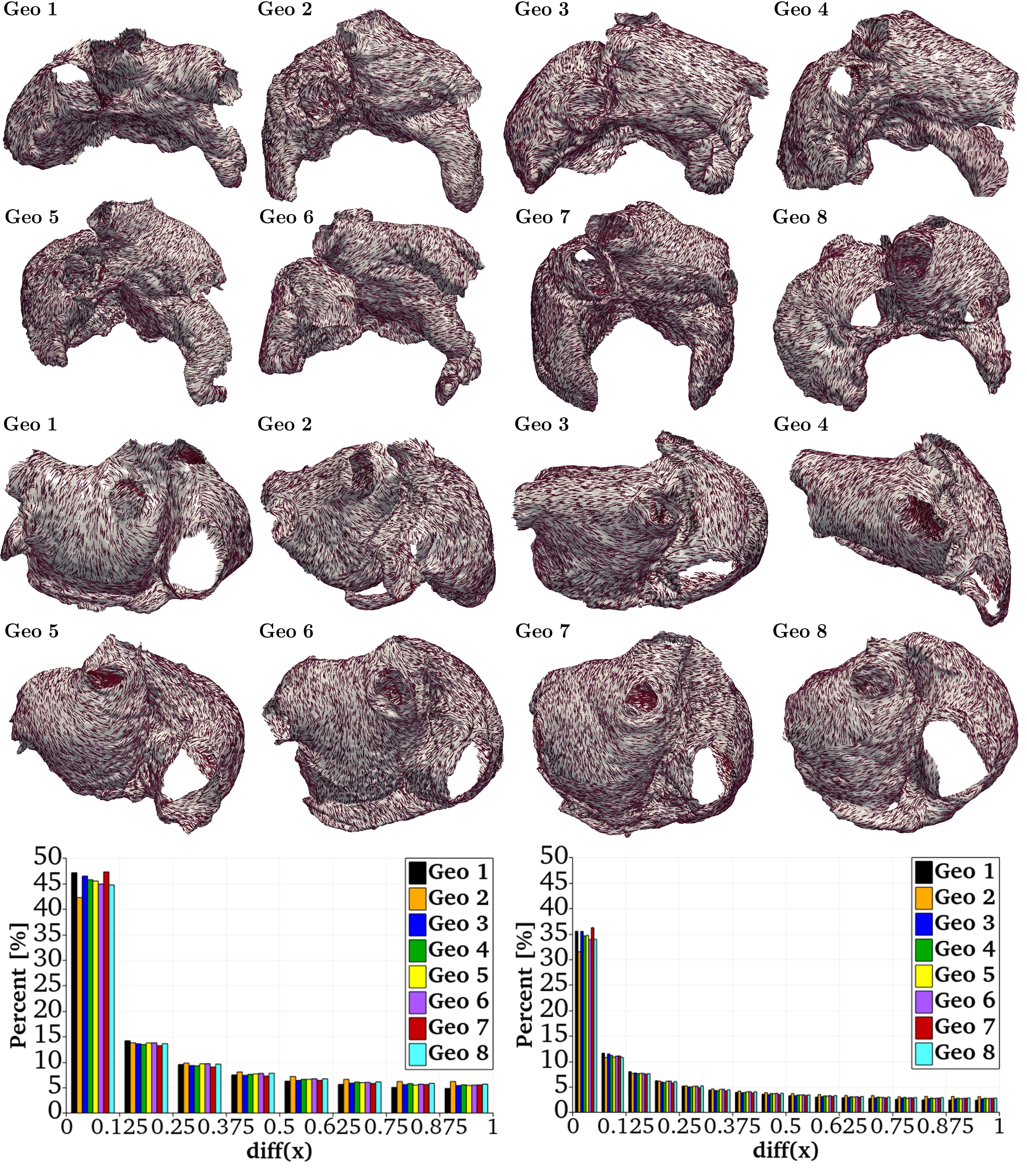}
    \caption{DTMRI epicardial fiber architecture recreated by the LDRBM assigning in each bundle the dominant angle retrieved by the LDRBM-DTMRI measuring procedure. Histograms showing the fiber orientation agreement distribution between LDRBM and DTMRI fibers: distribution of the $\text{diff}$ function with percentage values per histogram bin of 0.125 (left) and 0.0625 (right). The function diff, computed as $\text{diff}(\boldsymbol{x}) = 1 - |\boldsymbol{f}_\text{DTMRI}(\boldsymbol{x}) \cdot \boldsymbol{f}_\text{LDRBM}(\boldsymbol{x})|$, highlights the differences between LDRBM and DTMRI fibers. The percentages close to $\text{diff} \sim 0$ indicate the amount of fibers in good agreement with respect to the total fiber orientations.}
    \label{fig:fibers_LDRBM_epi}
\end{figure}
\begin{figure}[t!]
    \centering
    \includegraphics[width=1.0\textwidth]{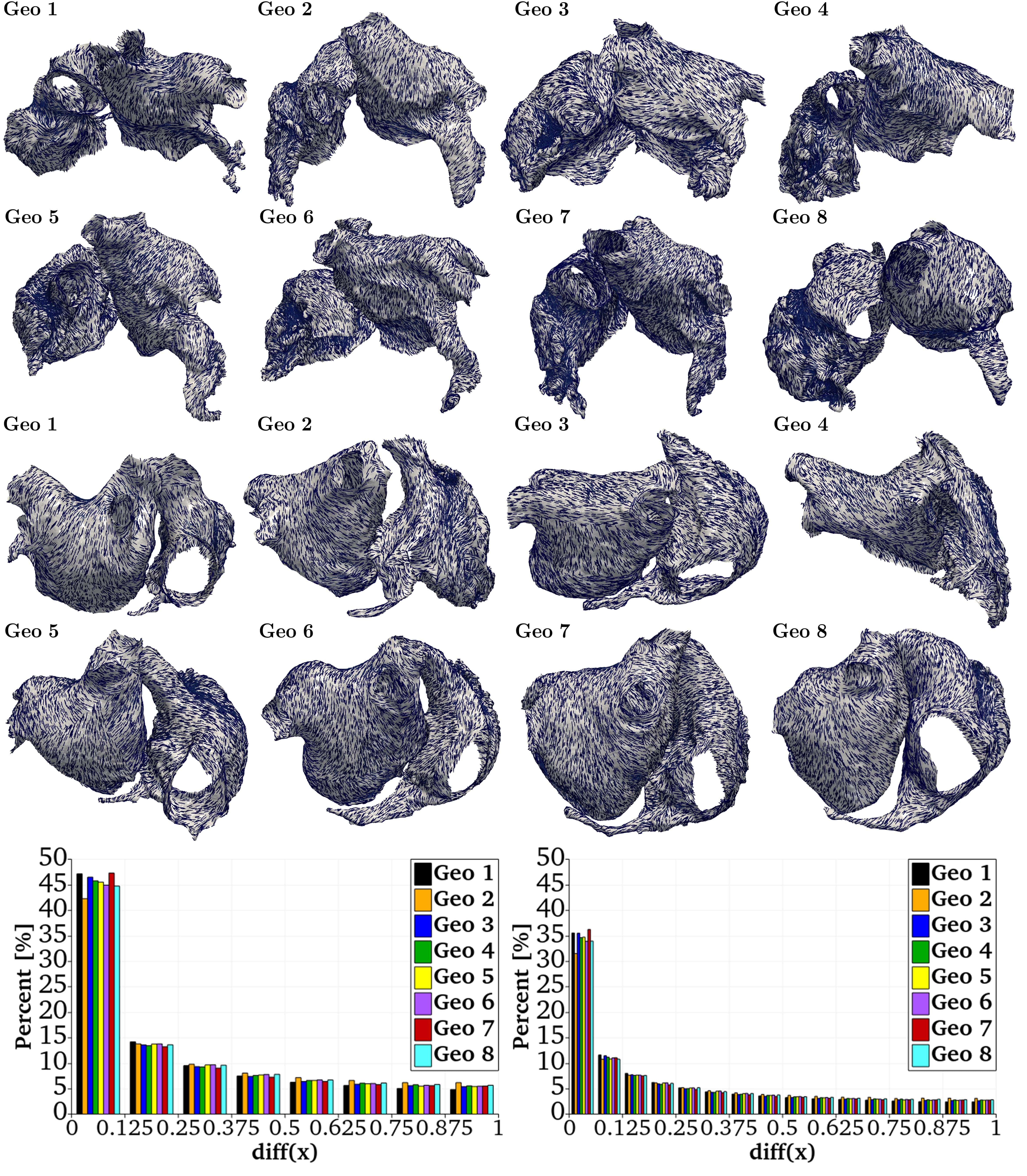}
    \caption{DTMRI endocardial fiber architecture recreated by the LDRBM assigning in each bundle the dominant angle retrieved by the LDRBM-DTMRI measuring procedure. Histograms showing the fiber orientation agreement distribution between LDRBM and DTMRI fibers: distribution of the $\text{diff}$ function with percentage values per histogram bin of 0.125 (left) and 0.0625 (right). The function diff, computed as $\text{diff}(\boldsymbol{x}) = 1 - |\boldsymbol{f}_\text{DTMRI}(\boldsymbol{x}) \cdot \boldsymbol{f}_\text{LDRBM}(\boldsymbol{x})|$, highlights the differences between LDRBM and DTMRI fibers.  The percentages close to $\text{diff} \sim 0$ indicate the amount of fibers in good agreement with respect to the total fiber orientations.}
    \label{fig:fibers_LDRBM_endo}
\end{figure}
\clearpage
\begin{figure}[t!]
    \centering
    \includegraphics[width=1\textwidth]{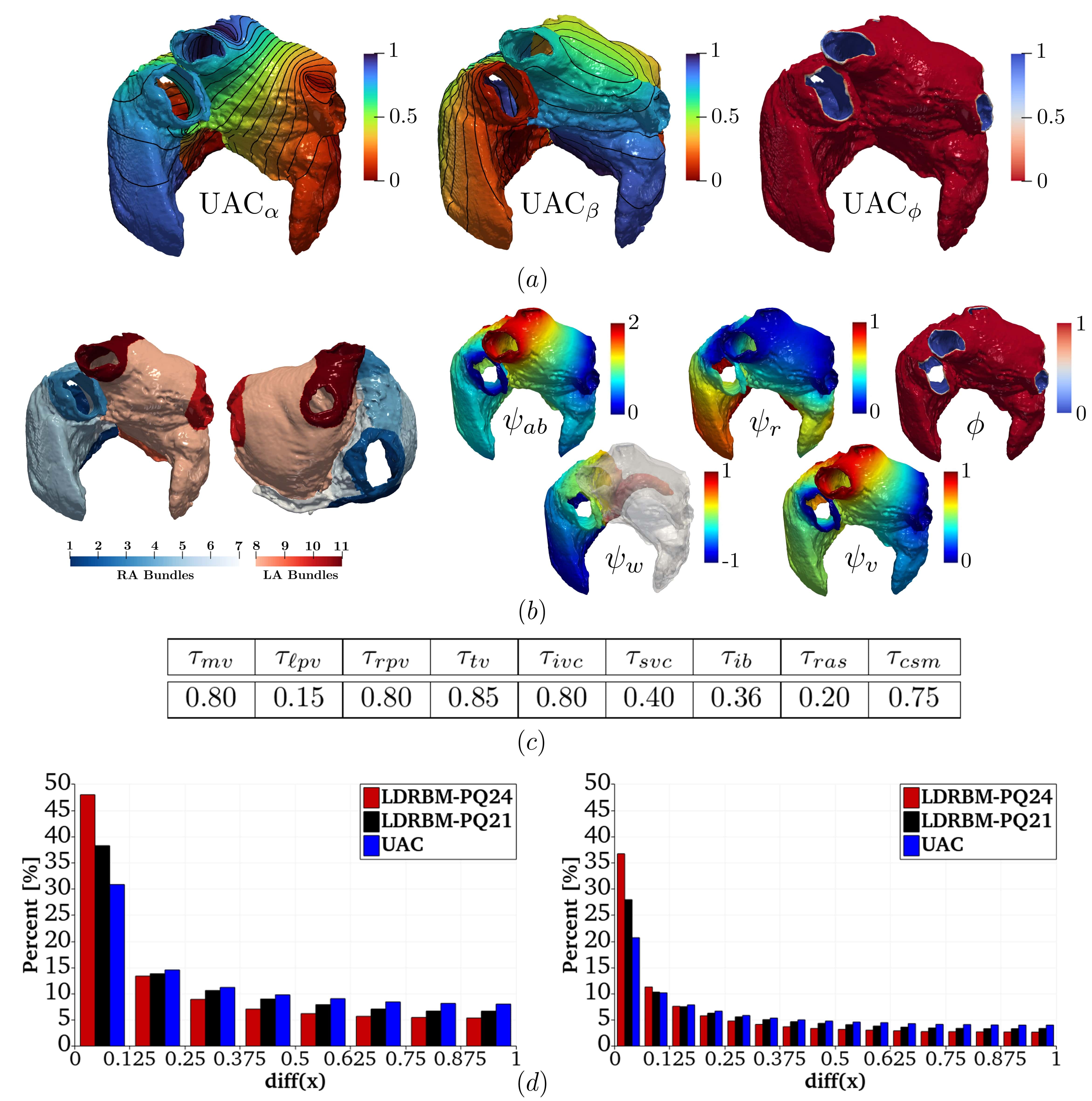}
    \caption{(a) Universal Atrial Coordinates (UAC) showing the three atrial coordinates. For each LA and RA surface, two atrial coordinates were defined: for RA a lateral-septal TV coordinate ($\text{UAC}_\alpha$) and an IVC-SVC coordinate ($\text{UAC}_\beta$); for the LA, a septal-lateral coordinate ($\text{UAC}_\alpha$) and a posterior-anterior coordinate ($\text{UAC}_\beta$); $\text{UAC}_\phi$ represents the transmural coordinate from the endocardium to the epicardium. Refer to \cite{roney2023constructing,roney2019universal} for furhter details.
    (b) Definition of the transmural distance $\phi$ and several intra-atrial distances $\psi_i$ for LDRBM$-$PQ21. These are obtained by solving a Laplace–Dirichlet problems with proper Dirichlet boundary conditions: $\psi_{ab}$ with three different boundary data prescribed on the atrial appendage, the rings of the caval veins and TV for RA, the pulmonary veins and MV ring for LA; $\psi_v$ represents the distance between the caval veins for RA and among the pulmonary veins for LA; $\psi_r$ stands for the distance between TV and the top of RA, and between MV and the union of the pulmonary veins rings for LA. Moreover, for RA $\psi_w$ is the distance between the free and the septum walls. Refer to \cite{piersanti2021modeling,piersanti2021phd} for furhter details.
    (c) LDRBM-PQ21 bundle parameters for LA and RA used in the comparison presented in Section~\ref{subsec:models_comparison}.
    (d)~Fiber disparities of different fiber models (LDRBM-PQ24, UAC~\cite{roney2021constructing} and LDRBM-PQ21~\cite{piersanti2021modeling}) relative to DTMRI data. The function diff, computed as as $\text{diff}(\boldsymbol{x}) = 1- |\boldsymbol{f}_\text{DTMRI}(\boldsymbol{x}) \cdot \boldsymbol{f}_i(\boldsymbol{x})|$, with $i=\text{LDRBM-PQ24}, \text{UAC}, \text{LDRBM-PQ21}$. Histograms showing the distribution of the $\text{diff}$ function with percentage values per histogram bin of 0.125 (left) and 0.0625 (right). The percentages close to $\text{diff} \sim 0$ indicate the amount of fibers in good agreement with respect to the DTMRI fiber orientations.}
    \label{fig:UAC_PQ21}
\end{figure}
\clearpage
    \end{appendices}
    
    \bibliographystyle{elsarticle-num}
    \bibliography{references}

\end{document}